\documentclass{aa}
\usepackage{lastpage}
\usepackage{graphicx}
\usepackage{natbib}
\usepackage{url}
\usepackage{color}
\usepackage{xcolor}
\usepackage{tikz}
\usepackage{txfonts}
\usepackage{orcidlink}
\usepackage{academicons}
\usepackage{xcolor}
\usepackage{graphicx}
\usepackage{epstopdf}
\usepackage{amssymb}
\usepackage{latexsym}
\usepackage{tikz-cd}
\newcommand{\aba}{\textcolor{blue}}
\usepackage{tikz}
\usetikzlibrary{matrix, fit, positioning,arrows.meta,arrows, shapes.geometric,backgrounds,shadows}
\tikzstyle{start}=[circle, rounded corners, minimum width=0.8cm, minimum height= 0cm, text centered, draw=black!3, fill=gray!5, text width = 1cm]
\tikzstyle{title}=[rectangle, rounded corners, minimum width=0.7cm, minimum height= 1cm, text centered, draw=black!5, fill=gray!2, text width = 3cm,drop shadow]
\tikzstyle{ref_title}=[rectangle, rounded corners, minimum width=0.8cm, minimum height= 1cm, text centered, draw=black!50, fill=gray!2, text width = 2cm,drop shadow]
\tikzstyle{start_title}=[rectangle, rounded corners, minimum width=0.8cm, minimum height= 1cm, text centered, draw=black!50, fill=gray, text width = 2cm,drop shadow]
\tikzstyle{estart}=[circle, rounded corners, minimum width=0.1cm, minimum height= 0.1cm, text centered, draw=black!0, fill=red!0, text width = 1cm]

\tikzstyle{startstop}=[rectangle, rounded corners, minimum width=2.8cm, minimum height= 1cm, text centered, draw=black!6, fill=red!2, text width = 2.5cm,drop shadow]
\tikzstyle{box_s}=[rectangle, rounded corners, minimum width=1.5cm, minimum height= 1cm, text centered, draw=black!6, fill=red!2, text width = 2cm,drop shadow]
\tikzstyle{box_s_k}=[rectangle, rounded corners, minimum width=1.5cm, minimum height= 1cm, text centered, draw=black!6, fill=red!15, text width = 2cm,drop shadow]

\tikzstyle{startstop_nb}=[rectangle, rounded corners, minimum width=2.8cm, minimum height=0.8cm, text centered, draw=black, fill=gray!0, text width = 2cm]
\tikzstyle{io}=[trapezium, trapezium left angle = 70, trapezium right angle =110,minimum width=2cm, minimum height= 1cm, text centered, draw=black, fill=blue!0,drop shadow]
\tikzstyle{process}=[rectangle, rounded corners,minimum width=3cm, minimum height= 0.7cm, text centered, draw=black, fill=orange!0,drop shadow]

\tikzstyle{process_TITLE}=[rectangle, rounded corners, minimum width=1cm, minimum height= 0.4cm, text centered, draw=black, fill=gray!0,drop shadow]

\tikzstyle{nomen}=[rectangle, rounded corners, minimum width=4cm, minimum height= 1cm,  draw=black!10, fill=yellow!2]
\tikzstyle{process_short}=[rectangle, rounded corners,minimum width=1cm, minimum height=0.5cm, text centered, draw=black, fill=orange!0,drop shadow]
\tikzstyle{decision}=[diamond, rounded corners,minimum width=2cm, minimum height= 0.7cm, text centered, draw=black!15, fill=green!0,drop shadow]
\tikzstyle{question}=[diamond,rounded corners, minimum width=2.5cm, minimum height= 0.7cm, text centered, draw=black, fill=blue!0,drop shadow]
\tikzstyle{decisionfwd}=[diamond, rounded corners,minimum width=2cm, minimum height= 0.7cm, text centered, draw=black, fill=green!0, drop shadow]
\tikzstyle{decision_short}=[diamond, rounded corners,minimum width=1cm, minimum height= 1cm, text centered, draw=black!2, fill=green!5, drop shadow]

\tikzstyle{bam}=[diamond, rounded corners,minimum width=1cm, minimum height= 1cm, text centered, draw=black!15, fill=green!5, drop shadow]

\tikzstyle{decision_short_input}=[diamond, minimum width=0.8cm, minimum height=0.05cm, text centered, draw=black, fill=gray!0, drop shadow]
\tikzstyle{arrow}=[thick, ->, >= stealth,line width=0.05cm]
\tikzstyle{arrow_ic}=[thick, ->, >= stealth,line width=0.05cm, draw=black!40]

\tikzstyle{arrow_d}=[dotted, ->, >= stealth]
\tikzstyle{arrow_pi}=[dashed, ->, >= stealth, line width=0.03cm]
\tikzstyle{arrow_nl}=[dashed, -, >= stealth, line width=0.03cm]
\tikzstyle{arrow_nls}=[thick, -, >= stealth, line width=0.03cm]
\tikzstyle{arrow_nlsa}=[thick, ->, >= stealth, line width=0.03cm,color=blue]
\tikzstyle{arrow_new}=[dotted, ->,>= stealth, color=red, line width = 0.05cm]
\tikzstyle{arrow_prep}=[dashed, ->,>= stealth, color=blue!50, line width = 0.04cm]
\tikzstyle{arrow_cat}=[dotted, ->,>= stealth, color=green!80, line width = 0.05cm]
\tikzstyle{arrow_prep_nh}=[dashed, -,>= stealth, color=blue, line width = 0.04cm]
\newcommand{\be}{\begin{equation}}
\newcommand{\dd}{{\rm d}}
\newcommand{\lp}{\left(}
\newcommand{\rp}{\right)}

\newcommand{\vr}{\textbf{r}}

\newcommand{\uv}{\textbf{u}}
\newcommand{\vv}{\textbf{v}}

\newcommand{\ee}{\end{equation}}

\newcommand{\vk}{\textbf{k}}

\newcommand{\vq}{\textbf{q}}

\usepackage[normalem]{ulem}

\begin{document}

%%%%%%%%%%%%%%%%%%%%%%%%%%%%%%%%%%%%%%%%%%%%%%%%%%
%%%%%%%%%%%%%%%%%%%%%%%%%%%%%%%%%%%%%%%%%%%%%%%%%
\title{ \textbf{DESI} mock challenge: Halo and galaxy catalogues with the bias assignment method}
\titlerunning{A. Balaguera-Antolínez et al.}
\author{Andr\'es Balaguera-Antol\'{\i}nez\inst{1,2} \orcidlink{0000-0001-5028-3035}, Francisco-Shu Kitaura\inst{1,2} \orcidlink{0000-0002-9994-759X}, Shadab Alam\inst{3,4}\orcidlink{0000-0002-3757-6359}, Chia-Hsun Chuang\inst{5}\orcidlink{0000-0002-3882-078X}, Yu Yu\inst{6}, Ginevra Favole\inst{1,2}\orcidlink{0000-0002-8218-563X}, Francesco Sinigaglia\inst{1,2,7} and 
  Cheng Zhao\inst{8},      David Brooks\inst{9}, Axel de la Macorra\inst{10}, Andreu Font-Ribera\inst{11}, Satya Gontcho A Gontcho\inst{12}, Klaus Honscheid\inst{13,14},  Robert Kehoe\inst{15}, Aron Meisner\inst{16}, Ramon Miquel\inst{11,17}, Gregory Tarlé\inst{18}, Mariana Vargas-Maga\~na\inst{10} and Zhimin Zhou\inst{19}}
\institute{
 Instituto de Astrof\'{\i}sica de Canarias, s/n, E-38205, La Laguna, Tenerife, Spain \email{balaguera@iac.es} \and
 Departamento de Astrof\'{\i}sica, Universidad de La Laguna, E-38206, La Laguna, Tenerife, Spain \and
Institue for Astronomy, Royal Observatory, University of Edinburgh, UK. \and
  Tata Institue of Fundamental Research, Homi Bhabha Road, Mumbai 400005, India \and
  Kavli Institute for Particle Astrophysics and Cosmology, Stanford University, 452 Lomita Mall, Stanford, CA, 94305, USA \and
  Shanghai Jiao Tong University, 800 Dongchuan Road, Shanghai, 200233, China\and
Department of Physics and Astronomy, Università degli Studi di Padova, Vicolo dell’Osservatorio 3, Padova, Italy \and
  Institute of Physics. Laboratory of astrophysics. École Polytechnique Fédérale de Lausanne, CH-1290. Switzerland\and
  Department of Physics \& Astronomy, University College London, Gower Street, London, WC1E 6BT, UK \and
  Instituto de F\'{\i}sica, Universidad Nacional Aut\'{o}noma de M\'{e}xico, Cd. de M\'{e}xico C.P. 04510, M\'{e}xico\and
  Institut de F\'{i}sica d’Altes Energies, The Barcelona Institute of Science and Technology, Campus UAB, 08193 Bellaterra Barcelona, Spain \and
  Lawrence Berkeley National Laboratory, 1 Cyclotron Road, Berkeley, CA 94720, USA \and
  Department of Physics, The Ohio State University, 191 West Woodruff Avenue, Columbus, OH 43210, USA\and
  Center for Cosmology and AstroParticle Physics, The Ohio State University, 191 West Woodruff Avenue, Columbus, OH 43210, USA  \and
  Department of Physics, Southern Methodist University, 3215 Daniel Avenue, Dallas, TX 75275, USA \and
   NSF's NOIRLab, 950 N. Cherry Ave., Tucson, AZ 85719, USA \and
  Instituci\'{o} Catalana de Recerca i Estudis Avan\c{c}ats, Passeig de Llu\'{\i}s Companys, 23, 08010 Barcelona, Spain \and
   University of Michigan, Ann Arbor, MI 48109, USA \and
   National Astronomical Observatories, Chinese Academy of Sciences, A20 Datun Rd., Chaoyang District, Beijing, 100012, P.R. China
 }
\authorrunning{Author}
\date{Received XXX; accepted ZZZ}
%%%%%%%%%%%%%%%%%%%%%%%%%%%%%%%%%%%%%%%%%%%%%%%%%%
%%%%%%%%%%%%%%%%%%%%%%%%%%%%%%%%%%%%%%%%%%%%%%%%%%
%\label{firstpage}
\abstract{
We present a novel approach to the construction of mock galaxy catalogues for large-scale structure analysis based on the distribution of dark matter halos obtained with effective bias models at the field level.}
{We aim to produce mock galaxy catalogues capable of generating accurate covariance matrices for a number of cosmological probes that are expected to be measured in current and forthcoming galaxy redshift surveys (e.g. two- and three-point statistics). The construction of the catalogues shown in this paper is part of a mock-comparison project within the Dark Energy Spectroscopic Instrument (DESI) collaboration.}
{We use the bias assignment method (\texttt{BAM}) to model the statistics of halo distribution through a learning algorithm using a few detailed $N$-body simulations, and approximated gravity solvers based on Lagrangian perturbation theory. We introduce cosmic-web-dependent corrections to modelling redshift-space distortions at the $N$-body level ---both in the halo and galaxy distributions---, as well as a multi-scale approach for accurate assignment of halo properties. Using specific models of halo occupation distributions to populate halos, we generate galaxy mocks with the expected number density and central-satellite fraction of emission-line galaxies, which are a key target of the DESI experiment.}
{\texttt{BAM} generates mock catalogues with per cent accuracy in a number of summary statistics, such as the abundance, the two- and three-point statistics of halo distributions, both in real and redshift space. In particular, the mock galaxy catalogues display $\sim 3\%-10\%$ accuracy in the multipoles of the power spectrum up to scales of $k\sim 0.4\,h^{-1}{\rm Mpc}$. We show that covariance matrices of two- and three-point statistics obtained with \texttt{BAM} display a similar structure to the reference simulation.}{\texttt{BAM} offers an efficient way to produce mock halo catalogues with accurate two- and three-point statistics, and is able to generate a variety of multi-tracer catalogues with precise covariance matrices of several cosmological probes. We discuss future developments of the algorithm towards mock production in DESI and other galaxy-redshift surveys.}
\maketitle

% 00000000000000000000000000000000000000000000000000000000000
\section{Introduction}
The cosmological volume spanned by the nearly $40$ million galaxies and quasars that are to be surveyed by the \emph{Dark Energy Spectroscopic Instrument} \citep[][]{2016arXiv161100036D} poses unprecedented challenges for both theoretical and numerical cosmology. 
  DESI is a robotic, fibre-fed, highly multiplexed spectroscopic surveyor operating on the Mayall 4 m telescope at Kitt Peak National Observatory \citep[][]{2022arXiv220510939A}. It can obtain simultaneous spectra of almost 5000 objects over a $\sim 3^{o}$ field \citep[][]{2016arXiv161100037D,2022arXiv220509014S,DESIcorrector2022}, and is currently conducting a five-year survey covering nearly one-third of the sky. DESI uses multiple supporting software pipelines and products, including significant imaging from the DESI Legacy Imaging Surveys \citep[][]{2017PASP..129f4101Z,2019AJ....157..168D,dr9} as well as an extensive spectroscopic reduction pipeline \citep[][]{spec2022}, a template-fitting pipeline to derive classifications and redshifts for each targeted source \citep[][]{redrock2022}, a pipeline aimed to assign fibres to targets \citep[][]{fba}, a pipeline to tile the survey and to plan and optimise observations as the campaign progresses \citep[][]{ops}, and a pipeline to select targets for spectroscopic follow-up \citep[][]{2022arXiv220808518M}. The DESI target selection relies on the public Legacy Surveys \citep{2019AJ....157..168D}, with preliminary target selection details published for the MWS \citep[][]{2020RNAAS...4..188A}, the LRGs sample  \citep{2020RNAAS...4..181Z}, BGS \citep{2020RNAAS...4..187R},  ELGs \citep{Raichoor_2020}, and QSOs \citep{2020RNAAS...4..179Y}. Specific target selection approaches for DESI are varied and extensive. In particular, it is important that we mention the work describing the DESI Survey Validation (SV) phase \citep{sv}, two papers describing the process through which truth tables were produced via
  visual inspection of target spectra acquired during the SV phase and how these are used to inform target selection for the DESI Main Survey \cite{2022arXiv220808517A,2022arXiv220808516L}, as well as a series of papers describing the selection of DESI bright-time and dark-time science targets \citep[MWS,][]{2022arXiv220808514C}; \citep[BGS,][]{2022arXiv220808512H}; \citep[LRG,][]{2022arXiv220808515Z}; \citep[ELG,][]{2022arXiv220808513R}; \citep[QSO][]{2022arXiv220808511C}. The Early DESI Data Release \citep{dr} and the Siena Galaxy Atlas \citep[SGA,][]{sga} are forecast for 2023. 
  
  The precision of the measurements of the statistical properties of the spatial distribution and weak-lensing signals to be obtained from such an unprecedented number of tracers  will shed light on the most intriguing features of the standard cosmological model; for example, the nature of dark energy \citep[e.g.][]{DESI, 2016arXiv161100036D} and primordial non-gaussianities \citep[see e.g.][]{2019arXiv190101581V,Alam_2021}. The accomplishment of these goals depends heavily on access to precise and accurate covariance matrices for the statistical analysis of several cosmological probes, such as clustering, weak-lensing signals, redshift-space distortions, and baryon acoustic oscillations \cite[e.g.][]{2013PhRvD..88f3537D,2013MNRAS.432.1928T,2014MNRAS.439.2531P,2015MNRAS.454.4326P,2016MNRAS.457..993P,2017MNRAS.472.4935H, 2018A&A...615A...1L,2019MNRAS.487.2701O}.

This paper is part of a mock challenge within the DESI collaboration \citep[see e.g.][]{2018ApJS..236...43G,2021arXiv211209138G,2022arXiv220206074D} which is  designed to establish a road map towards the construction of mock galaxy catalogues with per cent accuracy and precision in a number of statistical properties of the spatial distribution of galaxies. In particular, this article describes the application of a calibrated approach to producing mock catalogues, the so-called bias assignment method \citep[\texttt{BAM};][]{2019MNRAS.483L..58B}. 

While a number of predictive methods (such as \texttt{PThalos} \citep{2002MNRAS.329..629S,2013MNRAS.428.1036M}, \texttt{MoLUSC} \citep{2008ApJ...678..569S}, \texttt{PINOCCHIO} \citep[][]{2002MNRAS.331..587M,2013MNRAS.433.2389M, 2017JCAP...07..050M,2022JCAP...11..002B}, \texttt{COLA} \citep[][]{2013JCAP...06..036T,2016MNRAS.459.2118K,2018MNRAS.473.3051I}, \texttt{QPM} \citep{2014MNRAS.437.2594W}, \texttt{L-PICCOLA} \citep[][]{2015A&C....12..109H}, \texttt{FastPM} \citep{2016MNRAS.463.2273F}, \texttt{Peak-Patch} \citep[][]{1996ApJS..103....1B,1996ApJS..103...41B,1996ApJS..103...63B,2019MNRAS.483.2236S}, \texttt{CoVMOS} \cite[][]{2020A&A...633A..26B,2022arXiv221113590B}) and analytical approaches (such as the log-normal approaches; e.g. \citet[][]{1991MNRAS.248....1C,2016MNRAS.459.3693X, 2017JCAP...10..003A}) have been designed as an alternative to highly detailed \citep[e.g.][]{2005MNRAS.364.1105S, 2017ComAC...4....2P,  2020ApJS..248...32W,2021MNRAS.506.2871S}, albeit time-consuming, $N$-body simulations \citep[see e.g.][for a review of some of these methods]{2016Galax...4...53M}, the need for a step forward in terms of precision in the assessment of covariance matrices of cosmological observables has motivated the emergence of a new branch of techniques called {calibrated methods}; examples are  \texttt{HALOGEN} \cite[][]{2015MNRAS.450.1856A}, \texttt{EZmocks} \citep[][]{2015MNRAS.446.2621C}, and \texttt{PATCHY} \cite[][]{2014MNRAS.439L..21K}. The two latter methods were successfully used to generate accurate mock galaxy catalogues for clustering analysis \citep[see e.g.][]{2016MNRAS.456.4156K,2021MNRAS.503.1149Z} and were tested for analysis of forthcoming missions \citep[see][for a likelihood comparison of some of these methods]{2015MNRAS.452..686C,2019MNRAS.482.1786L,2018arXiv180609497B,2018MNRAS.tmp.2818C}. 

% *****************************************************************
% *****************************************************************
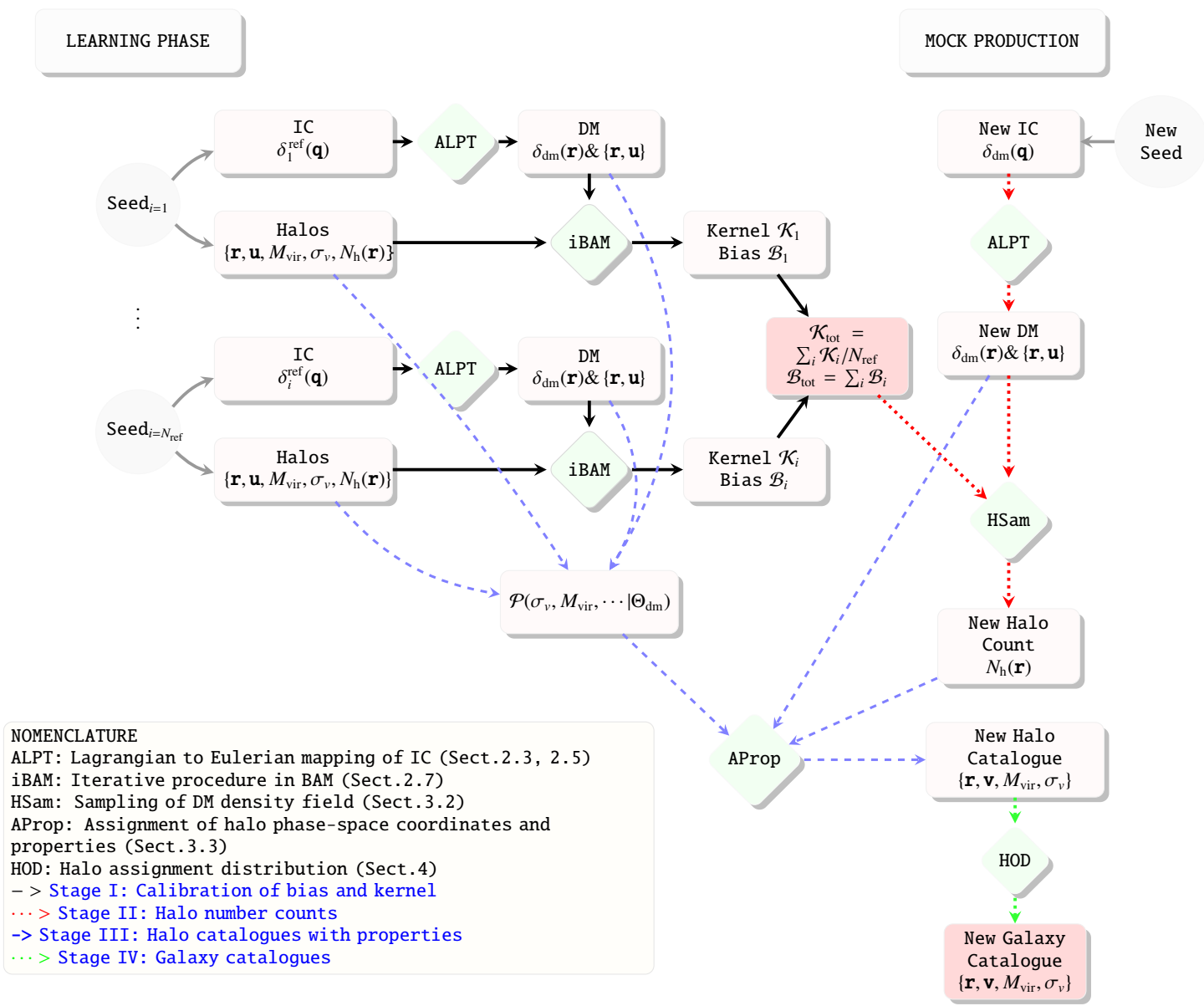
\begin{figure*}
\begin{tikzpicture}[font=\ttfamily\small,node distance=1.6cm]
\hspace{0cm}
\node(title)[title, xshift=5cm]{LEARNING PHASE};

\node(start)[start, below of = title, yshift=-1cm]{Seed$_{i=1}$};
\node(start1)[startstop, right of= start, xshift=1cm, yshift=1cm]{
IC\\ $\delta_{1}^{\,\rm ref}(\vq)$};
\draw[arrow_ic](start) to [bend left =15](start1);
\node(HREF)[startstop, below of = start1, xshift=0cm]{Halos\\ $ \{\vr,\uv,M_{\rm vir},\sigma_{v}, N_{\rm h}(\vr)$\}};
\draw[arrow_ic](start)to [bend right =15](HREF);
\node(DS1)[decision_short, right of = start1, xshift=0.8cm]{ALPT};
\draw[arrow](start1)--(DS1);
\node(DM)[box_s, right of = DS1, xshift=0.5cm]{DM \\ $\delta_{\rm dm}(\vr) \&\, \{\vr,\uv\}$};
\draw[arrow](DS1)--(DM);
\node(BAM)[bam, below of  = DM, xshift=0cm]{iBAM};
\draw[arrow](HREF)--(BAM);
\draw[arrow](DM)--(BAM);

\node(KERNEL)[box_s, right of = BAM, xshift=1cm]{ Kernel $\mathcal{K}_{1}$\\ Bias $\mathcal{B}_{1}$ };
\draw[arrow](BAM)--(KERNEL);

\node(starti)[estart, below of =start, yshift=-0.1cm]{\textbf{$\vdots$}};

\node(starti)[start, below of =start, yshift=-2cm]{Seed$_{i=N_{\rm ref}}$};
\node(start1i)[startstop, right of = starti,  xshift=1cm, yshift=1cm]{
IC\\ $\delta_{i}^{\rm ref}(\vq)$};
\draw[arrow_ic](starti)to [bend left =15](start1i);
\node(HREFi)[startstop, below of = start1i, xshift=0cm]{Halos\\ $\{\vr,\uv,M_{\rm vir},\sigma_{v},N_{\rm h}(\vr)\}$};
\draw[arrow_ic](starti)to [bend right =15](HREFi);
\node(DS1i)[decision_short, right of = start1i, xshift=0.8cm]{ALPT};
\draw[arrow](start1i)--(DS1i);
\node(DMi)[box_s, right of = DS1i, xshift=0.5cm]{DM \\ $\delta_{\rm dm}(\vr) \&\, \{\vr,\uv\}$};
\draw[arrow](DS1i)--(DMi);
\node(BAMi)[bam, below of  = DMi, xshift=0cm]{iBAM};
\draw[arrow](HREFi)--(BAMi);
\draw[arrow](DMi)--(BAMi);
\node(KERNELi)[box_s, right of = BAMi, xshift=1cm]{ Kernel $\mathcal{K}_{i}$ \\ Bias $\mathcal{B}_{i}$};
\draw[arrow](BAMi)--(KERNELi);

\node(KERNELs)[box_s_k, below of = KERNEL, xshift=1.3cm,yshift=-0.2cm]{ $\mathcal{K}_{\rm tot} = \sum_{i}\mathcal{K}_{i}/N_{\rm ref}$ \\ $\mathcal{B}_{\rm tot}= \sum_{i}\mathcal{B}_{i}$ };

\draw[arrow](KERNEL)--(KERNELs);
\draw[arrow](KERNELi)--(KERNELs);
\node(start_title)[title, left of = title,  xshift=15.2cm]{MOCK PRODUCTION};
\node(ICn)[box_s, right of = DM, xshift=5cm]{New IC \\ $\delta_{\rm dm}(\vq)$};
\node(startN)[start, right of = ICn, xshift=0.8cm]{New Seed};
\draw[arrow_ic](startN) --(ICn);

\node(DSn)[decision_short, below  of = ICn, xshift=0cm]{ALPT};
\node(DMn)[box_s, below of = DSn, xshift=0cm]{New DM \\ 
$\delta_{\rm dm}(\vr) \&\, \{\vr,\uv\}$};
\node(B)[decision_short, below of = DMn, yshift=-1.2cm]{HSam};
\draw[arrow_new](ICn)--(DSn);
\draw[arrow_new](DSn)--(DMn);
\draw[arrow_new](KERNELs)--(B);
\node(Hn)[box_s, below of = DMn, yshift=-3.2cm]{New Halo Count \\ $N_{\rm h}(\vr)$};
\draw[arrow_new](DMn)--(B);
\draw[arrow_new](B)--(Hn);
\node(PROP)[startstop, below of = BAMi, yshift=-0.5cm]{$\mathcal{P}(\sigma_{v},M_{\rm vir},\cdots|\Theta_{\rm dm})$};
%test
\draw[arrow_prep](DM) to [bend left=32](PROP);
\draw[arrow_prep](DMi) to [bend left=32](PROP);
\draw[arrow_prep](HREF)to [bend left=5](PROP);
\draw[arrow_prep](HREFi)to [bend right=20](PROP);
\node(ASSIGN)[decision_short, below of = KERNELi, yshift=-3cm]{AProp};
\draw[arrow_prep](Hn)--(ASSIGN);
\draw[arrow_prep](DMn)--(ASSIGN);
\draw[arrow_prep](PROP)--(ASSIGN);
\node(NCAT)[startstop, right of = ASSIGN, xshift=2.5cm]{New Halo Catalogue \\ $\{\vr,\vv,M_{\rm vir},\sigma_{v}\}$} ;

\draw[arrow_prep](ASSIGN)--(NCAT);

\node(HOD)[decision_short, below of = NCAT, xshift=0cm]{HOD} ;

\node(GALn)[box_s_k, below of = HOD, xshift=0cm]{New Galaxy Catalogue \\ $\{\vr,\vv,M_{\rm vir},\sigma_{v}\}$} ;
\draw[arrow_cat](NCAT)--(HOD);

\draw[arrow_cat](HOD)--(GALn);

% ------------------------------- Nomenclature
\node(NOMEN)[nomen, below of = starti, xshift=3cm, yshift=-5cm, text width=10cm]
{
NOMENCLATURE\\
\texttt{ALPT}: Lagrangian to Eulerian mapping of IC (Sect.\ref{sec:ic}, \ref{sec:gravity})\\
\texttt{iBAM}: Iterative procedure in \texttt{BAM} (Sect.\ref{sec:iter})\\
\texttt{HSam}: Sampling of DM density field (Sect.\ref{sec:counts})\\
\texttt{AProp}: Assignment of halo phase-space coordinates and properties (Sect.\ref{sec:coords})\\
\texttt{HOD}: Halo assignment distribution (Sect.\ref{sec:gals})\\
\textbf{$->$} \aba{  Stage I:  Calibration of bias and kernel}\\
\textcolor{red}{$\cdots>$} \aba{Stage II: Halo number counts}\\
\textcolor{blue}{\textbf{--->}} \aba{ Stage III: Halo catalogues with properties}\\
\textcolor{green}{\textbf{$\cdots>$}} \aba{Stage IV: Galaxy catalogues}\\
};

\end{tikzpicture}
\caption{Flow-chart representing the different stages involved in the generation of mock galaxy catalogues with \texttt{BAM} described in \S\ref{sec:ingredients}. The process is mainly divided into two sections: learning phase and mock production. In the learning phase, a number of kernels and halo biasses are calibrated from different realisations of the reference simulation and are stacked to generate one version of kernel and bias used in the mock production phase. The different colours in the arrows indicate the different stages involved in the process (e.g. calibration, generation of independent halo number counts, assignment of halo properties, construction of galaxy catalogues).}  
  \label{fig:bam_scheme}
\end{figure*}

%=======================number============================================
% *****************************************************************
\begin{figure*}
\includegraphics[trim = .2cm 0cm 0cm 0cm ,clip=true, width=1.02\textwidth]{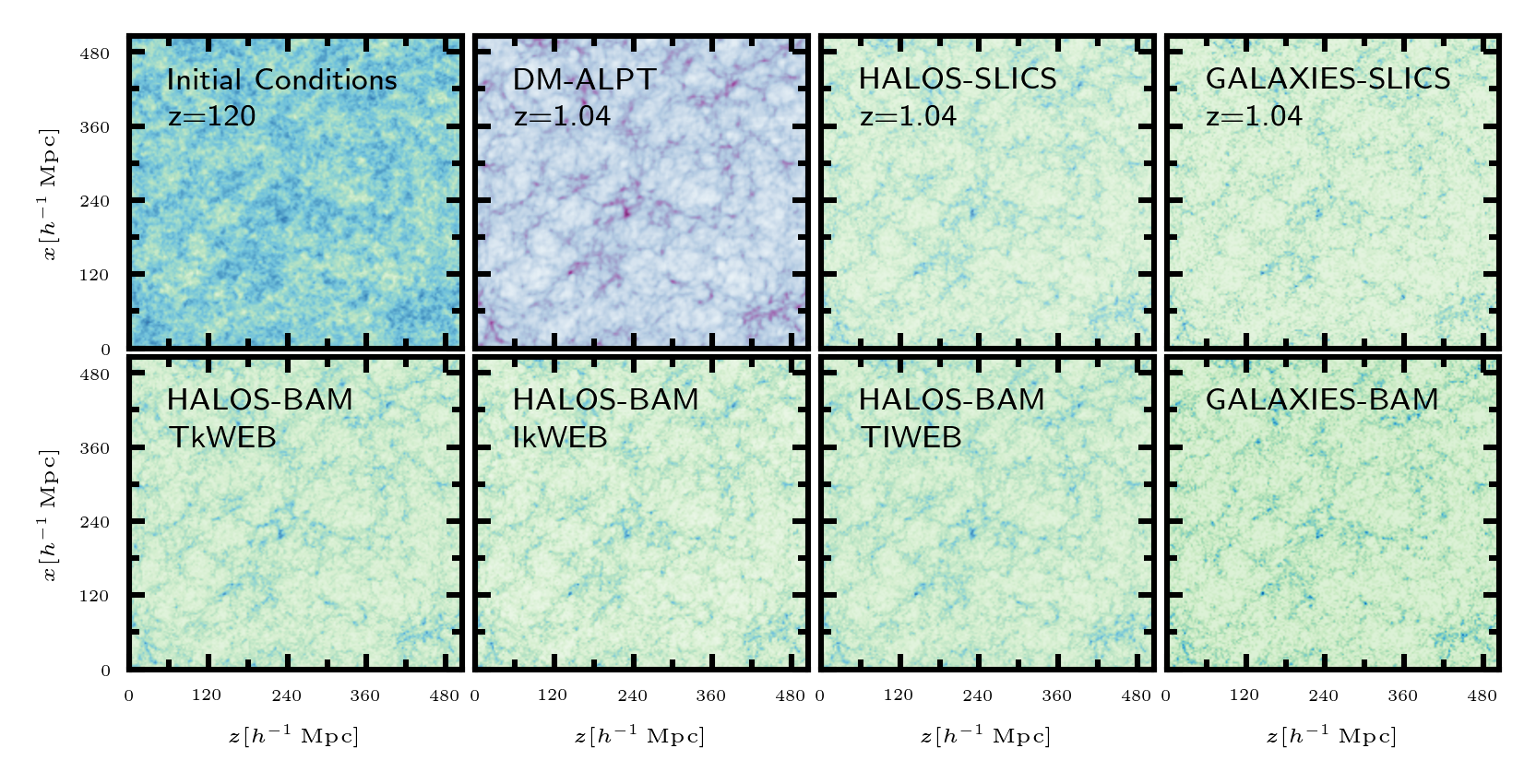}
\caption{Slices of $25$ Mpc $h^{-1}$ thick though different density fields involved in the calibration of the \texttt{BAM} products and its products. The bottom panel shows the reconstruction of the halo number density field using different models of halo bias (see \S\ref{sec:dmprops}). The rightmost column shows the galaxy density field from the reference  and from \texttt{BAM}, built by populating halo catalogues with a model of halo occupation distribution (see \S\ref{sec:gals}).}
\label{fig:slices_cal_bam}
\end{figure*}
% *****************************************************************

In recent years, machine-learning techniques have made an appearance in the cosmological scenario \citep[see e.g.][for a recent review]{2022arXiv220308056D} with a number of different goals and applications. Among others, these techniques have been used to learn the spatial distribution of dark matter tracers from a large number of detailed $N$-body simulations \cite[see e.g.][]{2021ApJ...915...71V,2021arXiv210702304K,2022arXiv220507898P}, generate corrections to the displacement field in Lagrangian perturbation theory \cite[e.g.][]{2019PNAS..11613825H}, increase the mass resolution of fast and computationally cheap simulations \citep[typically characterised by low mass resolutions, e.g.][]{2021PNAS..11822038L,2022arXiv220312669F}, learn the galaxy--dark matter connection from hydro-simulations \citep[e.g.][]{2019arXiv190205965Z}, and to provide a platform to obtain covariance matrices from fast and/or inaccurate sets of mocks \citep[see e.g.][]{2021MNRAS.503.1897C,2022arXiv220510881D}.  

\texttt{BAM} is the latest of a class of algorithms designed to produce mock galaxy catalogues. Its unique combination of physical content and learning scheme means that it can be regarded as both a calibrated method and a physically supervised machine-learning approach to the production of mock galaxy catalogues. The method represents a step forward in precision as well as efficiency, as it has been demonstrated to provide covariance matrices of the halo power spectrum with per cent accuracy (with respect to an $N$-body simulation) and at low cost in terms of computing time as well as in the number of training sets \citep[][]{2020MNRAS.491.2565B}. BAM has also been shown to be potentially useful to generating mock catalogues for Lyman-alpha and quasars by learning from hydro-dynamic simulations \citep[][]{2021ApJ...921...66S,2022ApJ...927..230S}.

In this work, we present a methodology that uses \texttt{BAM} to generate ensembles of halo catalogues with phase-space coordinates. The methodology implemented for constructing halo catalogues presented here improves on previous approaches by including more precise recipes for the peculiar velocities and intrinsic properties of halos (such as virial mass and velocity dispersion). 
Building on the set of halo catalogues, we implement a halo occupation distribution (HOD) framework \citep[e.g.][]{2002ApJ...576L.105C,2002PhR...372....1C,2002ApJ...575..587B,2004ApJ...609...35K} to populate these halos with galaxies, and in particular emission line galaxies (ELGs), which are a key target of the DESI galaxy redshift survey \citep[e.g.][]{Raichoor_2020}. The strategy that we envisage for \texttt{BAM} allows us to implement more approaches to populate dark matter halos with galaxies, such as the sub-halo abundance matching (SHAM) \cite[see e.g.][]{2004MNRAS.353..189V,2004ApJ...609...35K,2006ApJ...647..201C,2016MNRAS.461.3421F}, and to generate galaxy cluster catalogues \citep[see e.g.][]{2009MNRAS.395.1185C,2012MNRAS.425.2244B} based on different halo properties \citep[][]{2016MNRAS.460.2552H,2018ARA&A..56..435W}. This provides high flexibility at the time of producing mock catalogues containing a number of different dark matter tracers with the same underlying dark matter density field. This is optimal for multi-tracer analyses \cite[see e.g.][]{2012PhRvD..86j3513H,2013MNRAS.432..318A,2016MNRAS.455.3871A,2020RAA....20..158W,2021MNRAS.503.1149Z} as expected to be performed in many experiments. Indeed, \texttt{BAM} is expected to provide halo and mock galaxy catalogues for several ongoing galaxy-redshift surveys, such as DESI  \cite[][]{DESI}, EUCLID \cite[][]{Euclid},  J-PAS \citep{2014arXiv1403.5237B}, and the Nancy Grace Roman Space telescope \cite[][]{2015arXiv150303757S}. 

The outline of this paper is as follows. In \S\ref{sec:meth}, we describe the \texttt{BAM} approach to calibrating the halo bias. In \S\ref{sec:training}, we describe the reference $N$-body simulation and the different models of halo bias used in this work. Section \S\ref{sec:iter} depicts the methodology used to learn the halo bias while \S\ref{sec:mocks} is devoted to the construction of halo catalogues. In \S\ref{sec:gals}, we present the HOD model used to generate galaxy catalogues and describe the main statistical properties of the resulting ensemble. We end with conclusions and a list of potential developments designed to improve our method.

%\raggedbottom
\section{Description of the bias-assignment method}\label{sec:meth}

\subsection{The halo bias in \texttt{BAM}}\label{sec:hbias}
The halo bias (i.e. the link between the halo and dark matter distribution) is a quantity of paramount relevance to the understanding of halo, and subsequently galaxy, clustering, as it represents the midpoint between the distribution of light (galaxies) and the distribution of the underlying dark matter in the Universe. It is very well established that the bias of dark matter tracers needs to be modelled beyond the standard linear scale-independent scheme: scale dependencies induced by the process of halo formation and merging, the non-linear evolution of the dark matter density field \cite[see e.g.][and references therein]{1999ApJ...525..543M,2000ApJ...540...62S,2001MNRAS.320..289S,2007PhRvD..75f3512S,2007IJMPD..16..763Z,2010ApJ...724..878T,2011A&A...525A..98V,2012MNRAS.420.3469P,2013PhRvD..87h3002S,2015MNRAS.450.1486A,2017A&A...598A.103P,2018PhR...733....1D,2019MNRAS.482.1900H,2019arXiv191012452N}, and the discrete presentation of halo and matter density fields generalises the concept of halo bias to a non-local and stochastic quantity  \cite[see e.g.][]{1993ApJ...413..447F,1998ApJ...500L..79T,1999ApJ...520...24D,2000ApJ...544...63B,2005A&A...430..827S}. 
 
The \texttt{BAM} algorithm is designed to capture the aforementioned properties of the bias of dark matter tracers (halos in this case) at the field level by assuming that the number counts of dark matter halos in a cell of volume $\partial V$ depends on a set of properties $\{\Theta_{\rm dm}\}$ of the underlying dark matter (DM) density field evaluated on the same cells. This dependency is assumed to be represented by a probability distribution of halo occupation number $N_{\rm h}$ conditional to a set of $\mathcal{N}_{p}$ properties of the underlying dark matter field. Accordingly, we  represent the halo bias as a multi-dimensional histogram:
\begin{equation}
\label{eq:bias}
\mathcal{B}\left( N_{\rm h}|\Theta_{\rm dm}\right)_{\partial V}\equiv 
\frac{
\sum_{i=1}^{N_{\rm cells}}\mathbf{1}_{N_{h}}(N_{\rm h}(\vr_{i}))\prod_{\kappa=1}^{\mathcal{N}_{\rm p}}\mathbf{1}_{\gamma_{\kappa}}(\{\Theta_{\rm dm}(\vr_{i})\}_{\kappa})}
{\sum_{i=1}^{N_{\rm cells}}\prod_{\kappa=1}^{\mathcal{N}_{\rm p}}\mathbf{1}_{\gamma_{\kappa}}(\{\Theta_{\rm dm}(\vr_{i})\}_{\kappa})},
\end{equation}
where $\gamma_{\ell}\equiv [\{\Theta_{\rm dm}\}_{\ell}-\Delta_{\ell}/2,\{\Theta_{\rm dm}\}_{\ell}+\Delta_{\ell}/2 )$ represents the set of bins (of width $\Delta_{\ell}$) defined for the $\ell$-th property ($\ell=1,\cdots,\mathcal{N}_{p}$) of the density field, with $\mathbf{1}_{A}(x)$ as the \emph{indicator function}: $\mathbf{1}_{A}(x)=1$ if $x\in A$, and $0$ otherwise. The quantity $\mathcal{B}$ carries no information on the phases of the density fields, and therefore represents a statistical target that can be learned and mapped into a different realisation of the dark matter density field. Equation~(\ref{eq:bias}) approximates the true underlying halo bias, as it ignores key aspects such as the effects of the mass assignment and the correlation between pairs in different property bins, among others. The impact of these effects in the measurement of the halo bias is captured (and corrected for) within the iterative process, as discussed in \S\ref{sec:iter}.

%-------------------------------------------------------
\begin{figure*}
    \includegraphics[trim = 0cm 0cm 0cm 0cm ,clip=true,width=0.99\textwidth]{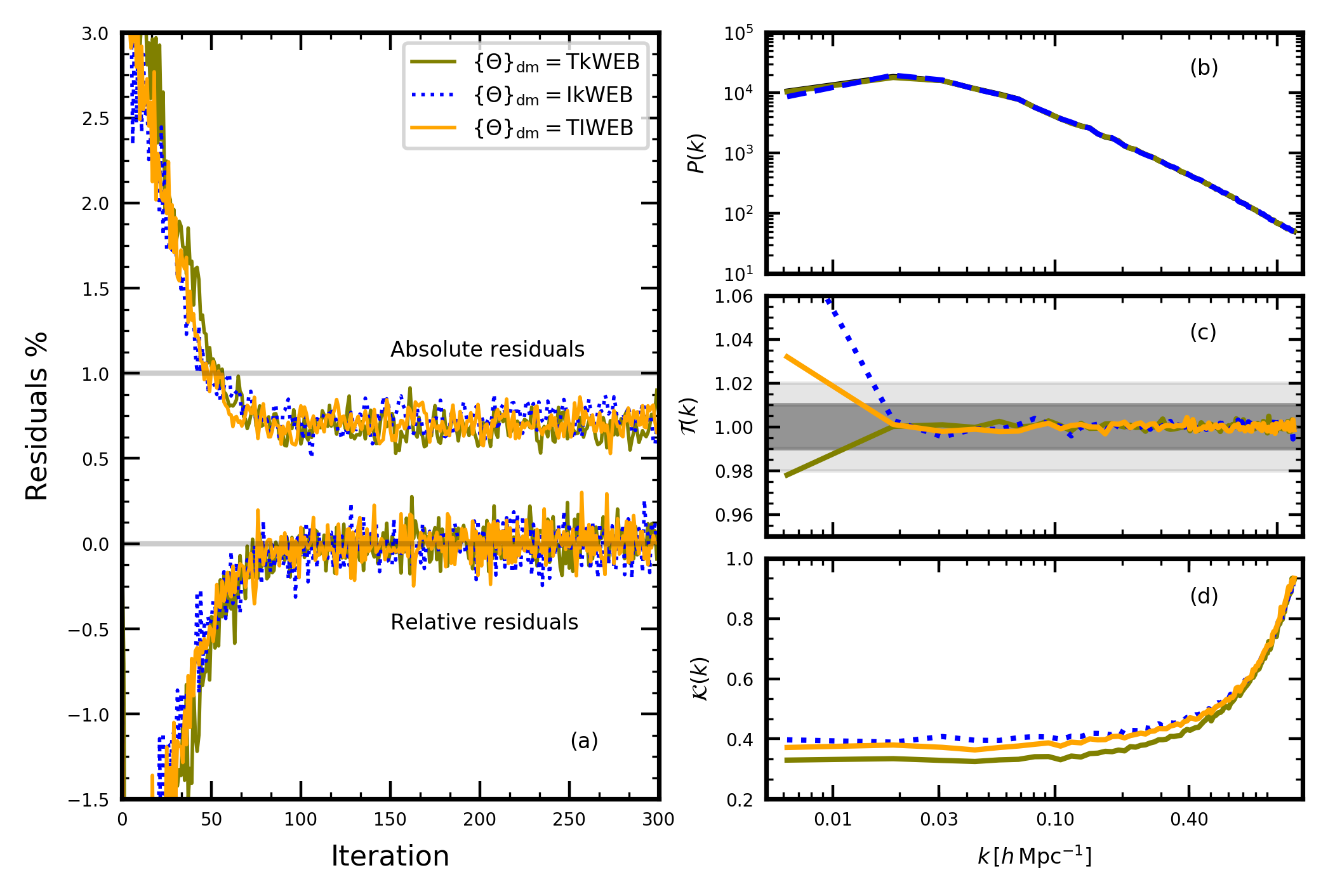}
\caption{Summary statistics of the calibration procedure in \texttt{BAM} based on one \texttt{SLICS} realisation. Panel (a): Residuals computed from the reference power spectrum and the halo power spectrum at different iterations within the calibration procedure. Absolute residuals (see Eq.~(\ref{eq:res})) show that the calibration leads to a precise ($<1\%$) reconstruction of the halo number counts and its two-point statistics, while relative values show that the deviation around the reference is randomly distributed, with a $\sim 0.15\%$ amplitude.  The different lines in each case show the behaviour under different models (\texttt{TkWEB, IkWEB, TIWEB}) of halo bias (see \S~\ref{sec:dmprops} for details). Panel (b) shows the power spectrum from the reconstructed halo number counts field in each of the halo models. Panel (c) shows the transfer function $\mathcal{T}_{i}(k)$ computed as in Eq.~(\ref{eq:transfer}), evaluated at the last iteration of the calibration procedure; the shaded area denotes the $3\%$ deviation around unity. Panel (d) shows the \texttt{BAM} kernel computed using Eq.~(\ref{eq:kernel}).}
\label{fig:calibration0}
\end{figure*}
%-------------------------------------------------------

\subsection{The ingredients of \texttt{BAM}}\label{sec:ingredients}

The \texttt{BAM} machinery relies on a number of ingredients, which are mainly related to properties and outputs of detailed $N$-body simulations. These can be enumerated as follows: 
\begin{enumerate}
    \item  {Initial conditions} (ICs) of a reference $N$-body simulation. These ICs are represented by an initial Gaussian random field built at a much lower resolution than that originally  used by the $N$-body run. A subset of these ICs corresponds to down-sampled versions of the original ensemble, evolved by the $N$-body code to redshifts at which dark matter halo catalogues are identified and used in this analysis.
    \item A set of a few {dark matter halo catalogues} containing phase-space coordinates as well as halo properties that will be used for the assignment of galaxies by means of, for example, HOD prescription. These halos correspond to the ICs whose initial seeds are the same as those in the subset described in point (1).
    \item An {approximated gravity solver} (or surrogate) that evolves ---in a fast way--- the low-resolution IC to the redshift of the tracer catalogue.
\end{enumerate}
Provided the above set of ingredients, the generation of mock galaxy catalogues in \texttt{BAM} is performed in four stages:
\begin{enumerate}
    \item  {Stage I: Calibration}: Learning process in which the halo bias (introduced in the previous section) and \texttt{BAM} kernel (introduced in \S\ref{sec:iter}) are calibrated using the two-point statistics of the reference as a target (or cost function).
    \item  {Stage II: Halo mock production}: Generation of independent halo number count fields through the sampling of independent dark matter density fields using halo bias. 
    \item {Stage III: Phase-space coordinates and properties}: (a) Assignment of position, (b) velocities, and (c) intrinsic properties to dark matter halos.
    \item  {Stage IV: Galaxy catalogues}: Implementation of a HOD model to populate dark matter halos with galaxies.
\end{enumerate}
We cover each of these steps throughout this article. To facilitate understanding of the processes involved in the method, we depict the different steps as a flow chart in Fig.\ref{fig:bam_scheme}.

%===================================================================

\subsection{Training set: Reference simulation and initial conditions}\label{sec:ic}
We use the \texttt{Scinet LIghtCone Simulations} (\texttt{SLICSs}) described by \citet{2018MNRAS.481.1337H}, which consist of an ensemble of cosmological $N$-body simulations run in a comoving box of $L_{\rm box}=505$ Mpc $h^{-1}$ per side, following the non-linear evolution of $1536^3$ particles initialised on a mesh of $3072^3$ points, from an initial redshift of $z_{\rm ini}=120$ down to $z=0$.

The original set of initial conditions of this simulation consist of about $1000$ realisations in the form of particle positions and velocities, which need to be converted to density fields. A fraction of this set is to be used in particle mesh codes \citep[such as \texttt{FastPM}][]{2016MNRAS.463.2273F} as part of the mock-comparison project in DESI \citep[][]{Variu2023}. Accordingly, the initial density fields are obtained from the displacement field $\Psi_{Z}(\tilde{\vq})$ by reversing the Zeldovich displacement \citep[][]{1970A&A.....5...84Z} as $\delta^{HR}_{IC}(\tilde{\vq})=-\nabla_{\tilde{\vq}}\cdot \Psi_{Z}(\tilde{\vq})$, where $\Psi_{Z}(\tilde{\vq})=\tilde{\vq}-\vq$ is computed from the particle positions $\tilde{\vq}$ relative to a regular distribution with coordinates $\vq$ on a $N_{HR}^{3}=1536^3$ lattice (which we refer to as {high-resolution}, HR).

For applications to \texttt{BAM}, we adopted a resolution of $N_{LR}^{3}=192^3$ cells \footnote{This same resolution is adopted by \texttt{EZmocks} for the same DESI comparison project.} and applied an ideal (real) low-pass filter in order to obtain low-resolution initial conditions. 
 The fiducial spatial resolution yields fields represented by a regular mesh with volume ${\partial V}\sim (2.6\, {\rm Mpc}\, h^{-1})^{3}$ and a Nyquist frequency of $\sim 1.2 h\,$Mpc$^{-1}$. For comparison against the reference simulation, and according to DESI scientific requirements, we adopt a maximum wavenumber of $\sim 0.4 h\,$Mpc$^{-1}$, which amounts to $\sim 30\%$ of the Nyquist frequency. At those scales, mass assignment effects inherent to the interpolation of halos in a mesh are expected to be negligible \citep[see e.g.][]{2005ApJ...620..559J}.

\subsection{The reference halo catalogues}\label{sec:training}
The corresponding halo catalogues from the \texttt{SLICS} consist of a set of virialised objects identified at $z=1.04$ with a spherical overdensity algorithm \cite[see e.g.][]{10.1093/mnras/stt1591}. We used a set of $80$ realisations of halo catalogues\footnote{This corresponds to the available number of initial conditions with the corresponding halo catalogues, and represents a reasonable number of realisations to compute covariance matrices for comparison purposes.} to assess the accuracy of our mocks in terms of two and three-point statistics. A subset (of a maximum of 27 randomly selected\footnote{The reason behind this figure is explained in \S\ref{sec:counts}.} references) was also used as part of the training set from which \texttt{BAM} learnt the halo bias (described in \S\ref{sec:iter}).
The mass resolution of the \texttt{SLICS} is $\sim 2.8\times 10^{9}M_{\odot}h^{-1}$. We selected dark matter halos with masses above $2\times 10^{11}M_{\odot}h^{-1}$, which agrees with the expected mass cut at which dark matter halos can host ELGs \citep[see e.g.][]{2020MNRAS.497..581A}. Number counts are generated over a mesh with our fiducial resolution of $192^{3}$ using the nearest grid-point mass assignment \citep{1988csup.bookH}.
Although the halo-finder algorithm allows the determination of different halo properties (e.g. mass, spin, concentration, velocity dispersion), the current set of reference catalogues involves only the virial mass and the velocity dispersion, along with halo coordinates and peculiar velocities, obtained from the position of the density used to identify each halo. These quantities are sufficient to apply an HOD prescription and populate halos with central and satellite galaxies. We note that the \texttt{BAM} method can be applied to reference halo catalogues built with different halo finders \cite[see e.g.][]{2020MNRAS.491.2565B}.

\subsection{Fast gravity solver}\label{sec:gravity}

\texttt{BAM} relies on a combined Lagrangian and Eulerian perturbation theory approach, dubbed augmented Lagrangian perturbation theory  \citep[\texttt{ALPT}; see ][]{doi:10.1093/mnrasl/slt101,2014MNRAS.439L..21K}, to map the initial conditions represented by Lagrangian coordinates $\vq$ (regularly spaced points at the redshift $z_{\rm ini}$) into final (Eulerian) comoving coordinates $\vr(z)$ via $\vr(z)=\vq+\Psi(\vq,z)$, where $\Psi(\vq,z)$ represents the displacement field. This displacement is assumed to be split into long and short-range components, $\Psi(\vq,z)=\Psi_{\rm short}(\vq,z)+\Psi_{\rm long}(\vq,z)$. 
\texttt{ALPT} implements the displacement field from second-order Lagrangian perturbation theory (\texttt{2LPT}) to model the large-scale (long-range) displacement \cite[see e.g.][]{10.1093/mnras/264.2.375,1995A&A...296..575B,2002PhR...367....1B}
\be
\Psi_{2LPT}(\vq,z)=-D^{(1)}(z)\nabla_{\vq}\phi^{(1)}(\vq)+D^{(2)}(z)\nabla_{\vq}\phi^{(2)}(\vq),
\ee
where $D^{(1)}(z)$ is the growth factor \citep[see e.g.][]{1977MNRAS.179..351H},
%\be\label{eq:growth}
%D^{(1)}(z=1/a-1)=\frac{5}{2}\Omega_{\rm %m}E(z)\int_{0}^{a}[a'E(a')]^{-3}\dd a',
%\ee
$D^{(2)}(z)\sim -(3/7)\Omega_{\rm m}^{-1/143}(D^{(1)}(z))^{2}$ \citep[][]{1995A&A...296..575B}. The potentials $\phi^{i}(\vq)$ are the solutions of the Poisson equations $\nabla_{\vq}^{2}\phi^{(i)}=\delta^{(i)}$, where $i=1$ is the linear density obtained in \S\ref{sec:ic}, and
\be\label{eq:delta2}
\delta^{(2)}=\sum_{i,j<i}\lp \partial_{ii}\phi^{(1)}\partial_{jj}\phi^{(1)}-(\partial_{ij}\phi^{(1)})^{2}\rp,
\ee
where we use the notation $\partial_{ij}\phi\equiv \partial^{2}\phi(\vq)/\partial q_{i}\partial q_{j}$. Equation (\ref{eq:delta2}) shows how \texttt{2LPT} takes into account the Hessian of the initial gravitational potential, and is therefore expected to develop the main features of the cosmic web on large scales. However, given that \texttt{2LPT} is not accurate on small scales \cite[see e.g.][]{doi:10.1093/mnrasl/slt101}, its displacement is filtered with a Gaussian kernel $\mathcal{G}_{s}(\vq)$, as $\Psi_{\rm long}(\vq,z)=\Psi_{2LPT}(\vq,z)\otimes \mathcal{G}_{s}(\vq)$, with a smoothing scale of $r_{s}=20$ Mpc$h^{-1}$ \footnote{We verified that smoothing scales in the range $10-20$ Mpc $h^{-1}$ lead to consistent results.}.

While \texttt{ALPT} models the large scales using Lagrangian perturbation theory, it relies on Eulerian perturbation theory to model the small-scale clustering signal. In particular, the short-range displacement is written as $\Psi_{\rm short}(\vq,z)=\lp1-\mathcal{G}_{s}(\vq)\rp\otimes \Psi_{sc}(\vq,z)$ where the displacement $\Psi_{sc}(\vq,z)$ is derived within the spherical collapse (SC) approximation \cite[see e.g.][]{1994ApJ...427...51B,2002PhR...367....1B},
$\psi_{sc}(\vq,z)=\nabla \cdot \Psi_{sc}(\vq,z)$ where $\psi_{sc}(\vq,z)$ is the solution to the Poison-like equation \citep[see e.g.][]{2006MNRAS.365..939M,2013MNRAS.428..141N} \be
\nabla^{2}\psi_{sc}(\vq,z)=3\lp \left[1-\frac{2}{3}D^{(1)}(z)\delta^{(1)}(\vq)\right]^{1/2}-1\rp.
\ee
The regular Lagrangian coordinates are then mapped into Eulerian coordinates using the total displacement:
\be\label{eq:disp}
\Psi_{ALPT}(\vq,z)=\Psi_{2LPT}(\vq,z)\otimes \mathcal{G}_{s}(\vq)+\lp1-\mathcal{G}_{s}(\vq)\rp\otimes \Psi_{sc}(\vq,z).
\ee
With dark matter particles evolved, a cloud-in-cell mass-assignment scheme is implemented to generate an approximated dark matter density field (A-DMDF) on the fiducial mesh. To improve the description of the non-linear dark matter field with a low number of particles, we implement the phase-space mapping technique \citep[][]{2012MNRAS.427...61A,Hahn:2013aa}. 

%-----------------------------------------------
\begin{figure*}
\includegraphics[trim = 0cm 0cm 0cm 0cm ,clip=true, width=1\textwidth]{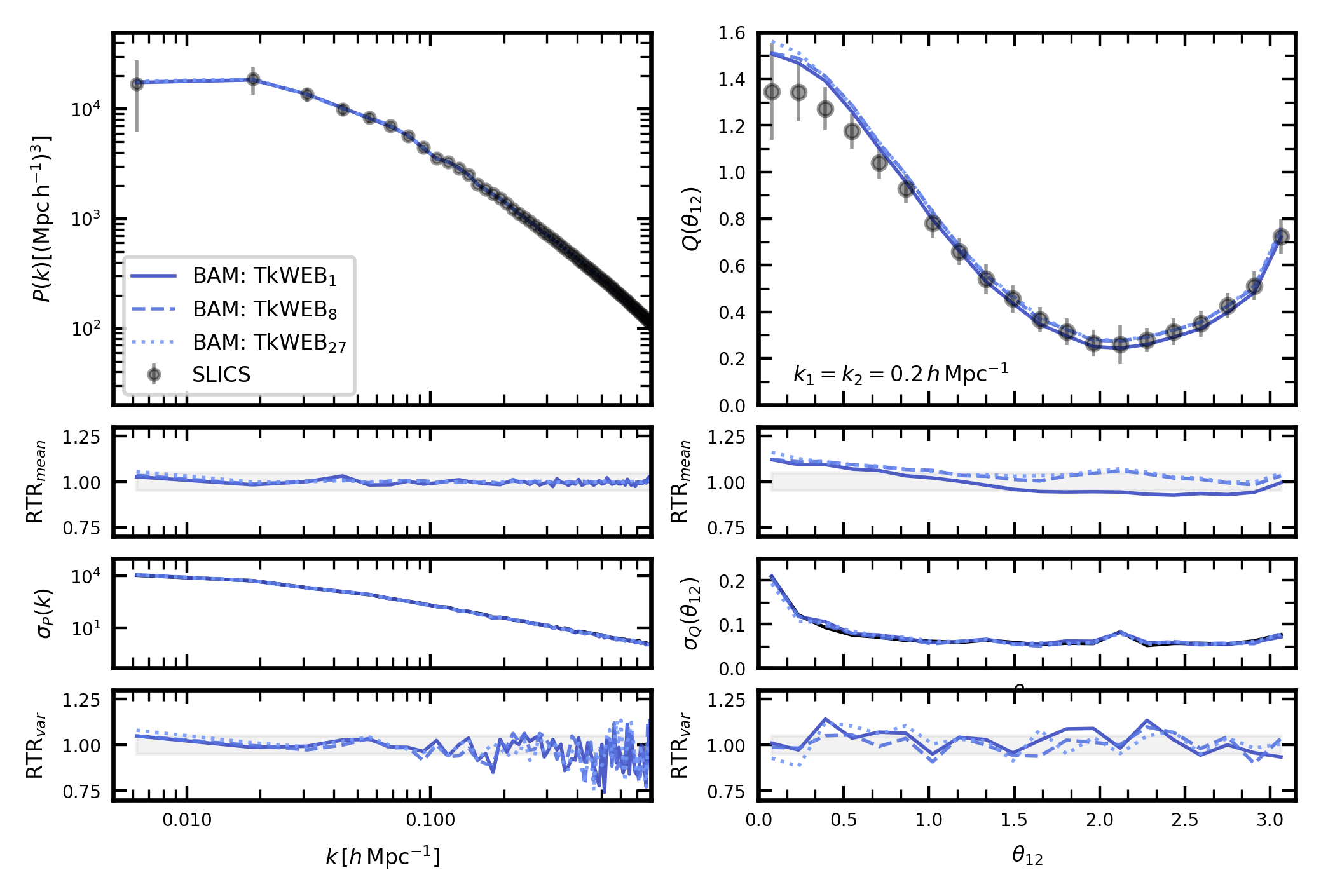}
\caption{Power spectrum (left) and reduced bispectrum (right, isosceles configurations) computed from sets of mock  halo number-count catalogues (of $80$ realisations each) obtained from the calibration of \texttt{BAM} using the \texttt{TkWEB} model as described in~\S\ref{sec:dmprops}. The first row shows the mean in each summary statistic. The second row shows the ratio of the mean statistics to that from the reference (RTR$_{mean}$). The third row shows the variance in the respective statistics, and the fourth their respective ratio to the variance from the reference ensemble (RTR$_{var}$). The shaded area in the second row denotes the $5\%$ deviation to unity.}
\label{fig:mean_tweb}
\end{figure*}
%-----------------------------------------------

We note that the method can in principle implement any approximated gravity solver (with the correct large-scale clustering signal), given that the \texttt{BAM} kernel is meant to correct for missing power towards small scales, with the aim being to generate the correct tracer power spectrum through a learning procedure (further discussed in \S~\ref{sec:iter}). Nevertheless, as long as we use the positions and velocities (see \S\ref{sec:coords}) from the dark matter particles computed from such approximated methods, along with the fact that the tidal field (see Eq.~(\ref{eq:delta2})) is a key ingredient of the method, the desired trade-off between precision, speed, and physical content means that we favour \texttt{2LPT} or \texttt{ALPT} over, for example, the Zeldovich approximation \cite[see][for a review on the Zeldovich approximation]{2014MNRAS.439.3630W}. Further developments designed to improve the precision without a significant increase in required computing time were recently presented by \citet[][]{2023arXiv230103648K} and will be implemented in the \texttt{BAM} machinery in future applications.

Finally, we highlight that the resulting mass of the dark matter particles ($\sim 10^{14}M_{\odot} h^{-1}$) used to define the cosmic web is  nearly five times larger than in the reference $N$-body simulation (see \S\ref{sec:training}). 

\subsection{Properties of the dark matter density field in \texttt{BAM}}\label{sec:dmprops}
\texttt{BAM} explicitly determines several properties $\{\Theta_{\rm dm}\}$ of the underlying DM density field upon which the occupation number of dark matter halos is assumed to depend, as explained in \S\ref{sec:meth}. In general, such properties can be nominally divided into local and non-local, depending on the quantity used to infer them. While as a local property, we can readily use the dark matter overdensity at each cell (obtained using a given mass-assignment scheme), non-local properties (also dubbed as environmental) can be extracted from quantities defined on scales larger than the cell volume, such as the tidal field tensor $\mathcal{T}_{ij}=\partial_{i}\partial_{j}\Phi$ (where $\Phi$ is the comoving gravitational potential satisfying the Poisson equation $\nabla^{2}\Phi=\delta_{\rm dm}$). In particular, previous implementations of \texttt{BAM} used the cosmic-web classification (\texttt{CWC}), which relies on the value of the eigenvalues $\lambda_{i}$ ($i=1,2,3$) of the tidal field \citep[see e.g.][]{2007MNRAS.375..489H, 2009arXiv0912.3448V, 2009MNRAS.396.1815F, 2016MNRAS.455..438A,2017ApJ...848...60Y,2018MNRAS.476.3631P} with respect to some arbitrary threshold $\lambda_{\rm th}$. Similarly, the information of the velocity shear of the DM particles \citep[see][]{1996Natur.380..603B,2018MNRAS.473.1195L,10.1093/mnras/stac671} and its eigenvalues can be used to characterise the halo occupation number. In this work, we restrict ourselves to the \texttt{CWC}.

The \texttt{CWC} allows us to define the behaviour of the halo number counts in knots (labelled $\hat{k}$, $\lambda_{1}>\lambda_{\rm th}$, $\lambda_{2}>\lambda_{\rm th}$, $\lambda_{3}>\lambda_{\rm th}$), filaments ($\hat{f}$, $\lambda_{1}<\lambda_{\rm th}$, $\lambda_{2}>\lambda_{\rm th}$, $\lambda_{3}>\lambda_{\rm th}$), sheets ($\hat{s}$, $\lambda_{1}<\lambda_{\rm th}$, $\lambda_{2}<\lambda_{\rm th}$, $\lambda_{3}>\lambda_{\rm th}$), and voids ($\hat{v}$, with $\lambda_{1}<\lambda_{\rm th}$, $\lambda_{2}<\lambda_{\rm th}$ and $\lambda_{3}<\lambda_{\rm th}$) \footnote{In this work, we use $\lambda_{\rm th}=0$.}. Furthermore, the \texttt{CWC} permits exploration of the dependency of halo occupancy on the mass $M_{k}$ of collapsing regions, defined as the number of dark matter particles in sets (regions) formed by cells classified as knots. These regions are identified through a friend-of-friend percolation algorithm \citep[][]{2015MNRAS.451.4266Z}. 

The set of properties (\texttt{CWC}$+M_{k}$) has been explored in previous \texttt{BAM} publications \citep[see e.g.][]{2020MNRAS.491.2565B}, where it was shown that it is key to reconstructing the halo number counts based on the dark matter density field. In the same context, \cite{10.1093/mnras/stac671} introduced the implementation of the invariants of the tidal field $I_{i}$ in the definition of halo bias used in \texttt{BAM} (where $I_{1}=\delta_{\rm dm}$, $I_{2}=\lambda_{1}\lambda_{2}+\lambda_{1}\lambda_{3}+\lambda_{2}\lambda_{3}$ and $I_{3}=\lambda_{1}\lambda_{2}\lambda_{3}$). This approach is designed to bridge a phenomenological description of the tidal field (e.g. with the \texttt{CWC}) and theoretical models of perturbation theory in which higher order terms can be written in terms of combinations of the eigenvalues of the tidal field.

In summary, we explore the following models for the reconstruction of halo density fields and the generation of mock catalogues:
\begin{itemize}
\item \texttt{TkWEB}: $\{\Theta_{\rm dm} \}\equiv \{ f_{1}(\delta_{\rm dm}), \hat{k},\hat{f},\hat{s},\hat{v},M_{k}\}$. Use the local density, cosmic-web types, and the mass of collapsing regions.
\item \texttt{IkWEB}: $\{\Theta_{\rm dm} \}\equiv \{ f_{1}(\delta_{\rm dm}), f_{2}(I_{2}), f_{3}(I_{3}),M_{k} \}$. Use the invariants of the tidal field and the mass of collapsing regions.
\item \texttt{TIWEB}: $\{\Theta_{\rm dm} \}\equiv \{ f_{1}(\delta_{\rm dm}), f_{2}(I_{2}), \hat{k},\hat{f},\hat{s},\hat{v},M_{k} \}$. Use the cosmic web classification and one invariant of the tidal field.
\end{itemize}
The functions $f_{i}(x)$ represent non-linear transformations designed to improve the extraction of the bias information in each variable $x=\{\delta_{\rm dm},I_{2},I_{3}\}$. We use $f_{1}(x)=\log(2+x)$ and $f_{2}(x)=f_{3}(x)=2(x^{\alpha}-\gamma)/(\eta-\gamma)-1$ with $\gamma\equiv {\rm min}(x^{\alpha})$, $\eta \equiv {\rm max}(x^{\alpha})$ and $\alpha$ a free parameter (fixed to $\sim 0.11$). The form of $f_{1}(x)$ has the usual form already used in \citep{2020MNRAS.491.2565B}, while the shape of $f_{2,3}(x)$ is designed to map the (large) dynamic range spanned by the invariants $I_{2,3}$ to the interval $[-1,1]$, thus simplifying its binning.
Other sets of properties, such as the eigenvalues of the tensor $\partial_{i}\partial_{j}\delta_{\rm dm}$ \citep[see e.g.][]{Peacock1985,1986ApJ...304...15B}, can also be applied to characterise the bias of dark matter tracers \citep[see e.g.][]{2021ApJ...921...66S}.

As previously mentioned, the physical motivation behind the choice of these models lies in the fact that local dark matter is not the only driver for halo clustering. Several works have already presented evidence of assembly bias in halos and galaxies \cite[see e.g.][]{1997MNRAS.286..795K, 10.1111/j.1365-2966.2004.07733.x, 2005MNRAS.363L..66G,2006ApJ...652...71W, 2007MNRAS.377L...5G,2007MNRAS.374.1303C, 2008MNRAS.387..921A,2008ApJ...687...12D,2010ApJ...708..469F,2017MNRAS.466.3834L,2017JCAP...03..059L,2017ApJ...848L...2M,2018MNRAS.474.5143M,2018MNRAS.476.4877M,2019MNRAS.484.1133C,2021MNRAS.502.3242X}. This type of bias not only includes the clustering of halos as a function of their intrinsic properties but also as a function of their environment \citep[see e.g.][]{2017ApJ...848...60Y, 2018MNRAS.473.3941F}, a dependency that can be covered with approaches such as the \texttt{TkWEB} model. Furthermore, the inclusion of the mass of collapsing regions allows us to include short-range non-local bias, focusing on regions with a distinct (collapsing) dynamical state.

% =============================================================
\begin{figure}
\includegraphics[trim = 0cm 0cm 0cm 0cm ,clip=true, width=0.52\textwidth]{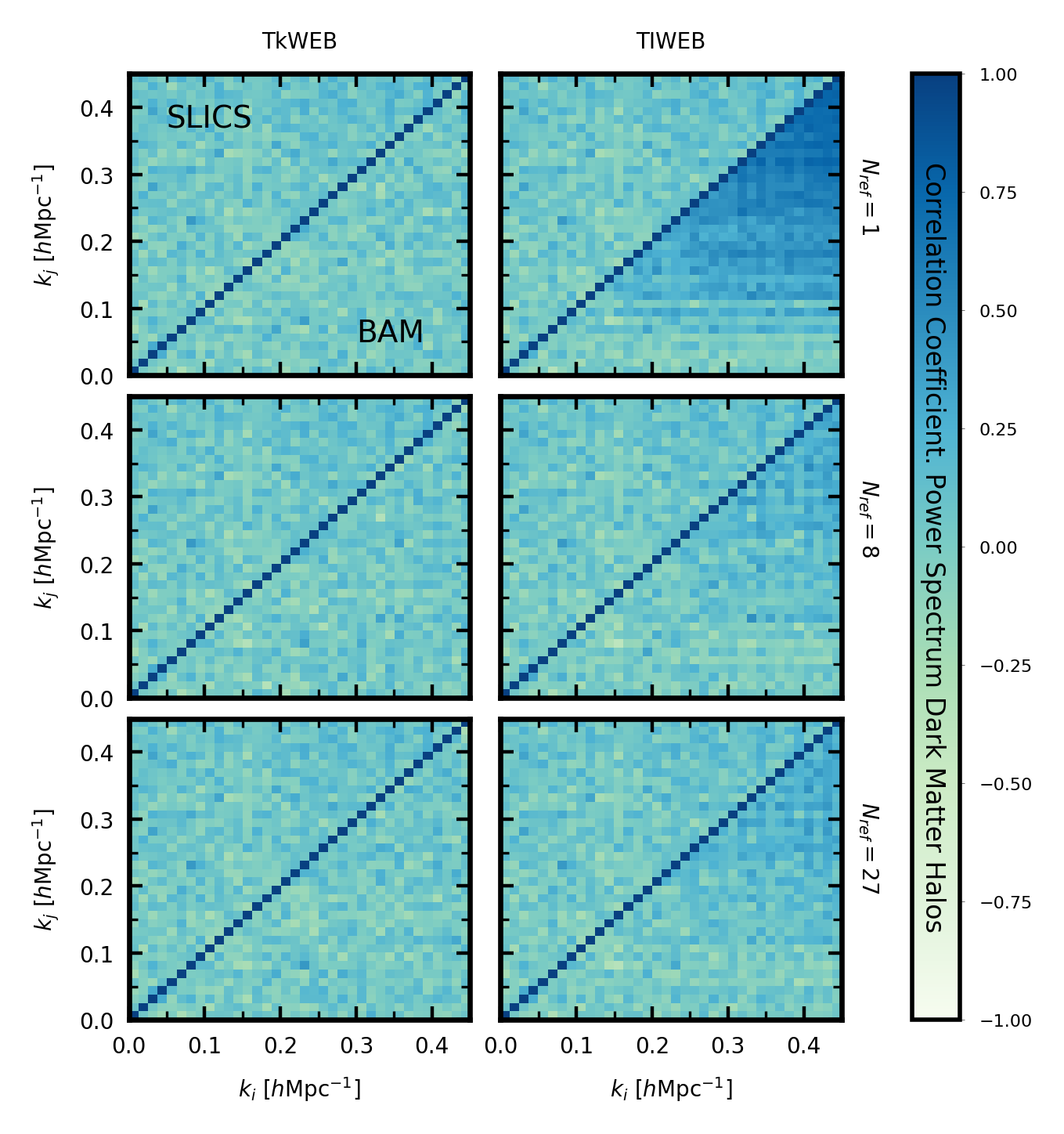}
\caption{Correlation coefficients of the halo power spectrum obtained from a set of $80$ realisations of halo number-counts using the \texttt{SLICS} and \texttt{BAM} mock catalogues, calibrated from the different number of references $N_{ref}$ and two characterisations of the properties of the dark matter density field (\texttt{TkWEB} and \texttt{TIWEB}).}
\label{fig:corr_power_models}
\end{figure}
% =============================================================
% =============================================================
\begin{figure}
\includegraphics[trim = 0cm 0cm 0cm 0cm ,clip=true, width=0.51\textwidth]{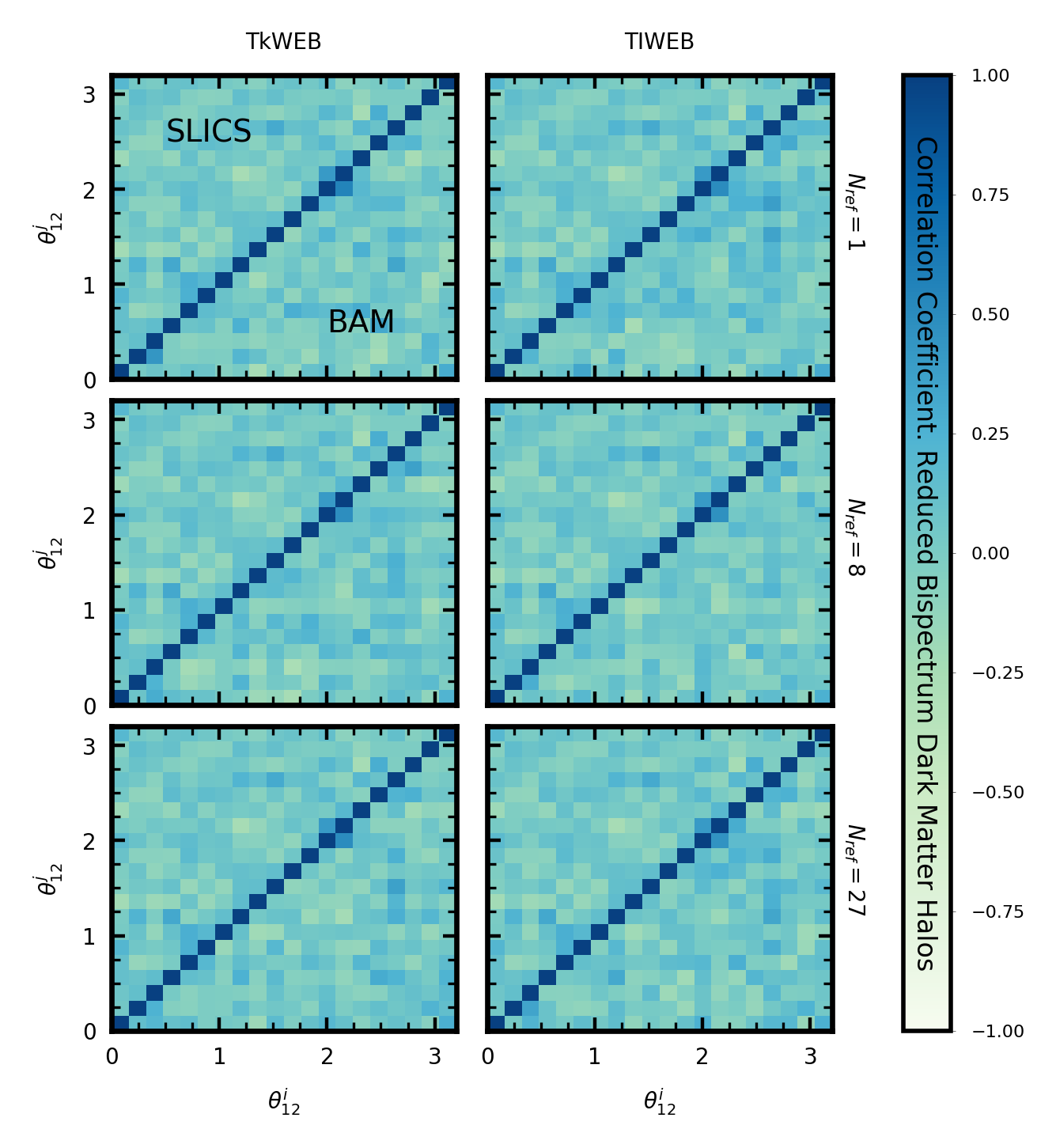}
\caption{Same as Fig.\ref{fig:corr_power_models} but for the reduced bispectrum of dark matter halos using isosceles configuration $k_{1}=k_{2}=0.2\,h$ Mpc$^{-1}$.}
\label{fig:corr_bisp_models}
\end{figure}
% =============================================================
% =============================================================
\begin{figure}
\includegraphics[trim = .2cm 0cm 0cm 0cm ,clip=true, width=0.51\textwidth]{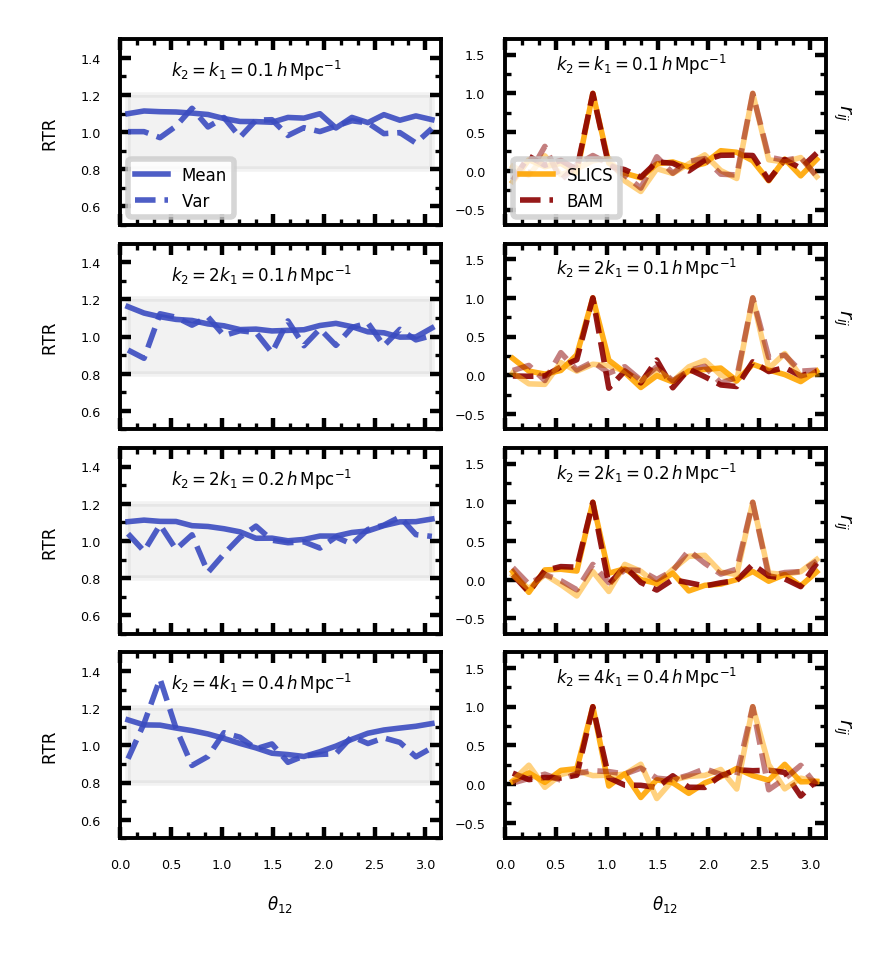}
\caption{Comparison of the signal of reduced bispectrum obtained from $80$ mock catalogues generated with \texttt{BAM} (using the \texttt{TkWEB} model) and the same signal from the reference set, for several triangle configurations, which are specified in each panel. The left column shows the ratio to the reference (RTF) of the mean (solid lines) and variance (dashed lines); the shaded areas in the panels of this column denote a $20\%$ deviation from unity. The right column shows two elements of the correlation coefficients of the  bispectrum $r_{ij}$.}
\label{fig:corr_bisp_confs}
\end{figure}
% =============================================================

On the other hand, implementing the invariants of the tidal field (i.e. the \texttt{IkWEB} model) allows us to assess the halo bias of Eq.~(\ref{eq:bias}) in a more complete fashion. This can be understood from the degree of arbitrariness arising in the framework of the \texttt{CWC}, whose characterisation depends on the parameter $\lambda_{\rm th}$. The invariants of the tidal field do not suffer from this freedom and therefore contain all the information in the cosmic-web decomposition. Also, and similarly important, the connection between the invariants of the tidal field and the different terms present in a perturbative approach \citep[see e.g.][]{2009JCAP...08..020M,10.1093/mnras/stac671} allows \texttt{BAM} to explicitly include a non-negligible signal of non-local bias up to third order in perturbation theory, a signal that is expected to be measured in forthcoming experiments \cite[see e.g.][]{2022PhRvD.105l3518G}.
Finally, the \texttt{TIWEB} model is designed to use the information from the \texttt{TkWEB}, adding the information from one invariant of the tidal field or a function thereof. One such function is the so-called {tidal anisotropy parameter}
defined as a function of the eigenvalues of the tidal field as $\alpha^{2} \propto (\lambda_{1}-\lambda_{2})^{2}+(\lambda_{1}-\lambda_{3})^{2}+(\lambda_{3}-\lambda_{2})^{2}$ \citep[e.g.][]{2018MNRAS.476.3631P}. This property is used in the assignment procedure for intrinsic halo properties, which is explained in \S\ref{sec:hprops}.

Along with the models, the total number of bins adopted to discretise the information in the different dark matter properties (e.g. $f_{1}(\delta_{\rm dm}), M_{k}$) is also important when assessing whether the process can fall into an over-fitting regime. This can be quantified by computing the ratio $\eta$ between the total number of bins and the total number of spatial cells used in the field description of halos and dark matter. After a series of numerical tests (mainly focused on the ideal number of dark matter property bins needed to achieve the convergence of the method as explained in \S\ref{sec:iter}), we obtain $\eta \sim 0.02, 2.2,$ and $0.9,$ for \texttt{TkWEB,IkWEB} and \texttt{TIWEB} respectively. This implies that the \texttt{IkWEB} model is likely to incur overfitting \footnote{Typically, for the \texttt{TkWEB} model, we use approximately $200$ bins for the local density $f_{1}(\delta_{\rm dm})$, about $200$ bins for $M_{k}$ , and the four types of cosmic web.}. This situation does not arise during the calibration procedure because the kernel and bias are applied to the same dark matter field from which these quantities are obtained. However, when implementing these products on independent dark matter density fields (as described in \S\ref{sec:counts}), the \texttt{IkWEB} model will be more sensitive to any difference in the dark matter distribution of the new field with respect to the reference. In that case, the algorithm generates biased estimates of the halo power spectrum for some realisations (e.g. those with density peaks not present in the reference), which leads to mode coupling in the covariance matrix of the power spectrum.

% =============================================================

\subsection{Learning phase: Iterative procedure and calibration of halo bias}\label{sec:iter}
The learning procedure in \texttt{BAM} is designed to generate two main outputs, namely, (i) the so-called \texttt{BAM}-kernel, and (ii) the corresponding (multi-dimensional) halo bias introduced by Eq.~(\ref{eq:bias}). The  role of the halo bias is to assign the number of tracers in cells according to the underlying dark matter properties, keeping track of all the statistical anisotropies of the latter. The role of the kernel is twofold: it corrects for any effective large-scale contribution from non-local bias dependencies not accounted for in the set $\{\Theta_{\rm dm}\}$; and it also corrects for any aliasing effects caused by the representation of the DM field and the halo distribution on a mesh with respect to the original halo-finding algorithm used to construct the reference catalogue.

Let us now describe the procedure developed in the so-called learning phase of \texttt{BAM}. The main scope of the process is to modify the A-DMDF such that, when sampling it using the halo bias obtained with Eq.~(\ref{eq:bias}), we reconstruct the statistics of halo number counts to per cent precision up to the Nyquist frequency. Let us now focus on the $i$-th iteration: At this stage, the algorithm starts determining the properties $\{\Theta^{i}_{\rm dm}\}$ from a dark matter density field obtained as the convolution of the input A-DMDF with the so-called \texttt{BAM}-kernel (to be defined below) $\mathcal{K}$ \footnote{We use the same symbol for its Fourier or configuration space representation.}, which in turn is the result of the previous iteration: 
 \be\label{eq:conv}
\tilde{\delta}_{\rm dm}^{i}(\vr)\equiv \lp  \mathcal{K}_{i-1}\otimes \delta_{\rm dm}\rp(\vr)
,\ee
where the kernel is a Dirac's delta function for the first iteration, and remains spherically symmetric in subsequent iterations. With this new density field, the halo bias $\mathcal{B}(N_{\rm h}^{\rm ref}|\Theta_{\rm dm}^{i})$ is measured using  Eq.~(\ref{eq:bias}), and is then used to sample the density field $\tilde{\delta}^{i}_{\rm dm}$ to obtain a new version of halo number counts (which we also refer to as the {reconstructed field}):
\be\label{eq:sam1}
\{N_{h}^{i}(\vr)\}  \curvearrowleft \mathcal{B}\lp N_{\rm h}^{\rm ref}    \, \mid \, \{\Theta_{\rm dm}\} = \{ \Theta^{i}_{ \rm dm}(\vr)\} \rp_{\partial V}.
\ee
The main statistical property adopted as the target for the \texttt{BAM} algorithm is the halo power spectrum. This is obtained as an spherical average of the 3D Fourier transform of the new halo number count field, $N^{i}_{h}(\vk)$, performed in shells identified with a wavenumber $k_{n}$,
\be\label{eq:mpow}
P_{i}(k_{n})=\frac{1}{N_{n}}\sum_{\vk \in \Delta k_{n}}|N^{i}_{h}(\vk)|^{2}-S,
\ee
where $N_{n}$ is the number of Fourier modes in the $n$-th shell (of width $\Delta k_{n}$), and the sum incorporates all vector modes with magnitude in that shell. Here, $S=1/\bar{n}$ is the Poisson shot noise \citep[][]{1980lssu.book.....P} of the reference halo catalogue. We then define a power transfer function, 
\be \label{eq:transfer0}
\mathcal{T}_{i}(k_{n})\equiv 
\frac{P_{\rm ref}(k_{n})}{P_{i}(k_{n})},
\ee
where $P_{\rm ref}(k_{n})$ denotes the power spectrum of the reference halo catalogue (measured as in Eq.~(\ref{eq:mpow})). The sampling procedure of Eq.~(\ref{eq:sam1}) is performed such that the new \texttt{HDF} not only contains the same number of objects as that of the reference but also shares its number-count statistics (number-count distribution function).

For each spherical shell in Fourier space, \texttt{BAM} implements a Metropolis-Hasting algorithm \citep[see e.g.][and references therein]{2009arXiv0906.0664H} to accept or reject the corresponding value of the transfer function defined in Eq.~(\ref{eq:transfer0}). As metric, \texttt{BAM} uses the quadratic difference between the mock and reference power spectra in units of the Gaussian variance \cite[see e.g.][]{dodelson:2003} of the latter (the standardised Euclidean distance). That is, we define a mode-by-mode likelihood of the form
\be\label{eq:like}
\mathcal{L}_{i}(k_{n})\equiv {\rm exp} \lp -\frac{(P_{i}(k_{n})-P_{\rm ref}(k_{n}))^{2}}{4(P_{\rm ref}(k_{n})+S)^{2}} N_{n}\rp.
\ee
The algorithm maximises the function $\mathcal{L}_{i}(k_{n})$ by accepting the new power spectrum ---and therefore the corresponding transfer function $\mathcal{T}_{i}(k_{n})$--- with a probability ${\rm min}\lp 1, \mathcal{L}_{i}(k_{n})/\mathcal{L}_{i-1}(k_{n})\rp$. If the transfer at a given mode is not accepted, the algorithm retains the previously accepted value. To express this fact, we define a set of weights $\omega_{i}(k_{n})$ constructed according to the rejection criteria:
\be \label{eq:transfer}
\omega_{i}(k_{n})\equiv 
\begin{cases}
\mathcal{T}_{i}(k_{n}) & {\text{if accepted,}} \\ 
1 & {\text{if rejected.}}\\
\end{cases}
\ee
These weights are used to define (and to update) the \texttt{BAM}-kernel in Fourier space (which is isotropic, being only a function of $k_{n}$), by making products of the weights at the current step with those from the preceding iterations (at the same shell $k_{n}$):
\be \label{eq:kernel}
\mathcal{K}_{i}(k_{n})=\prod_{\ell =1}^{\ell=i-1}\omega_{\ell}(k_{n}).
\ee
We use this new version of the kernel to convolve the input DM field, as in Eq.~(\ref{eq:conv}), from which a new iteration follows (in practice, the convolution with the kernel is done in Fourier-space).
We note that the transfer function defined in Eq.(\ref{eq:transfer}) is applied to the dark matter density field, and not directly to the field we are trying to reconstruct. Hence, there is no explicit need to define it under a squared root \citep[see e.g.][]{doi:10.1093/mnras/254.2.315}.

% ==========================================================
\begin{figure}
    \includegraphics[trim = .15cm 0cm 0cm 0cm, clip=true,width=0.5\textwidth]{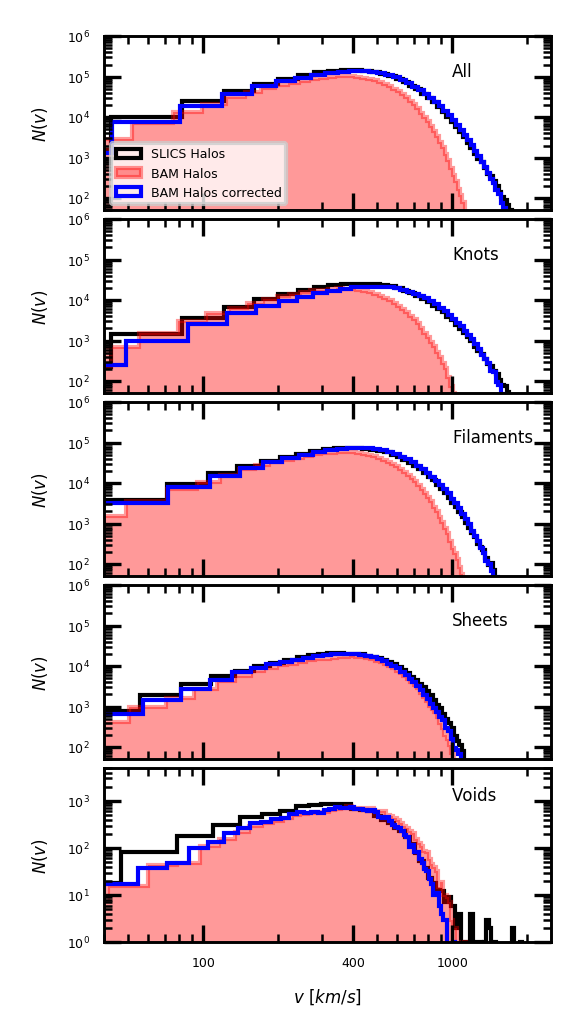}
\caption{Example of the distribution of halo peculiar velocities $v=|\vec{v}|$ in one reference (solid black histogram) and one \texttt{BAM} (solid blue and filled histogram) halo catalogue, in different types of cosmic web (see~\S\ref{sec:coords}). In all panels, the {corrected} histogram is obtained after applying the isotropic velocity correction described in \S\ref{sec:coords}.}
\label{fig:vel_dist}
\end{figure}
% ==========================================================

The learning (or calibration) phase is considered to converge when the absolute residuals $R$, defined as
\be \label{eq:res}
R_{i}[\%]\equiv \frac{100}{N_{\rm F}} \sum_{n=1}^{n=N_{\rm F}}|\mathcal{T}_{i}(k_{n}) -1 |,
\ee
(where $N_{\rm F}$ is the number of spherical shells used to measure the power spectrum) reach the threshold $\sim 1\%$. We used $300$ iterations, although with about $ 150$ iterations, the calibration has already reached the sub-per cent residuals. In terms of computing time, at a $32$-thread workstation, the calibration procedure (with 300 iterations) takes $\sim 1$ hour.  

In order to verify that the outputs of the iterative process are independent of the realisation used as a reference, we repeated this procedure for a number of reference simulations (IC plus corresponding halo catalogues) available in the \texttt{SLICS} set. Figure~\ref{fig:slices_cal_bam}  shows slices through the different density fields involved in the calibration procedure performed with one randomly selected reference simulation. In particular, we show the halo number counts on a mesh (second row) reconstructed using the three halo bias models described in \S\ref{sec:dmprops}. 

Figure \ref{fig:calibration0} shows the summary statistics arising from the products of the iterative stage, using different models for the multidimensional halo bias of Eq.~(\ref{eq:bias}), and using one reference simulation. All the models shown are in a position to generate sub-per cent residuals in the calibration (see panel (a)) with reconstructed power spectra (panel (b)) within a $5\%$  difference with respect to the reference (up to the Nyquist frequency). 
The models of halo bias shown display minor differences in their performances when explored at the level of the reconstructed power spectrum, as can be inferred from panel (c) of the previously mentioned figure, where we show the ratio of the power spectrum from the reconstructed field to that measured from the reference. It is only on the first Fourier mode that the differences with respect to the references are above $2\%$, while for the rest of the probed Fourier modes and up to the Nyquist frequency, the differences oscillate around $\sim 0.6\%$. We explicitly verified that similar trends are obtained when another realisation is used to perform the calibration. It is key to note that the fluctuations on large scales are not only a consequence of the small volume but are also linked to the stochastic nature of halo bias as expressed by Eq.(\ref{eq:bias}).

The differences among the implemented models of halo bias can be observed in the shape of their corresponding kernel, as shown in panel (d) of Fig.~\ref{fig:calibration0}. 
We note that the definition of the kernel implies that it does not explicitly encode any information on the anisotropies of the halo density field, which are clearly present in the large-scale distribution in the form of a filamentary structure. Indeed, the kernel in configuration space is fully symmetric, although the patterns can change according to the model of halo bias (and the type of mass-assignment scheme). The information on the anisotropies in the 3D halo distribution is instead statistically encoded in the halo bias, and as such, the model (i.e. the set of properties $\{\Theta\}$) is key to reproducing higher order statistics, as we show below. 

In general, the overall shape of the kernel agrees within all tested models: a constant amplitude towards large scales, with a scale dependency on small scales. The difference in the large-scale amplitude encodes the different content of information on the assembly bias \footnote{In this context, we use {`halo assembly bias'} to refer to the clustering signal that is dependent on all possible effects not included within the description of halo bias of \S~\ref{sec:dmprops}.}, and is accounted for as long as different non-local terms are included. That is, the higher the amount of information on halo bias, the closer the kernel is to unity on large scales. The constant amplitude of the kernel towards large scales is a property that can be used to generate mock catalogues on larger cosmological volumes. This will be the subject of forthcoming publications.
% ======================================================================================================
\section{Construction of halo catalogues}\label{sec:mocks}
In this section, we describe the steps followed to generate halo number counts on a cubic mesh starting from independent dark matter density fields and using the outputs described in \S\ref{sec:iter}. To compare the summary statistics of the mocks produced within \texttt{BAM} with those from the reference, we make all comparisons with a set of $N_{\rm sim}=80$ mocks. We test different models of halo bias, and based on the performance of the summary statistics obtained from the mocks constructed with these models, we adopt one of them to generate the final set of mock galaxy catalogues.

%------------------------------------------------------------------------------
%------------------------------------------------------------------------------
\begin{figure}
\begin{tikzpicture}[>=latex,scale=1.3][font=\ttfamily\small,node distance=1.6cm]
\hspace{0cm}
  \shade[ball color=gray!50!] (0,-2) coordinate(DM) circle (0.3) ;
  \shade[ball color=red!30!] (5,0) coordinate(R) circle (.3) ;
  \shade[ball color=green!30!] (3,-0.8) coordinate(Rc) circle (.3) ;
%  \draw[thick,dashed] (0,-2) -- (5,0) ; %joins DM with Rcol
  \draw[thick,solid,->] (5,0) -- (5.5,1.8) ; % mimis vel from Random
  \draw (5.1,1.3) node[above]{$\vec{v}_{\rm r}$} ;
%  \draw[thick,solid,->] (5,0) -- (5.7,0.9) ;  %vel of Random original
  \draw[thick,solid,->, color=red] (5,0) -- (5.2,0.7) ;  %vel of Random original
  \draw (5.5,0.4) node[above]{\textcolor{red}{$\vec{v}^{0}_{\rm r}$}} ;
  \draw[thick,dotted,->] (3.1,-0.6) -- (3.5,1.0) ; %mimics vel of Ran at Rancol position
  \draw (3.2,0.6) node[above]{$\vec{v}_{\rm r}$} ;
  \draw[thick,solid,->] (3,-0.8) -- (1.8,0) ;  %vel of Random col
  \draw[thick,solid,->] (0,-2) -- (-0.1,-0.8) ;  %vel of dm
  \draw (0.2,-0.9) node[above]{$\vec{v}_{\rm dm}$} ;
%  \draw[thick,solid,->] (0,-2) -- (-0.1,-0.4) ;  %vel of dm original
  \draw[thick,solid,->,color=red] (0,-2) -- (-0.16,-0.15) ;  %vel of dm original
  \draw (0.1,-0.3) node[above]{\textcolor{red}{$\vec{v}^{0}_{\rm dm}$}} ;
  \draw[arrow_nlsa](R)--(Rc);
  \draw (3.1,-0.5) arc (50:119:.6);
  \draw (2.6,-0.2) node[right]{$\beta$} ;
  \draw (3.7,-0.5) arc (30:90:.6);
 \draw (3.6,-0.2) node[right]{$\alpha$} ;
 \draw (3.2,-1.5) node[above]{Random-collapsed} ;
  \draw (4.5,-0.5) node[right]{Random} ;
  \draw[thick,dashed,->] (1,-1.6) -- (0,-2.0) ; %mimics vel of Ran at Rancol position
  \draw[thick,dashed,->] (1.6,-1.38) -- (3,-0.8) ; %mimics vel of Ran at Rancol position
  \draw (1.,-1.5) node[right]{$f_{\rm col}d_{r}$} ;
  \draw (1.6,0.0) node[above]{$\vec{v}_{\rm r}'$} ;
  \draw (0,-2.3) node[below]{Closest DM} ;
  \draw (3.68,-1.7) node[below]{$|\vec{v}_{\rm r}|=\gamma[\delta_{\rm dm}(\vec{r}_{\rm ran})]|\vec{v}^{0}_{\rm r}|$} ;
  \draw (3.2,-2.0) node[below]{$|\vec{v}_{\rm r}'|=\zeta|\vec{v}_{\rm r}|$} ;
  \draw (3.8, -2.3) node[below]{$|\vec{v}_{\rm dm}|=\gamma[\delta_{\rm dm}(\vec{r}_{\rm dm})]|\vec{v}^{0}_{\rm dm}|$} ;
 \end{tikzpicture}
\caption{Subgrid modelling for the assignment of coordinates in phase space.  The coordinates of the random tracers are modulated by the fraction $f_{\rm col}$ such that if $d_{r}$ denotes the separation between the random particle and its closest dark matter particle, the new separation is $f_{\rm col}d_{r}$. Velocities are modified in two steps: (i) an isotropic density-dependent correction $\gamma(\delta_{\rm dm})$ is applied to the velocities of both random tracers and dark matter tracers ($\vec{v}_{\rm r} \to \vec{v}'_{\rm r}$), and (ii), after collapsing the random tracers, their velocities are modified using the parameters $\beta$ (direction) and $\zeta$ (modulo). The angles $\alpha$ and $\beta$ are defined on the plane generated by the vector joining random particles with their closest dark matter particle and the velocity of the random particle $\vr_{r}$.}
  \label{fig:col_scheme}
\end{figure}
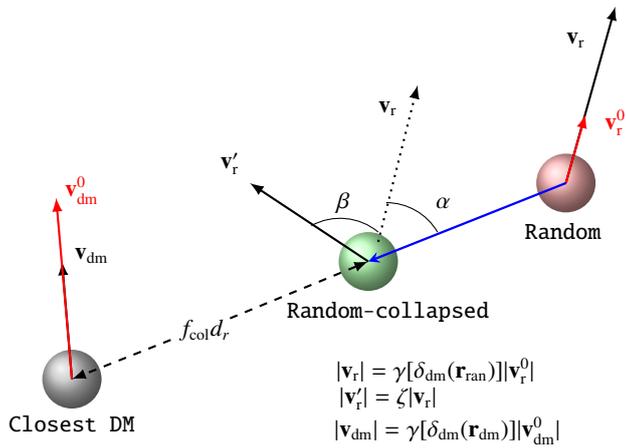

\subsection{The halo bias and kernel}\label{sec:using}
Based on the set of $N_{\rm sim}$ initial conditions described in \S\ref{sec:ic}, we generated the same number of realisations of (approximated) dark matter density fields $\delta_{\rm dm}^{j}$ ($j=1,\cdots, N_{\rm sim})$ using the methods described in \S\ref{sec:gravity}. These are convolved with the \texttt{BAM}-kernel (obtained from the learning phase, \S\ref{sec:iter}) to generate a new dark matter density field $\tilde{\delta}_{\rm dm}^{j}\equiv \mathcal{K}\otimes \delta_{\rm dm}^{j} $, after which the non-local properties (e.g. 
types of cosmic web) of the resulting field $\tilde{\delta}_{\rm dm}^{j}$ are determined. According to these properties, the algorithm populates these dark matter fields with a number of haloes in cells sampling as
\begin{equation} 
\label{eq:mock_crea}
\{ N_{h}^{j}(\vr) \} \curvearrowleft 
\mathcal{B}\lp N_{\rm h}^{{\rm ref}}\, \mid \, \{ \tilde{\Theta}_{\rm dm}\} = \{ \tilde{\Theta}_{\rm dm}^{j}(\vr)\}\rp_{\partial V}.
\end{equation}
A previous analysis with \texttt{BAM} \citep[][]{2019MNRAS.483L..58B} showed that when using reference simulations probing larger cosmological volumes (e.g. approximately three times that of the \texttt{SLICS}), only one realisation (one member of the reference set) is sufficient to generate an ensemble of number counts with precise summary statistics (up to the four-point statistics). 
However, numerical tests with the current setup have shown that this procedure, which is based on one single calibration (i.e. based on a single realisation), can suffer from effects that are due to the relatively small volume of the reference simulation (cosmic variance). To circumvent this, we generalise the sampling procedure of Eq.~(\ref{eq:mock_crea}) and allow each dark matter density field to be sampled with the bias and kernel independently inferred from one or more reference catalogues. That is, we calibrated $N_{ref}$ halo bias and kernels from the same number of \texttt{SLICS} references (as shown in \S\ref{sec:iter}) and constructed a total bias by `stacking' the independent halo bias, along with a kernel, obtained as the average from those of each reference:
\begin{eqnarray}
\label{eq:btot}
\mathcal{B}_{\rm tot}(N_{h}|\{\Theta\})&=&\sum_{j=1}^{N_{ref}}\mathcal{B}_{j} \lp N_{h}|\{\Theta\} \rp _{\partial V},  \nonumber \\
\mathcal{K}_{\rm tot}(k_{n})&=&\frac{1}{N_{ref}}\sum_{j=1}^{N_{ref}}\mathcal{K}_{j}(k_{n}).
\end{eqnarray}
Adding the results of different calibrations as expressed by Eq.~(\ref{eq:btot}) is equivalent to increasing the volume of the reference simulation (keeping the same minimum tracer mass and spatial resolution) to an effective value $V_{\rm eff}=N_{ref}^{1/3}L_{\rm box}^{3}$. However, we note that the stacked version of the halo bias $\mathcal{B}$ will differ from that measured from an $N$-body simulation probing the volume $V_{\rm eff}$ (with the same initial conditions) because of the absence of super-sample modes \cite[e.g.][]{2006MNRAS.371.1205R,2013PhRvD..87l3504T} in the halo bias, with this difference manifesting as an underestimation of high-density peaks, leading to biased estimates of covariance matrices of clustering probes. This is not a problem for the present case because we apply the kernel and bias of Eq.~(\ref{eq:btot}) to the DM field with the volume of the reference simulation.

%================================================================
\begin{figure}
\includegraphics[trim = 0.2cm 0cm -0.3cm 0cm, clip=true, width=0.51\textwidth]{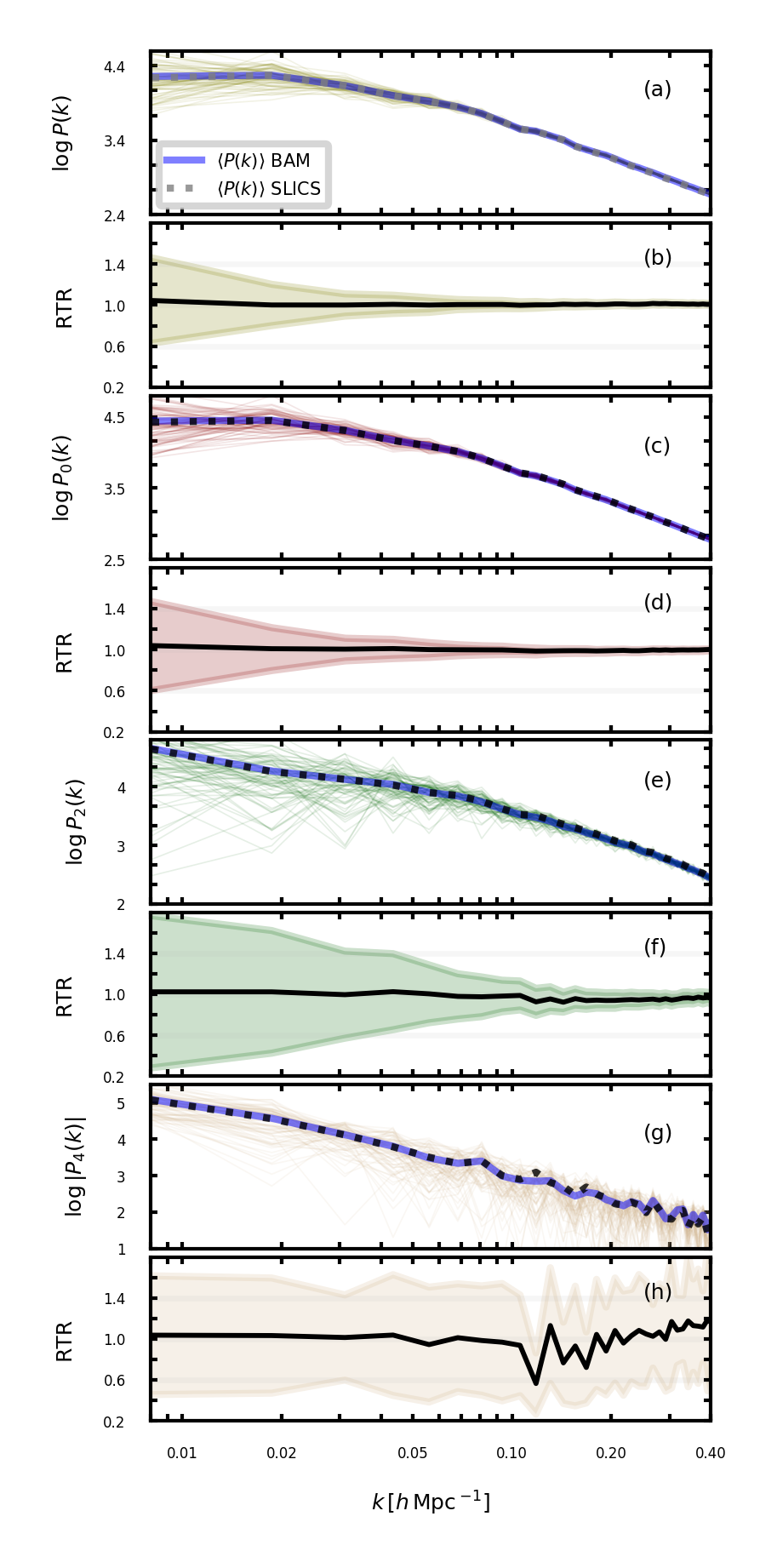}
\caption{Power spectrum of halo catalogues in real space $P(k)$ (panels (a) and (b)) and redshift space, with the latter represented by the monopole $P_{0}(k)$ (panels (c) and (d)), the quadrupole $P_{2}(k)$ (panels (e) and (f)), and the hexadecapole $P_{4}(k)$ (panels (g) and (h)). Panels (a), (c), (e), and (g) show the mean power spectrum from the $80$ \texttt{SLICS} realisations (grey dashed line) and the mean from the same number of \texttt{BAM} mocks (solid blue lines). Panels (b), (d), and (f) show the ratio (RTR) of the \texttt{BAM} mean spectrum to that of the reference. The shaded areas denote the standard deviation computed from the mean and variance of each set.}
\label{fig:powerhalos_mean}
\end{figure}
%================================================================

% ==========================================================
\begin{figure*}
\includegraphics[trim = 0cm 0cm 0cm 0cm ,clip=true, width=1.05\textwidth]{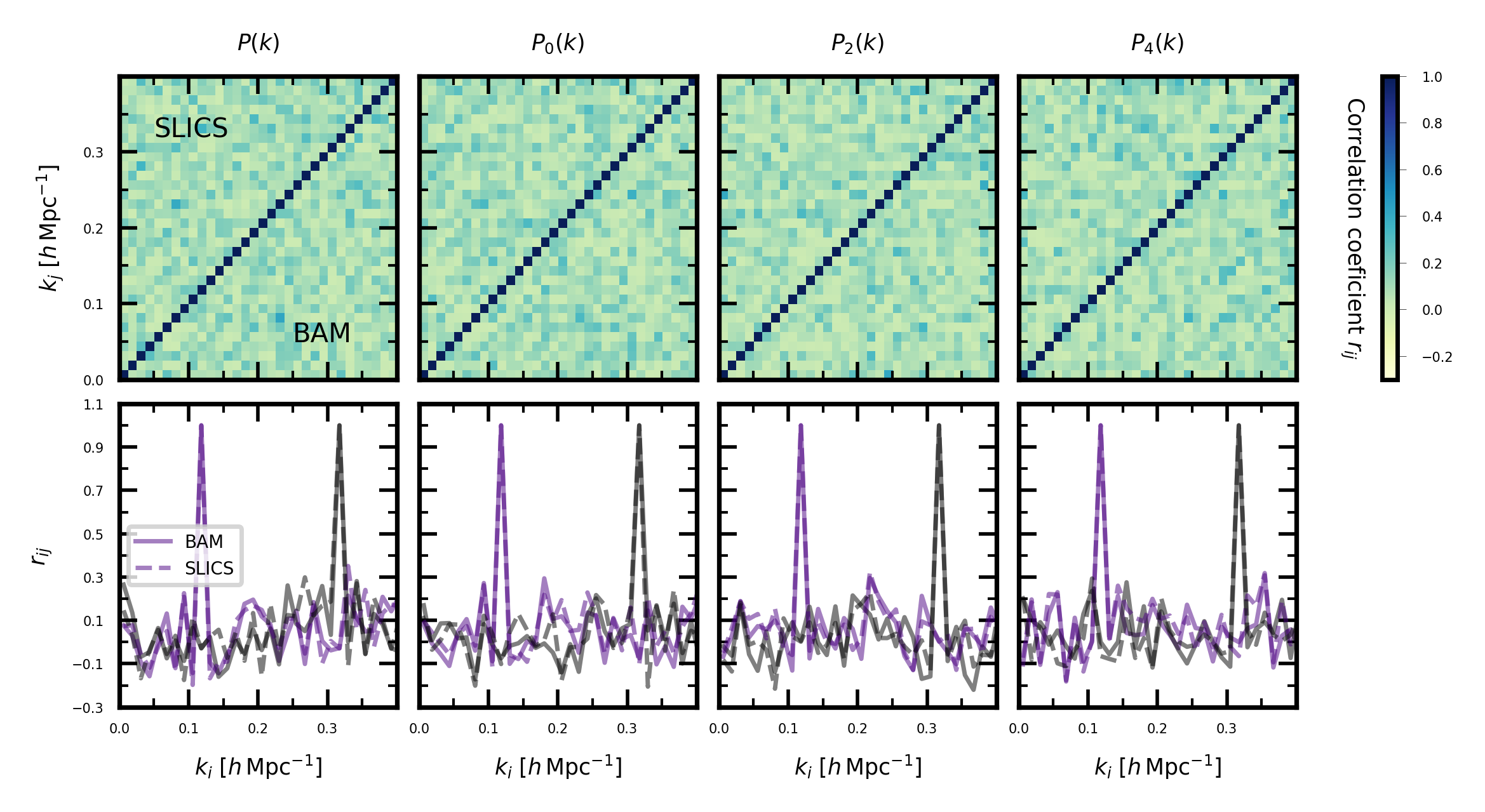}
\caption{Correlation matrix of halo power spectrum with \texttt{BAM}. Top row: Correlation matrix $r_{ij}=C_{ij}/\sqrt{C_{ii}C_{jj}}$ (where $C_{ij}$ is the covariance matrix) obtained from the \texttt{BAM} mock halo  catalogues and the \texttt{SLICS} references computed in real space and redshift space, the latter expressed through the monopole $P_{0}(k)$, the quadrupole $P_{2}(k),$ and the hexadecapole $P_{4}(k)$. Bottom row: Examples of elements of the correlation matrix $r_{ij}$ at two different wavenumbers $k_{j}\sim 0.1$ and $\sim 0.32\, h$ Mpc$^{-1}$ from the two sets of halo catalogues.}
    \label{fig:cova_power_rss_tweb}
\end{figure*}
% =======================================================
% --------------------------------------------------------
\begin{figure*}
    \includegraphics[trim = 0cm 0cm 0cm 0cm, clip=true, width=1.0\textwidth]{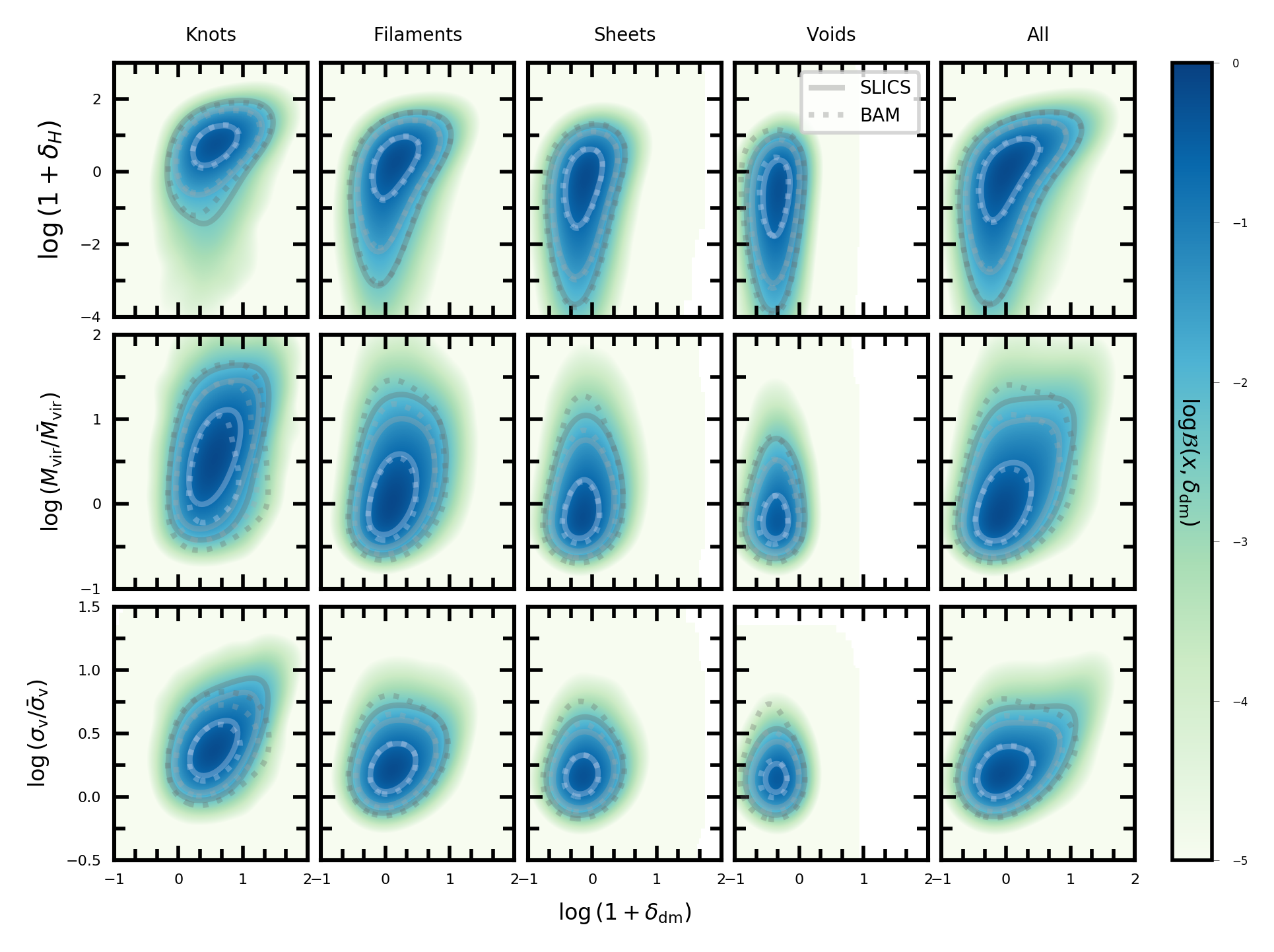}
\caption{Joint probability distribution $\mathcal{B}(x,\delta_{\rm dm})$ of halo properties $x$ (number counts, virial mass and velocity dispersion) and the underlying dark matter density, interpolated on a $192^{3}$ mesh using a CIC mass-assignment scheme and for different cosmic-web environments. Solid lines (and coloured regions) denote contours enclosing $98\%$ and $68\%$ of the total number of cells in a reference \texttt{SLICS} simulation. Dotted lines represent the same quantity obtained from one \texttt{BAM} halo catalogue.}
\label{fig:halo_prop}
\end{figure*}
% --------------------------------------------------------

In \citet[][]{2019MNRAS.483L..58B}, it was also demonstrated that the implementation of a kernel along with its corresponding halo bias (i.e. the set of outputs obtained from the learning phase with a given IC) is key to delivering halo fields with accurate summary statistics, in particular, the covariance matrix of the power spectrum. Therefore, the implementation of Eq.~(\ref{eq:btot}) can be a potential source of inaccuracy because the resulting kernel $\mathcal{K}_{\rm tot}$ has not necessarily attached a halo bias represented by $\mathcal{B}_{\rm tot}$. Instead, it is closer to what we can obtain using a reference simulation with {fixed-amplitude} initial conditions \citep[][]{2016MNRAS.462L...1A}; that is, Eq.~(\ref{eq:btot}) is designed to suppress cosmic variance in the kernel while keeping it in the total halo bias. Accordingly, these two quantities are not physically (statistically) compatible because the abundance of massive halos (or high-density regions) present in the total bias is sensitive to the amount of cosmic variance of the corresponding IC \citep[see e.g.][]{2013MNRAS.435.2065H,2016MNRAS.455..438A}, which is  the same cosmic variance that an averaged kernel is designed to suppress. Keeping this in mind, we implemented Eq.~(\ref{eq:btot}) to assess whether or not increasing the effective volume can provide better statistics at the number-counts level. We discuss the results in the following section.

In Appendix \S\ref{sec:unit}, we present the performance of \texttt{BAM} using larger cosmological simulations that were generated with an IC that was in turn generated with variance-suppressing methods \citep[][]{2019MNRAS.487...48C,2018ApJS..236...43G,2021MNRAS.508.4017M}. In forthcoming publications, we shall address this subject in more detail.

\subsection{Stage II: Generation of halo number counts}\label{sec:counts}

According to the discussion of the previous section, we generated $N_{\rm sim}$ halo number-count fields, increasing the effective volume by a factor of $2$ and $3$, that is, using $N_{ref}=2^{3}$ and $N_{ref}=3^{3}$ calibrations obtained from the same number of references.
To obtain a global picture of the performance of the different characterisations of the halo bias, we repeated this procedure for all the models proposed in \S\ref{sec:dmprops}. As an example, Figure \ref{fig:mean_tweb} 
shows a comparison between the summary statistics of $80$ \texttt{BAM} mocks and the same number of realisations from the reference set. This shows that \texttt{BAM} can generate mock catalogues whose mean and variance of halo power spectrum are in $5\%$ agreement with respect to the same statistics obtained from the reference set. We verified that similar results are obtained with the \texttt{TiWEB} model.

%------------------------------------------------------------------------------

To further assess the level of accuracy with respect to the same statistics from the reference, we use the three-point statistics in Fourier space. In particular, we explore the reduced bi-spectrum (or hierarchical three-point amplitude) $Q(\theta_{12}|k_{1},k_{2})$ \citep[][]{1980lssu.book.....P}, where $\theta_{12}$ is the cosine of the angle between the sides $k_{1}$ and $k_{2}$. We use estimates of the bispectrum to assess the precision of the method \footnote{We used the code \texttt{bispect} \url{https://github.com/cheng-zhao/bispec}} \cite[see e.g.][]{2012MNRAS.420.3469P,2012JCAP...02..047G}, using an isosceles configuration with $k_{1}=k_{2}=0.2 h\,$Mpc$^{-1}$  as an example. We remind the reader that this quantity is not constrained in the calibration procedure and can therefore be used as a yardstick to determine which of the models (or amount of effective volume) provides the best scenario to generate mock catalogues in the form of halo number counts. In the case of the \texttt{TkWEB} model (Fig.~\ref{fig:mean_tweb}), the signal of the reduced bispectrum is mostly within $5\%$ of that of the reference, except for low values of $\theta_{12}$, where the difference can be of the order of $10\%$. The variance of the bispectrum for such a configuration is also within $5\%-10\%$ of that of the reference. We verified that the results based on the \texttt{TIWEB} model show the same general trend.

The correlation matrix $r_{ij}=C_{ij}/\sqrt{C_{ii}C_{jj}}$ (where $C_{ij}$ is the covariance matrix) of the statistics under inspection (power spectrum in this case) for different halo bias models is shown in Fig.~\ref{fig:corr_power_models} along with a number of references used as training sets. In general, we can conclude that the \texttt{TkWEB} model generates correlation coefficients that are in good agreement with those from the reference. The \texttt{TIWEB} model displays extra coupling, which tends to decrease as the number of training references increases, which emphasises the need for larger cosmological volumes when one or two realisations are expected to be used as a training set and more detailed models are to be used. We have similarly verified that (as anticipated in \S\ref{sec:dmprops}) the \texttt{IkWEB} model displays strong mode coupling towards small scales even with $N_{ref}=27$, and we therefore discard it for the present applications. Such extra couplings are likely to be a consequence of the overfitting regime in which this model has been applied (as shown in \S\ref{sec:dmprops}), enhanced by the lack of compatibility between kernel and bias, as discussed in \S\ref{sec:using}. 

In terms of three-point statistics, Fig.~\ref{fig:corr_bisp_models} reveals that both the \texttt{TkWEB} and \texttt{TIWEB} models can generate sets of number counts whose noise in the  correlation matrix of the reduced bispectrum (for isosceles configurations) qualitatively agrees with that observed from the reference simulation, especially $N_{ref}=27$. Figure \ref{fig:corr_bisp_confs} complements the presentation of the performance of the statistical properties of the halo mocks by showing the behaviour of the reduced bispectrum ---in several configurations (using the \texttt{TkWEB} with $N_{ref}=27$)---in response to the corresponding signal from the reference: the left column shows ratios of the mean (solid lines) and variance (dashed lines) of the \texttt{BAM} ensemble to the results from the \texttt{SLICS}; the \texttt{BAM} mocks reproduce the mean reduced bispectrum, with average deviations (computed over the $\theta_{12}$-range) of $\sim 7\%$, while the variance shows an average deviation of $\sim 2\%$ with respect to the reference. We expect that the implementation of improved gravity solvers, which provide a more accurate description of the underlying DM field \citep[e.g.][]{2023arXiv230103648K}, will help to reduce the difference in the mean signal.  

The right column of Fig.~\ref{fig:corr_bisp_confs} shows two elements of the correlation matrix of the reduced bispectrum as obtained from the reference (solid lines) and the \texttt{BAM} set (dashed lines), showing that in general, the \texttt{BAM} approach is able to replicate the noise in the correlation matrix of three-point statistics (in real space).

Based on these results, we adopt the \texttt{TkWEB} model to generate independent realisations of halo number counts; this model will be used to generate the final set of halo catalogues as described in the following section. We used $N_{ref}=27$ references, but note that this particular model is already good enough to allow us to use the calibration from only one reference simulation. 

\subsection{Stage III-a: Assignment of halo coordinates}\label{sec:coords}
To transform the set of number counts obtained in \S\ref{sec:counts} into an ensemble of discrete tracers, we assign coordinates and velocities following the approach of \citet[][]{2016MNRAS.456.4156K}, which consists in using the phase-space coordinates of dark matter particles generated by the approximated gravity solver (\S\ref{sec:gravity}). The sampling of the halo number counts field is complemented with a set of random tracers (e.g. tracers with random coordinates within each cell), which are used when, at a given cell, the number of halos requested is larger than the available number of dark matter particles. The fraction of such random tracers depends on the redshift of the reference, and for the current setup represents $\sim  20\%$ of the total number of tracers. 

We use dark matter particles to sample the halo number count field in an attempt to maintain a precise clustering signal on scales below the fiducial cell size. As the randomly distributed tracers impact the shape of the power spectrum when analysed at higher Nyquist frequencies, \texttt{BAM} introduces a subgrid modelling based on the collapse of the random tracers towards their closest DM particles. This collapse is modulated by a fraction $f_{\rm col}$ of the separation between each random tracer and its nearest DM particle. That is, for a separation between a random particle and its closest DM particle $d_{\rm r}$, we displace the former towards the latter such that their new separation is the product $f_{\rm col}d_{\rm r}$. This is depicted in Fig.~\ref{fig:col_scheme}. Numerical experiments (some of which are discussed in \S\ref{sec:unit}) have shown that this parameter depends on the redshift and the nature of the approximated gravity solver \citep[][]{Bala2023}. For the current setup,  $f_{\rm col}\sim 0.35$ provides a good description of the halo power spectrum. Furthermore, the parameter $f_{\rm col}$ can be generalised to depend on halo properties once these are assigned (\S\ref{sec:hprops}). 

% --------------------------------------------------------
\begin{figure}
    \includegraphics[trim = 0.2cm 0cm 0cm 0cm, clip=true,width=0.5\textwidth]{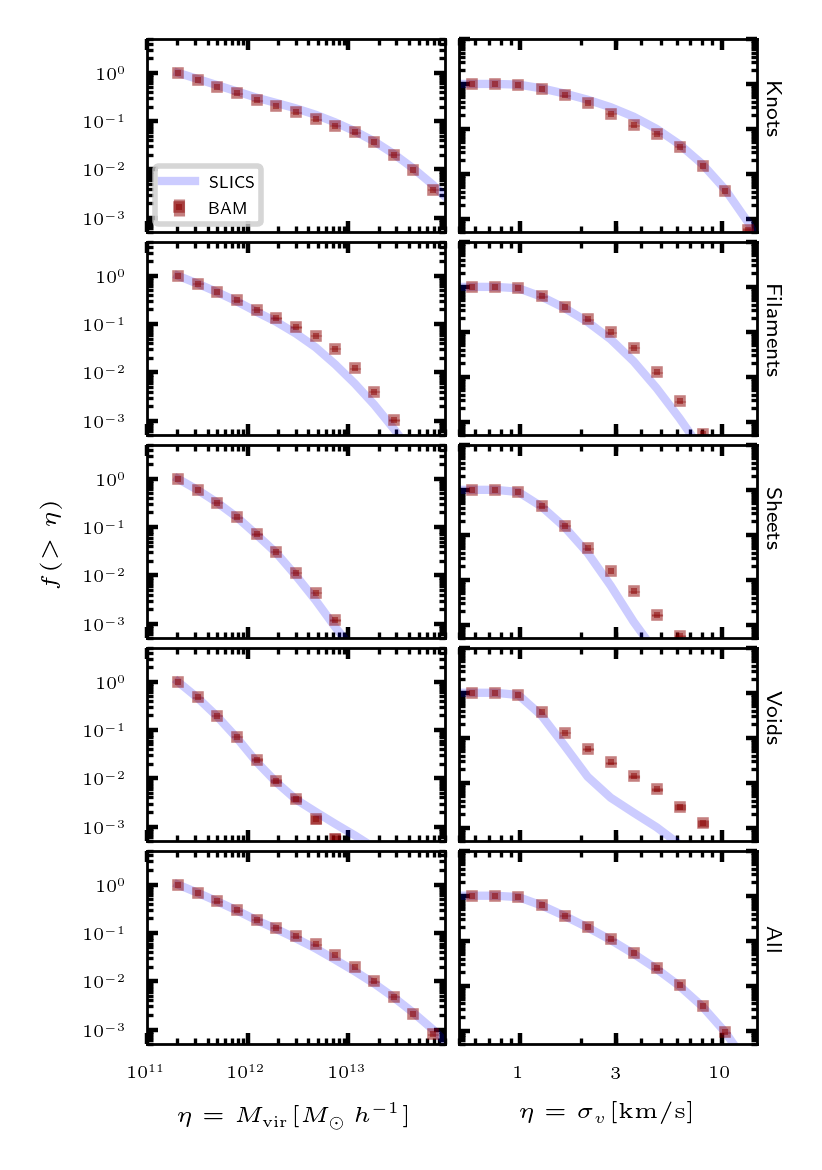}
\caption{Cumulative halo abundance as a function of virial mass $M_{\rm vir}$ (left column) and velocity dispersion $\sigma_{v}$ (right column)  obtained in different cosmic-web types (normalised to the number of halos in each cosmic-web type). Error bars indicate the mean and standard deviations computed from $80$ realisations.}
\label{fig:abundance}
\end{figure}
% --------------------------------------------------------
The panel (a) of Fig.~\ref{fig:powerhalos_mean} shows the mean real-space power spectrum obtained from an ensemble of $80$ \texttt{BAM} halo catalogues with coordinates assigned as previously described. Each realisation is embedded in a $400^{3}$ cubic mesh using the triangular-shaped-cloud interpolation scheme \citep{1988csup.bookH} \footnote{All power spectra are computed using the \texttt{COPOWER} code, \url{https://github.com/balaguera/COPOWER}}. Panel (b) shows that the accuracy of the mean power from the \texttt{BAM} (with respect to the \texttt{SLICS}) is below $3\%$ up to $k\sim 0.4\,h\,$Mpc$^{-1}$.
%------------------------------------------------------------------------------
\begin{figure*}
\includegraphics[trim = 0cm 0cm 0cm 0cm, clip=true, width=1.0\textwidth]{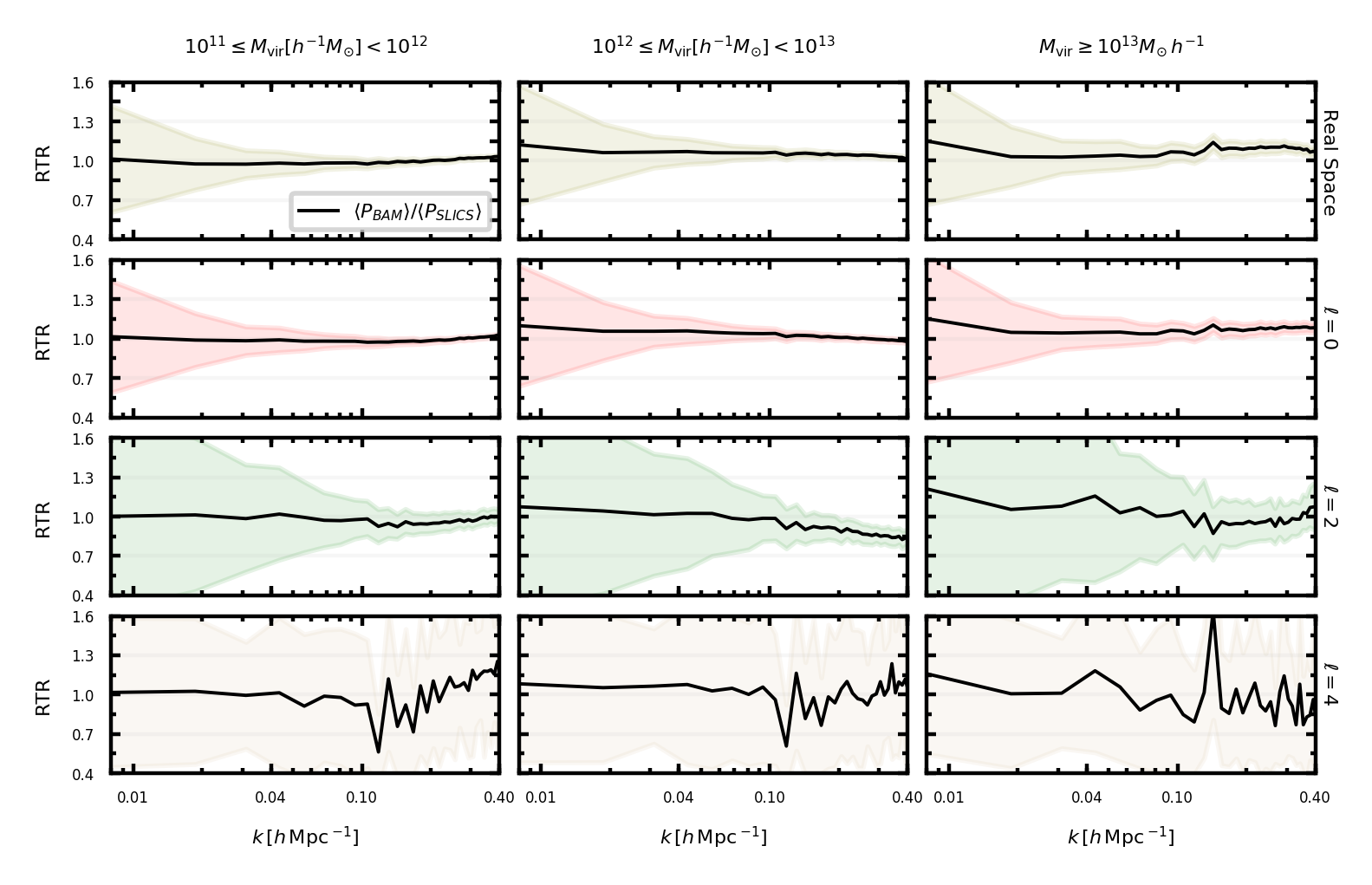}
\caption{Ratio (ratio-to-reference) between the mean power spectrum from the set of $80$ \texttt{BAM} mock halo  catalogues and that obtained from the same number of \texttt{SLICS} catalogues, both in real and redshift space (monopole $\ell=0$, quadrupole $\ell=2$, hexadecapole $\ell=4$), in three bins of halo virial mass. The shaded areas denote the $1\sigma$ region (standard deviation) computed from the means and their respective errors.}
\label{fig:powerhalos}
\end{figure*}
%------------------------------------------------------------------------------

%==============================================================================================================
%==============================================================================================================
%==============================================================================================================
\subsection{Stage III-b: Assignment of halo velocities}\label{sec:vels}
The displacement obtained with $\texttt{ALPT}$ (see Eq.~(\ref{eq:disp})) provides the velocities  of dark matter particles at their Eulerian coordinates $\vr:$
\be\label{eq:vdisp}
\vv_{ALPT}(\vr,z)=\vv_{2LPT}(\vq,z)\otimes \mathcal{G}_{s}(\vq)+\lp 1-\mathcal{G}_{s}(\vq,)\rp\otimes \vv_{sc}(\vq,z),
\ee
where the \texttt{2LPT} velocity field is written as \citep[see e.g.][]{10.1093/mnras/264.2.375,2014MNRAS.439L..21K}
\be\nonumber
\vv_{2LPT}(\vq,z)=\left[-f^{(1)}D^{(1)}\nabla_{\vq}\phi^{(1)}(\vq)+f^{(2)}D^{(2)}\nabla_{\vq}\phi^{(2)}(\vq)\right]Ha.
\ee
In this expression, $f^{(i)}\equiv f^{(i)}(z)={\rm d}\ln D^{(i)}(a)/{\rm d}\ln a$ are the growth indices computed as $f^{(1)}(z)\sim \Omega_{\rm mat}(z)^{5/9}$ and $f^{(2)}(z)\sim 2\Omega_{\rm mat}(z)^{6/11}$ \citep[see e.g.][]{1991MNRAS.251..128L}. The velocity field associated with the SC model is analogously derived as $\vv_{sc}(\vq,z)=\nabla \psi_{SC}(\vq,z),$ where $\psi_{SC}(\vq,z)$ is the solution of the Poisson equation
\be\nonumber
\nabla^{2}\psi_{SC}(\vq,z)=-f^{(1)}(z)H(z)aD^{(1)(z)}\delta^{(1)}(\vq) \left(1-\frac{2}{3}D^{(1)}(z)\delta^{(1)}(\vq) \right)^{-1/2}.
\ee
We assign peculiar velocities to dark matter halos in two steps. First, we generate a velocity field in Eulerian space using the velocities computed with Eq.~(\ref{eq:vdisp}) and implement an NGP interpolation scheme. It is well known that this kind of approach introduces sampling artefacts due to the fact that it relies on the particles to generate the velocity field \citep[see e.g.][]{2013PhRvD..88j3510Z,2015PhRvD..91d3522Z}, which means cells without tracers are incorrectly assigned a null velocity. Alternatives such as the `{nearest point' }\citep[see e.g.][]{2015PhRvD..91d3522Z,2018ApJ...861...58C} or more sophisticated algorithms such as the Delaunay Tesselation \cite[see e.g.][]{2007MNRAS.382....2R} or the Kriging scheme \cite[see e.g.][]{2015PhRvD..92h3527Y} are designed to reduce the spurious bias introduced by these sampling artefacts. We implement a hybrid approach and use NGP as the primary method, assigning  
to empty cells the average velocity computed from the first neighbour cells. A second step consists of a trilinear interpolation of the resulting velocity field at the position of both dark matter particles and random tracers (introduced in the previous section).

%------------------------------------------------------------------------------
\begin{figure*}
\includegraphics[trim = 0cm 0cm 0cm 0cm ,clip=true, width=0.335\textwidth]{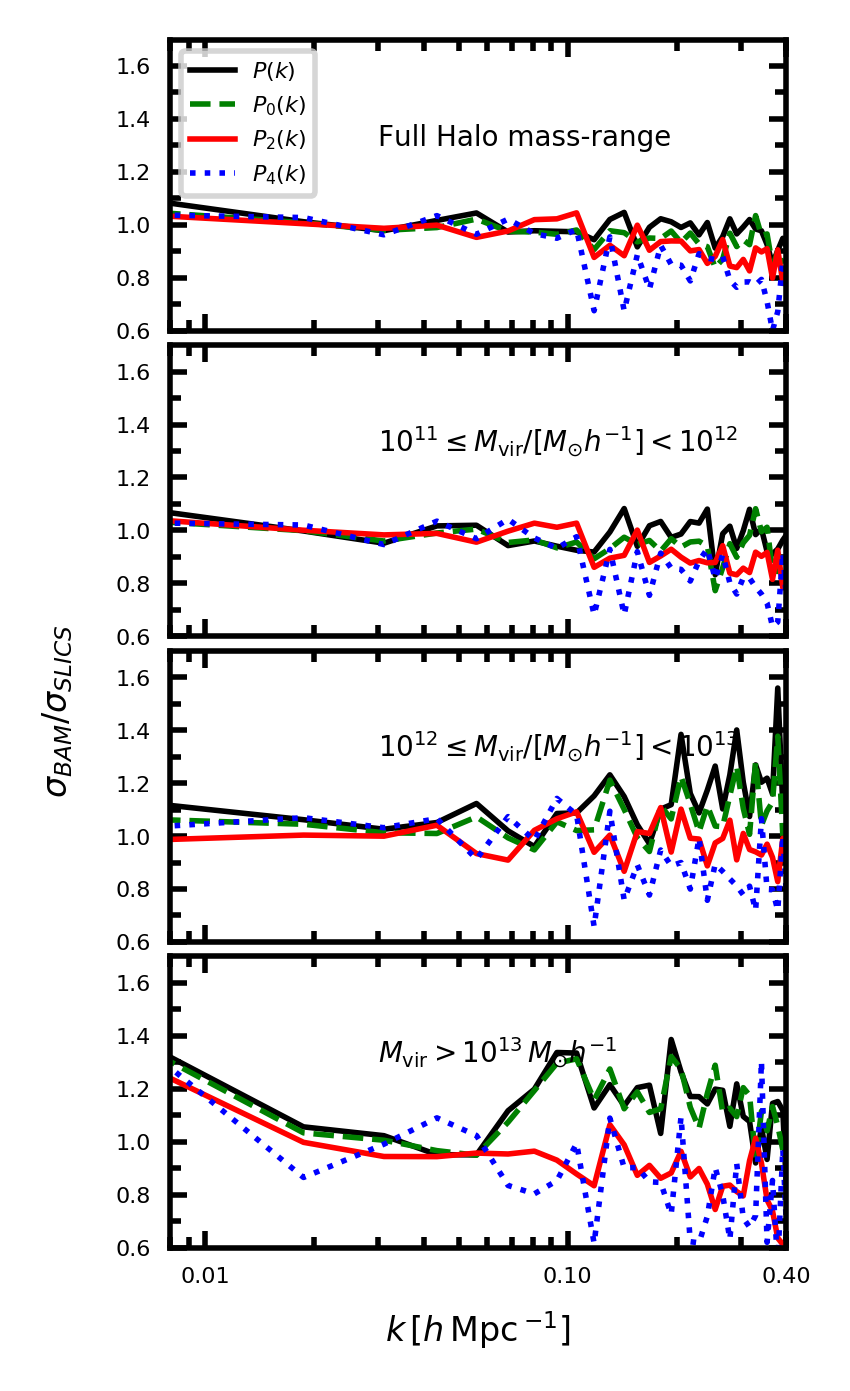}
\includegraphics[trim = 0cm 0cm 0cm 0cm ,clip=true, width=0.62\textwidth]{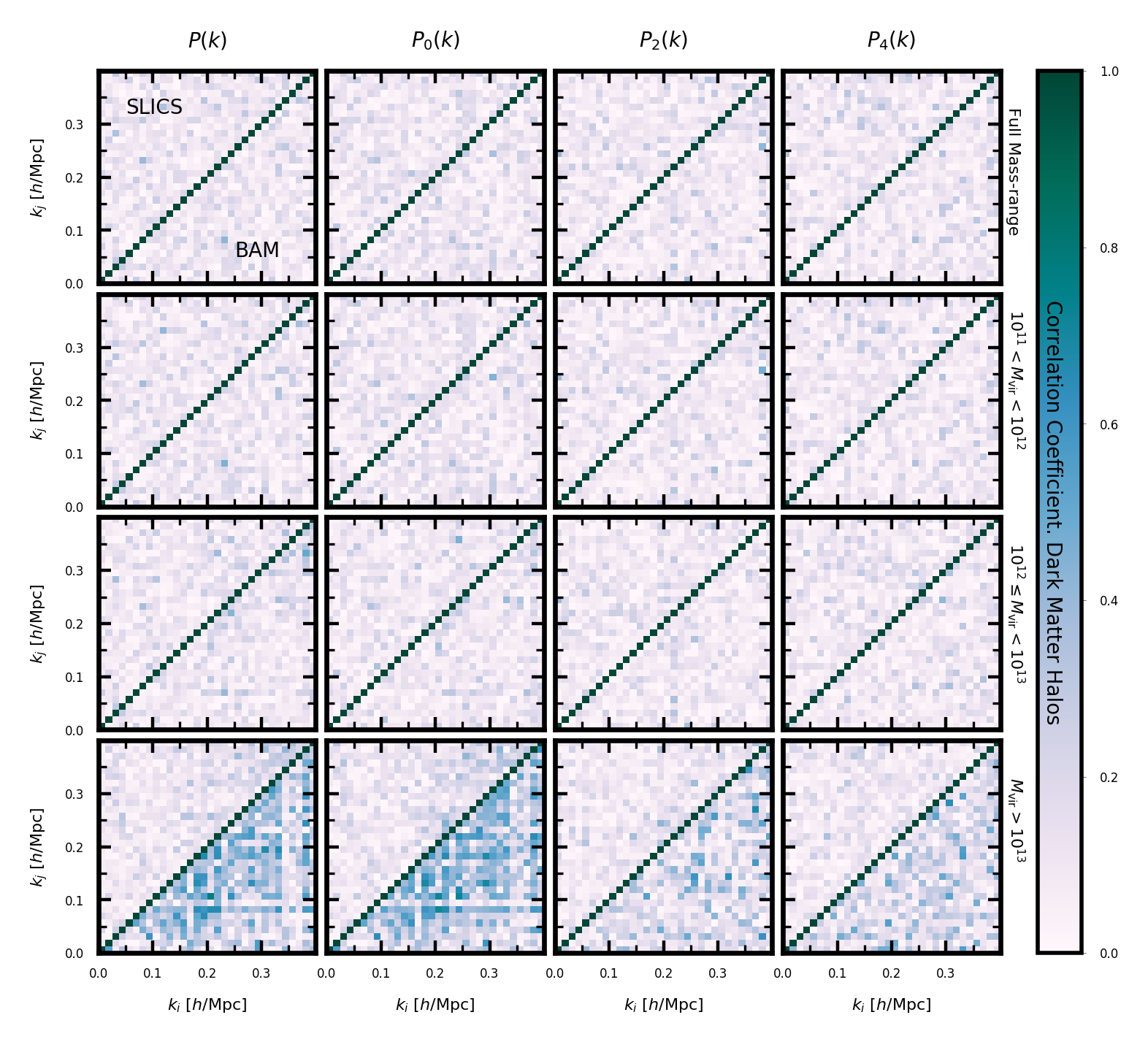}
\caption{Correlation matrix of power spectrum. Left column: Ratio between the variance of the power spectrum from the \texttt{BAM} mocks and that measured from the \texttt{SLICS} in  different halo-mass bins. Right column: Correlation coefficients obtained from the \texttt{BAM} mock halo catalogues and the \texttt{SLICS} references computed in real and redshift space, with the latter expressed through the monopole $P_{0}(k)$, the quadrupole $P_{2}(k),$ and the hexadecapole $P_{4}(k)$. Three bins of halo mass are shown (rows).}
    \label{fig:cova_power_rss}
\end{figure*}
%------------------------------------------------------------------------------

Figure~\ref{fig:vel_dist} shows an example of the resulting distribution of the modulus of the halo peculiar velocity $v=|\vec{v}|$ from one realisation of \texttt{SLICS} and \texttt{BAM} sets (sharing the same seed). One strong feature arising from this comparison is the difference in the abundance (in terms of $v$) towards high velocities: above $\sim 300$ km$/s$ the abundance from the \texttt{BAM} halos (i.e. \texttt{ALPT}) is, in general, underestimated with respect to the reference. To correct this deviation, we introduce an isotropic correction to the $i$-th component of the velocity of  each particle in the form $v_{i}^{0}(\vr)\to v_{i}(\vr)=\gamma(\vr) v^{0}_{i}(\vr)$, where $\gamma(\vr)=(1+\delta_{\rm dm}(\vr))^{\alpha}$. Numerical experiments have revealed that $\alpha\sim 0.2$ leads to good agreement in the halo velocity distribution, as is also presented in Fig.~\ref{fig:vel_dist}.
We verified that this correction is indeed needed to obtain the good agreement between the clustering signal of the \texttt{BAM} mocks and that from the reference. We speculate that the origin of this correction is linked to the lack of small-scale modelling of coherent flows in \texttt{ALPT} combined with the resolution used in the analysis. A more detailed analysis (exploring e.g. redshift and cosmology dependencies) will be presented in future publications.

The second correction to the velocities is applied as part of the subgrid modelling, focusing again on the random tracers (as depicted in Fig.\ref{fig:col_scheme}). In this case, along with the collapse towards the closest dark matter tracer (discussed in the previous section), we induce a rotation (or collapse) of the random velocities, modifying its orientation (through an angle $\beta$) and magnitude (through a parameter $\zeta$),  which can be a function of the tracer properties or local density. In this work, we empirically set this parameter to $\beta=(1-f_{\rm coll})(\pi-\alpha)$ if $\alpha < \pi$, and $\beta=(1-f_{\rm coll})(\alpha-\pi)$ if $\alpha \geq \pi$, with $\zeta=1$; that is,  we apply a rotation to the velocity vector to align it with the axis connecting the random particle and the dark matter particle, keeping its magnitude fixed. We verified that the effect of $\beta\neq 0$ helps to improve the signal in redshift space towards small scales and leave a thorough study of their impact in the velocity field to a future study.

The performance of the velocity assignment in terms of the two-point statistics is presented in panels (c) to (h) of Fig.~\ref{fig:powerhalos_mean}, where we show the mean halo power spectrum in redshift space. This signal is obtained by transforming the halo coordinates $(x,y,z)$ along a line-of-sight axis (taken to be one of the three Cartesian coordinates, e.g. the $z$-direction) using the distant observer approximation to its redshift coordinate via $z\to s=z+v_{z}/(aH(a))$ \citep[see e.g.][]{1987MNRAS.227....1K}. The clustering in this space is summarised through the Legendre decomposition which, according to the distant observer approximation, can be measured as \cite[see e.g.][]{1998ASSL..231..185H}
\be\label{pdec}
P_{\ell}(k_{i})=\frac{2\ell+1}{2}
\sum_{\vk \in \Delta k_{j}}|N_{\rm h}(\vk)|^{2} \mathcal{L}_{\ell}(\mu_{k})-\delta^{K}_{\ell 0}S,
\ee
where the sum denotes averages in spherical shells, $\mu_{k}=\cos \vk \cdot \hat{z}=k_{z}$, $|N_{\rm h}(\vk)|^{2}$ is the three-dimensional halo power spectrum, $\mathcal{L}_{\ell}(x)$ is the Legendre polynomial of order $\ell,$ and $S$ is the Poisson shot noise (as in Eq.~\ref{eq:mpow}). We measure the monopole ($\ell=0$), the quadrupole ($\ell=2$), and the hexadecapole ($\ell=4$) as main statistical probes of redshift-space distortions. 

%------------------------------------------------------------------------------
\begin{figure}
\includegraphics[trim=0.2cm 0cm 0cm 0cm, clip=true,width=0.5\textwidth]{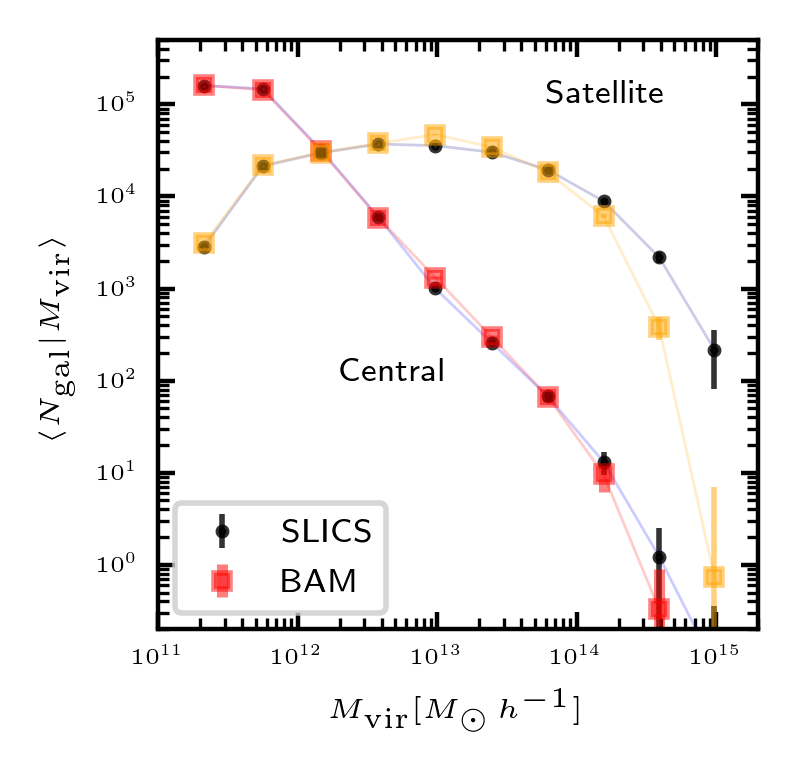}
\caption{Mean number of galaxies (central and satellite) from the \texttt{BAM} and \texttt{SLICS} ensembles as a function of host halo mass. The points with error bars denote the mean and variance from each ensemble.}
\label{fig:abundance_gals}
\end{figure}
%------------------------------------------------------------------------------
In general, the redshift-space power spectrum probed on scales up to $k\sim 0.4 h\,$Mpc$^{-1}$ agrees within the $1\sigma$ uncertainty region with that of the reference, as is demonstrated by panels (c) to (h) in Fig.~\ref{fig:powerhalos_mean}. 
Figure \ref{fig:cova_power_rss_tweb} shows the correlation coefficient of the halo power spectrum in real and redshift space computed from the halo distribution. The bottom panels of Fig. 11 show two elements of the correlation matrix, which reveal good agreement between the two compared sets, both in terms of the width of the correlation coefficients and the underlying noise. We verified that this agreement is also observed when using \texttt{BAM} realisations with seeds different from those of the reference set. 

%==============================================================================================================

\subsection{Stage III-c: Assignment of halo properties}\label{sec:hprops}
Given that the \texttt{BAM} mock halo catalogues are to be considered as the building blocks of galaxy catalogues (though with the  implementation of an HOD model), the \texttt{BAM} algorithm pays special attention to the assignment of halo properties such as the virial mass $M_{\rm vir}$ and the velocity dispersion $\sigma_{\rm v}$. This step is indeed a critical and far-from-trivial task within the construction of \texttt{BAM} mock catalogues \citep[][]{Bala2023}, given that a simultaneous generation of precise clustering and halo properties would imply the assessment of the distribution of pairs in all possible bins of halo properties, a computation which goes openly against the need for speed in the generation of mock catalogues.

The procedure encoded in \texttt{BAM} finds its motivations in early methods developed by \citet[][]{2015MNRAS.451.4266Z} \citep[see also][]{2015MNRAS.446.2621C}, which were envisaged to generate luminous red galaxy catalogues \citep[see e.g][]{2016MNRAS.456.4156K, 2016MNRAS.460.1173R}. Those algorithms used the properties of the underlying dark matter density field, as in \texttt{BAM}. However, \texttt{BAM} takes the method to a greater level of detail, in which more properties of the dark matter and dark matter tracers are considered.

The assignment procedure in \texttt{BAM} (see Fig.\ref{fig:bam_scheme}) relies on a hierarchical approach in which a `{main property}' is defined and assigned, followed by the assignment of secondary properties using the scaling relation with respect to the main property. To determine the main property, different options can be considered; for example, selecting the halo property with the tightest correlation with the underlying dark matter density field, or choosing the halo property that drives the main dependencies in the HOD framework. The latter option would lead us to treat the virial mass $M_{\rm vir}$ as the main property (which mainly determines galaxy number-count statistics in the HOD framework), followed by the velocity dispersion (which dictates the redshift space distribution of satellite galaxies). Nevertheless, given that \texttt{BAM} is designed to explore environmental dependencies (defined through the dark matter density field), we select the first option. Accordingly, while the correlation between virial mass and local dark matter density is $\sim 28\%$, the velocity dispersion displays tighter correlations ($\sim 58\%$) with the local dark matter density. This is not surprising, as quantities directly derived from the dynamical properties of the dark matter particles in halos trace the depth of the potential wells  very well (see Appendix \ref{ap:hier}) and are less prone to ambiguities typical of the definition of the mass of a dark matter halo \citep[see e.g.][]{2011MNRAS.416.2388S,2019ApJ...887...17Z}. In Appendix \ref{sec:multi}, we describe the methodology implemented for the assignment of properties within \texttt{BAM}.  Figure~\ref{fig:halo_prop} shows the correlation between different halo properties with respect to the underlying dark matter density field, again for different cosmic-web types. We verified that the scaling relations between virial mass and velocity dispersion show an acceptable level of agreement with those from the reference set.
%==============================================================================================================
%==============================================================================================================
%==============================================================================================================

\subsection{Results}\label{sec:results}
Fig.~\ref{fig:abundance} shows the halo abundance as a function of the two assigned halo properties from a set of $80$ realisations of \texttt{BAM} and the same number of \texttt{SLICS} simulations. We see good agreement  in general, but this agreement partially breaks down for tracers with high velocity dispersion in low-density regions (voids) where \texttt{BAM} overestimates the abundance in terms of that particular property. 

Similarly, we also verified that the multi-scaling approach generates better results than a direct assignment of properties. Although this approach naturally goes towards solving the problem of halo exclusion, it relies on the specification of the different thresholds, and we checked that the precision of the mean power spectrum of halos is sensitive to such figures, especially on the high-mass halo population. New alternatives are being explored and will be presented in forthcoming publications.

%------------------------------------------------------------------------------
\begin{figure*}
\includegraphics[trim = 0cm 0cm 0cm 0cm,clip=true, width=1.0\textwidth]{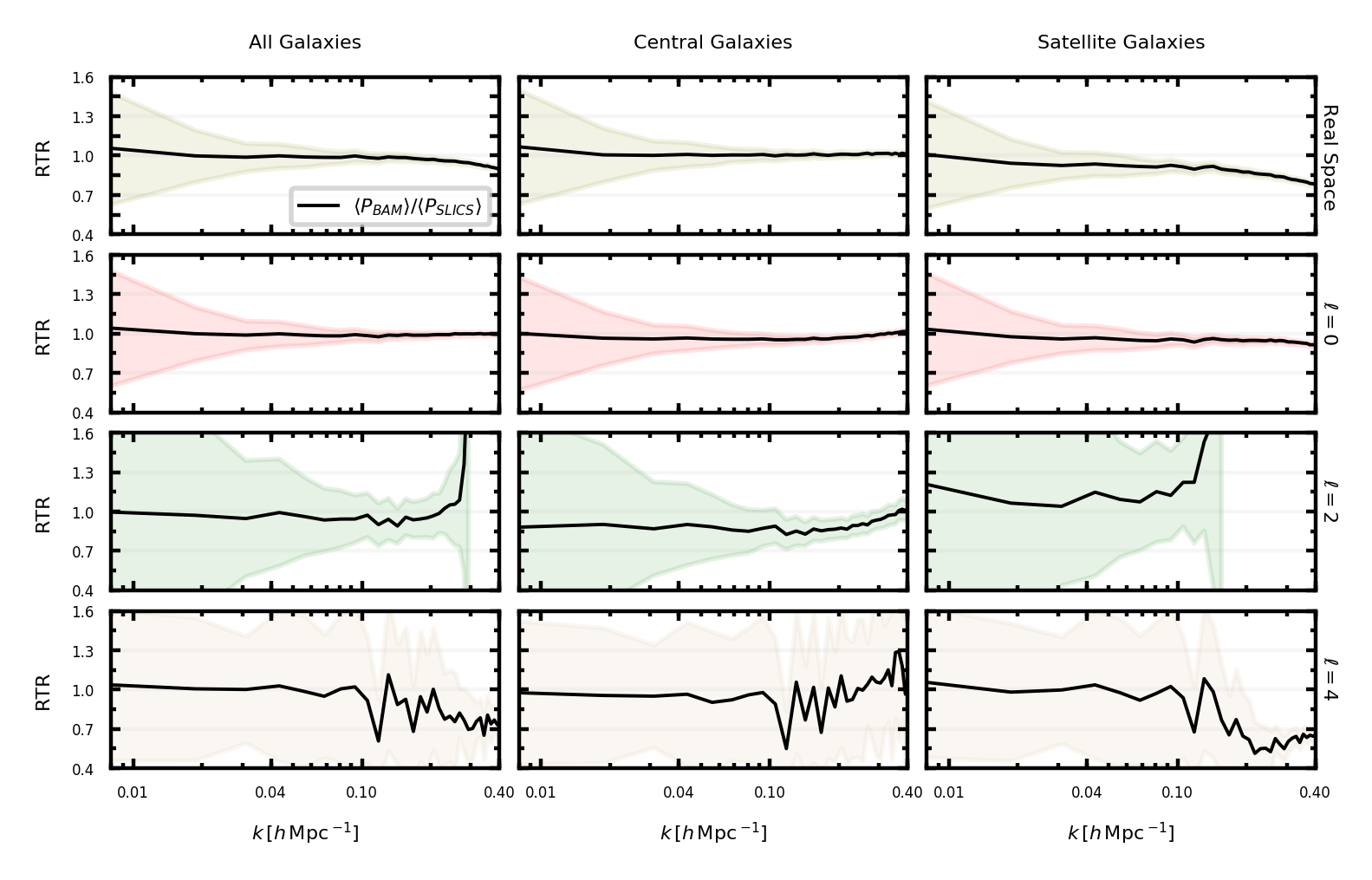}
\caption{Same as Fig. \ref{fig:powerhalos} but for the mock galaxy catalogues generated as explained in \S\ref{sec:gals}.}
    \label{fig:power_rss_gal}
\end{figure*}
%------------------------------------------------------------------------------

Figure \ref{fig:powerhalos} shows the ratio between the mean power spectra from the \texttt{BAM} set and that from the \texttt{SLICS}, both in real and redshift space and in three disjoint halo-mass bins. The area around that curve denotes the standard deviation. In general, the trend shown as a function of the halo mass is similar in the two sets of mock catalogues. However, closer inspection reveals $\sim 5\%$ deviations towards small scales both in real and redshift space, and in particular, for the most massive halos.

Figure~\ref{fig:cova_power_rss} shows the variance (left column) and correlation matrix (right column) of the halo power spectrum in real and redshift space for the full halo population. We can read from this figure that in real space, the \texttt{BAM} mocks display a closer correlation between modes on small scales compared to the reference simulation. The situation is mildly better in redshift space, where indeed the correlation matrix for the quadrupole agrees to a greater  extent with the \texttt{SLICS} simulations. The most significant discrepancy appears when exploring the clustering as a function of the halo mass. In particular, high-mass halos display covariance matrices of the power spectrum with a strong mode coupling. Such coupling comes from realisations where the power spectra deviate considerably from the expected mean of the ensemble (i.e. that from the reference suite). 
These discrepancies originate from two different aspects of the \texttt{BAM} approach. On one hand, the assignment of halo properties turns out to be complex, especially for massive objects, where effects such as halo exclusion (see Appendix A) are not fully modelled. On the other hand, the deviations seen in redshift space are inherited from those in real space, plus any remaining deviation from the true halo velocity field from the velocity field generated from the \texttt{ALPT}. These two aspects mean that there is room for improvement in the assignment of coordinates, velocities, and properties of the halo population, especially towards small scales.

With the procedures described in the previous sections, we generated $770$ mock halo catalogues based on the same number of initial conditions. One of the great advantages of the method is the small computing-time requirements: the generation of this set of halo catalogues was achieved in $\sim 2$ days ($\sim 4$ minutes per mock) using a work station with $128$ threads and $256$ Gb of random access memory (RAM). %(ram-memory consumption per mock of the process is $\sim 50$ Gb). 

%==============================================================================================================
%==============================================================================================================
\section{Stage IV: Construction of galaxy catalogues}\label{sec:gals}

One of the main advantages of \texttt{BAM} is its capability to provide catalogues of different dark matter tracers, all sharing the same underlying dark matter and halo distribution. 
This is key to providing covariance matrices for multi-tracer analysis \cite[see e.g. ][]{2012PhRvD..86j3513H,2013MNRAS.432..318A,2016MNRAS.455.3871A,2020RAA....20..158W,2021MNRAS.503.1149Z}. While this is not a unique feature of this method \citep[see e.g.][]{2021MNRAS.503.1149Z}, it represents an improvement over approaches that need to be calibrated with a particular galaxy population.

To assign galaxies to the dark matter halos of \texttt{BAM}, we implemented the HOD prescription based on the \emph{high-mass quenched} model \cite[see e.g.][]{2020MNRAS.497..581A}, which describes the abundance of emission line galaxies (ELGs) in dark matter halos \cite[see e.g.][]{2018MNRAS.474.4024G}. This model suppresses the probability for the central ELG galaxies to be found in very massive dark matter halos. In particular, the probability that a halo of mass $M_{\rm vir}$  hosts a central ELG is expressed as 
\be\label{eq:hdoc}
\langle N_{\rm c}|M_{\rm vir}\rangle= A\phi(M_{\rm vir})\Phi(\mathcal{M})  +\frac{1}{2Q}\left[1+{\rm erf} \lp \log_{10} \left[\frac{M_{\rm vir}}{M_{c}}\right]^{1/\sigma_{M}} \rp \right]
,\ee
where $\phi(x)=\mathcal{N}(\log_{10}M_{c},\sigma_{M})$, $\mathcal{M}\equiv \gamma M_{\rm vir}$, $\Phi(x)=\int_{-\infty}^{x}\phi(x)\dd x$ and $A=2(p_{\rm max}-1/Q)/{\rm max}(2\phi(x)\Phi(x))$. Here, $Q$ denotes the quenching efficiency, $p_{\rm max}$ controls the saturation level of occupancy, and $M_{c}$ is the cut-off mass for ELGs, which determines the maximum of the occupation distribution. The number of ELG satellites is generated from a Poisson realisation with a mean modelled as a power law with a lower cut-off:
\be\label{eq:hdos}
\langle N_{s}|M_{\rm vir}\rangle = \lp\frac{M_{\rm vir}-\kappa M_{c}}{M_{1}}\rp^{\alpha},
\ee
where $M_{1}$ is a characteristic satellite mass, while the parameter $\kappa$ defines a cut-off mass (in units of $M_{\rm c}$) below which the occupancy of satellites drops to zero. The satellites are distributed within dark matter halos following an NFW density profile \citep[][]{1996ApJ...462..563N}. On the other hand, the random components of the velocities of the satellite galaxies are derived from a normal distribution, $v^{s}\curvearrowleft \mathcal{N}(0,\sigma_{\rm v})$. The total velocity of the satellite galaxies is then given by the sum of the halo peculiar velocities of the parent halos and the random component. The velocities of the central galaxies are the same as those of their parent halos.

The HOD prescription of Eqs.~(\ref{eq:hdoc}) and (\ref{eq:hdos}) has been simultaneously applied to the \texttt{SLICS} and \texttt{BAM} halos to obtain their respective galaxy catalogues \citep[see ][for the set of parameter $\{Q,\gamma,M_{c},\kappa,\alpha,\sigma_{M},M_{1}\}$]{2020MNRAS.497..581A}. The performance of the \texttt{BAM} galaxy mocks in terms of the galaxy occupation distribution is presented in Fig.~\ref{fig:abundance_gals}, where we observe how the larger deviations with respect to the reference are embodied in the satellite population at the high-halo mass end ($\sim 60\%$ difference at $M_{\rm vir}\sim 4\times 10^{14}\,M_{\odot}h^{-1}$, mass scale below which $\geq 99 \%$ of the sample is contained). 

%------------------------------------------------------------------------------
\begin{figure*}
\includegraphics[trim = 0cm 0cm 0cm 0cm ,clip=true, width=0.335\textwidth]{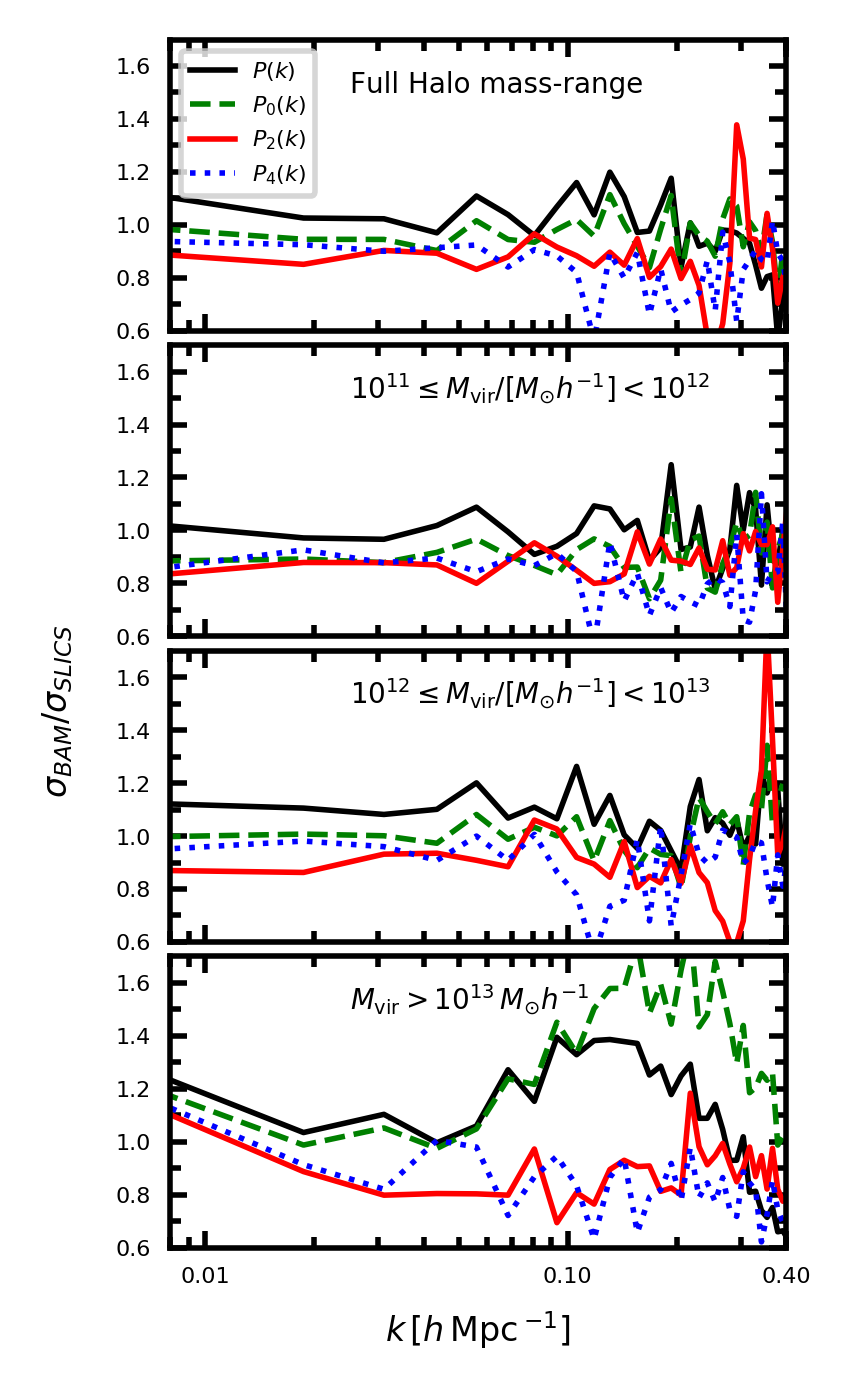}
\includegraphics[trim = 0cm 0cm 0cm 0cm ,clip=true, width=0.62\textwidth]{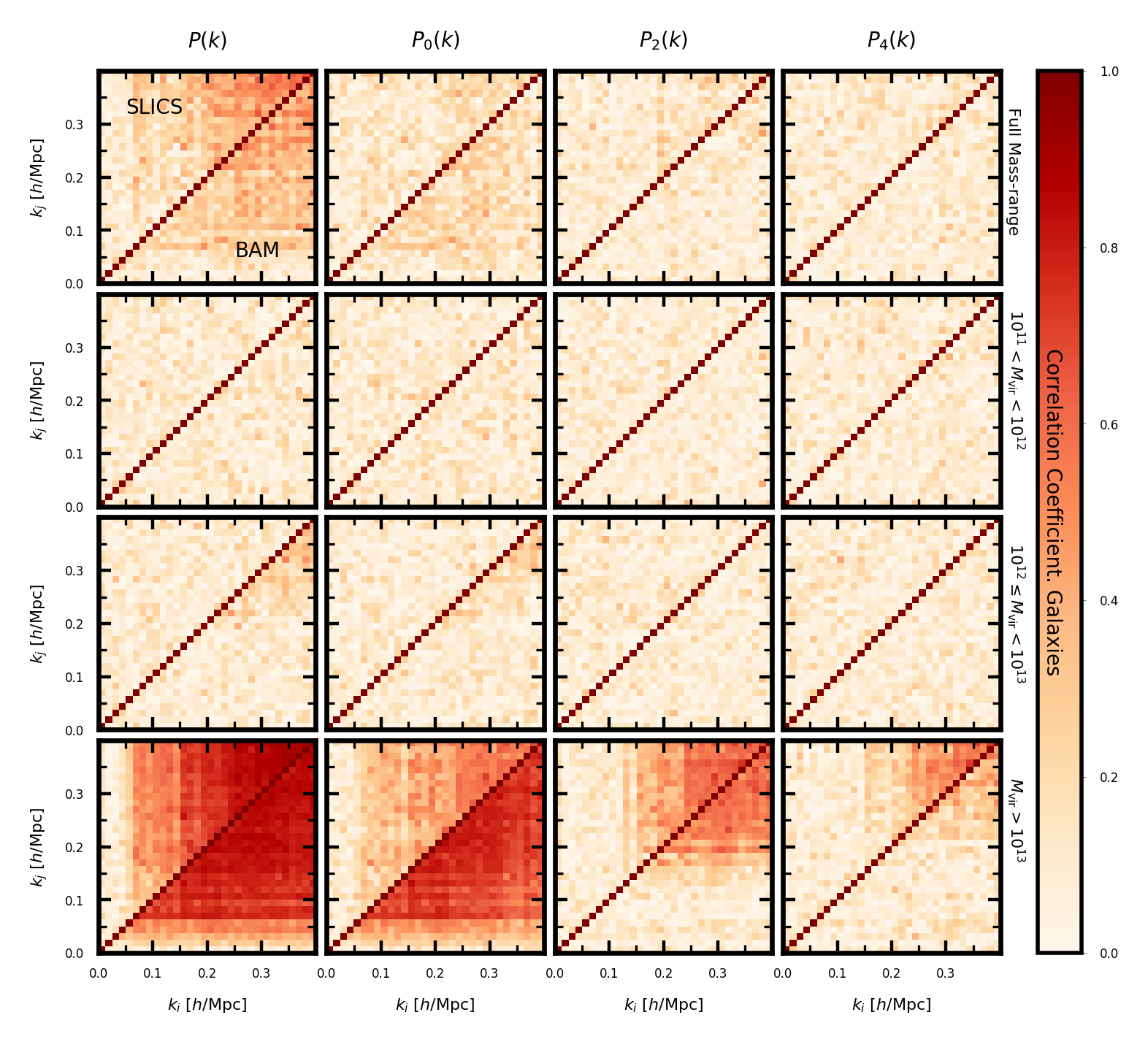}
\caption{Same as Fig.~\ref{fig:cova_power_rss} but for the galaxy population.}
    \label{fig:cova_power_rss_gals}
\end{figure*}
%------------------------------------------------------------------------------
%------------------------------------------------------------------------------
\begin{figure*}
\includegraphics[trim = 0cm 0cm 0cm 0cm ,clip=true, width=0.335\textwidth]{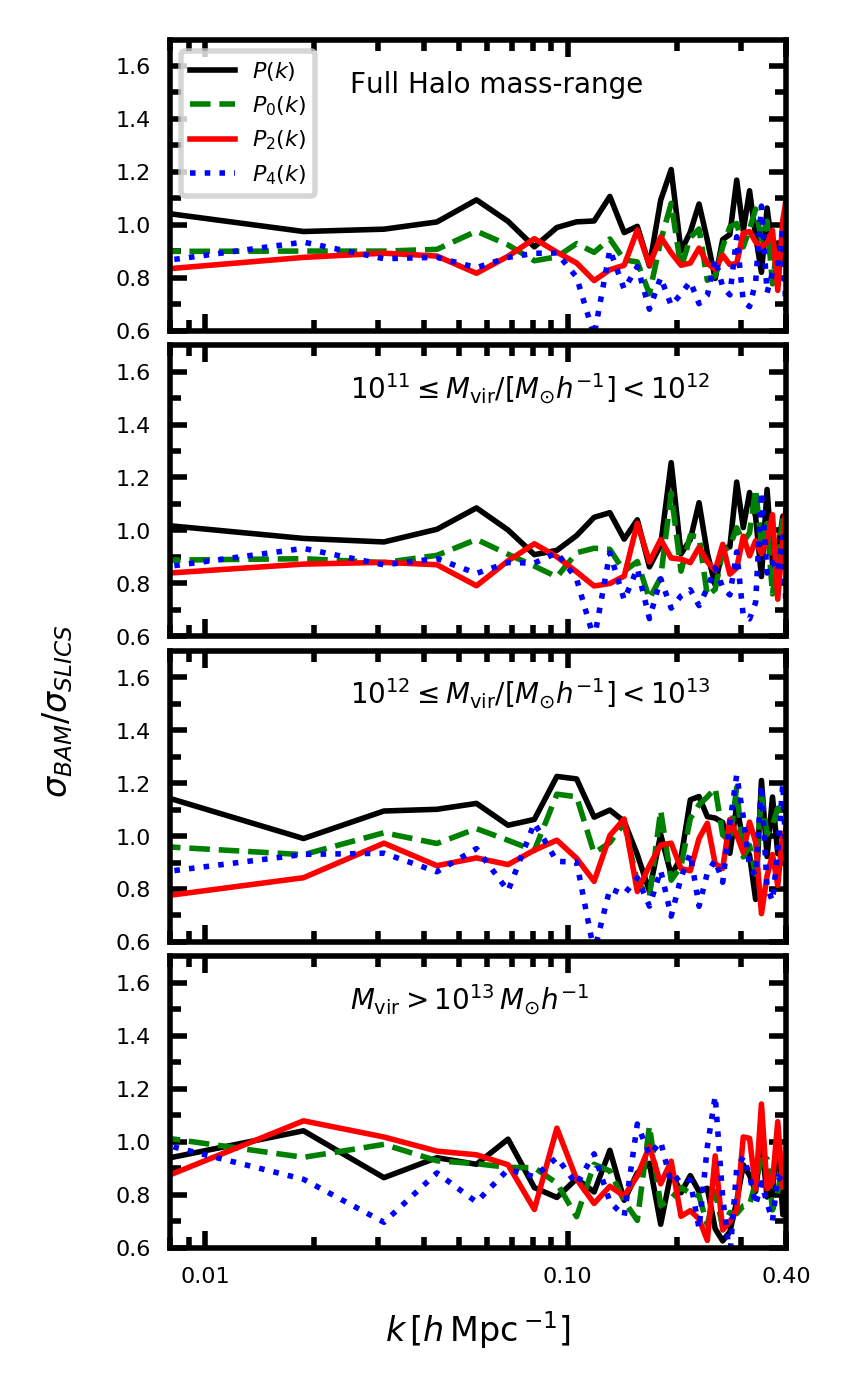}
\includegraphics[trim = 0cm 0cm 0cm 0cm ,clip=true, width=0.62\textwidth]{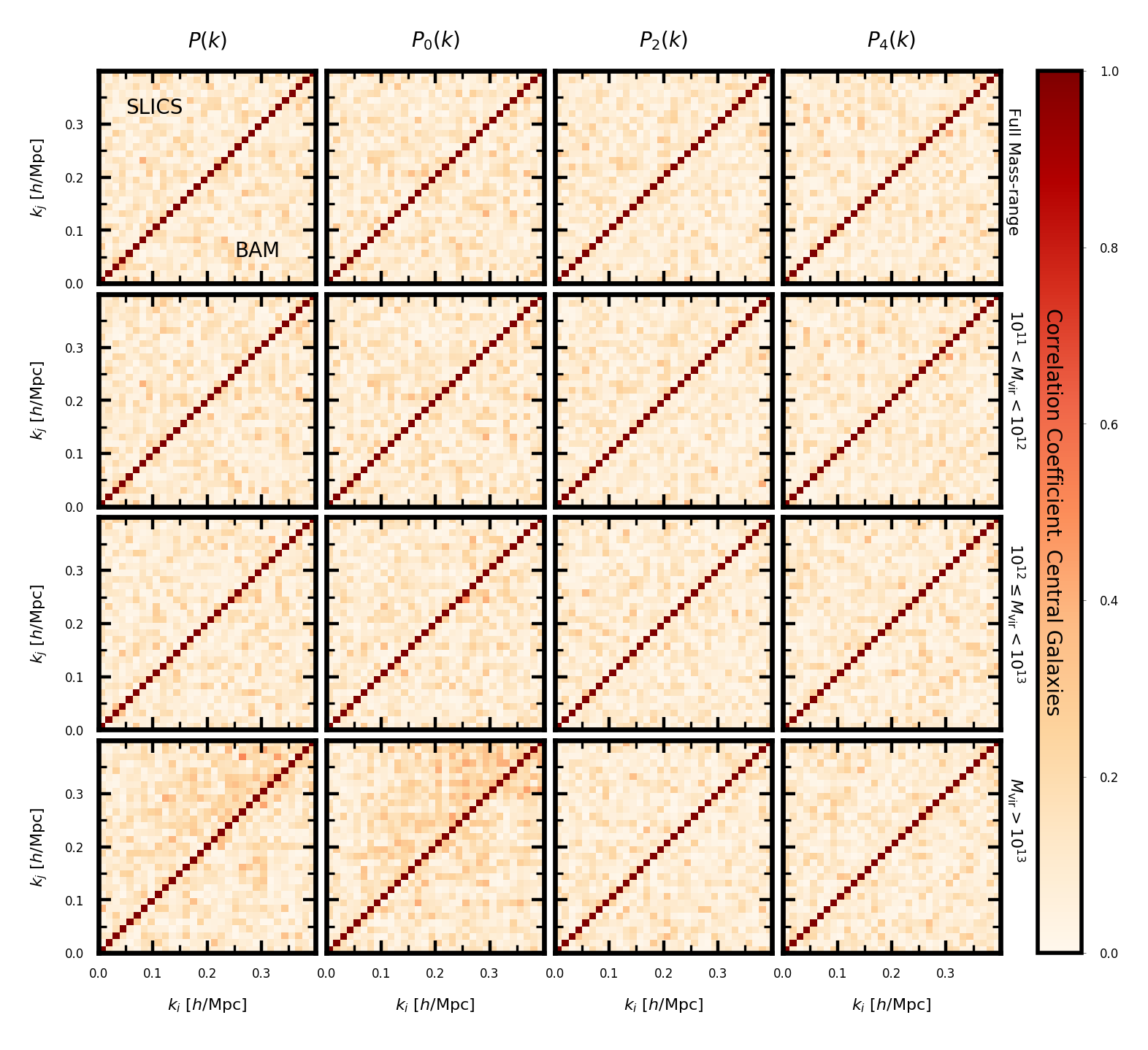}
\caption{Same as Fig.~\ref{fig:cova_power_rss} but for the central galaxy population.}
    \label{fig:cova_power_rss_bcg}
\end{figure*}
%------------------------------------------------------------------------------

Let us now summarise the performance of the mock galaxy catalogues produced with \texttt{BAM} through the assessment of different statistical probes.
\emph{Fourier space:} Figure \ref{fig:power_rss_gal} shows the ratio of the mean power spectrum of galaxies in \texttt{BAM} with respect to the signal from the same type of populations in the reference simulation, both in real and redshift space. The most noticeable difference in real space comes from the clustering of satellites, which on small scales directly probes the density profile of the dark matter halo \cite[see e.g.][]{2002PhR...372....1C}. As the two sets of mocks share the same density profile (by construction), the differences in the clustering pattern can be traced back to the assignment of halo masses, as this is a key property shaping the mass--concentration relation.

Similarly inherited from the parent halo population, the galaxy redshift-space power spectrum displays a lack of power towards small scales. We note that, contrary to the halo population in which such deviations can be solely tackled from the velocity field of the approximated gravity solver, at this stage we must also add ---in addition to the velocity field of the dark matter particles---  information on the halo properties used to derive galaxy coordinates in phase space.

\begin{figure*}
\includegraphics[trim = 0cm 0cm 0cm 0cm ,clip=true, width=0.335\textwidth]{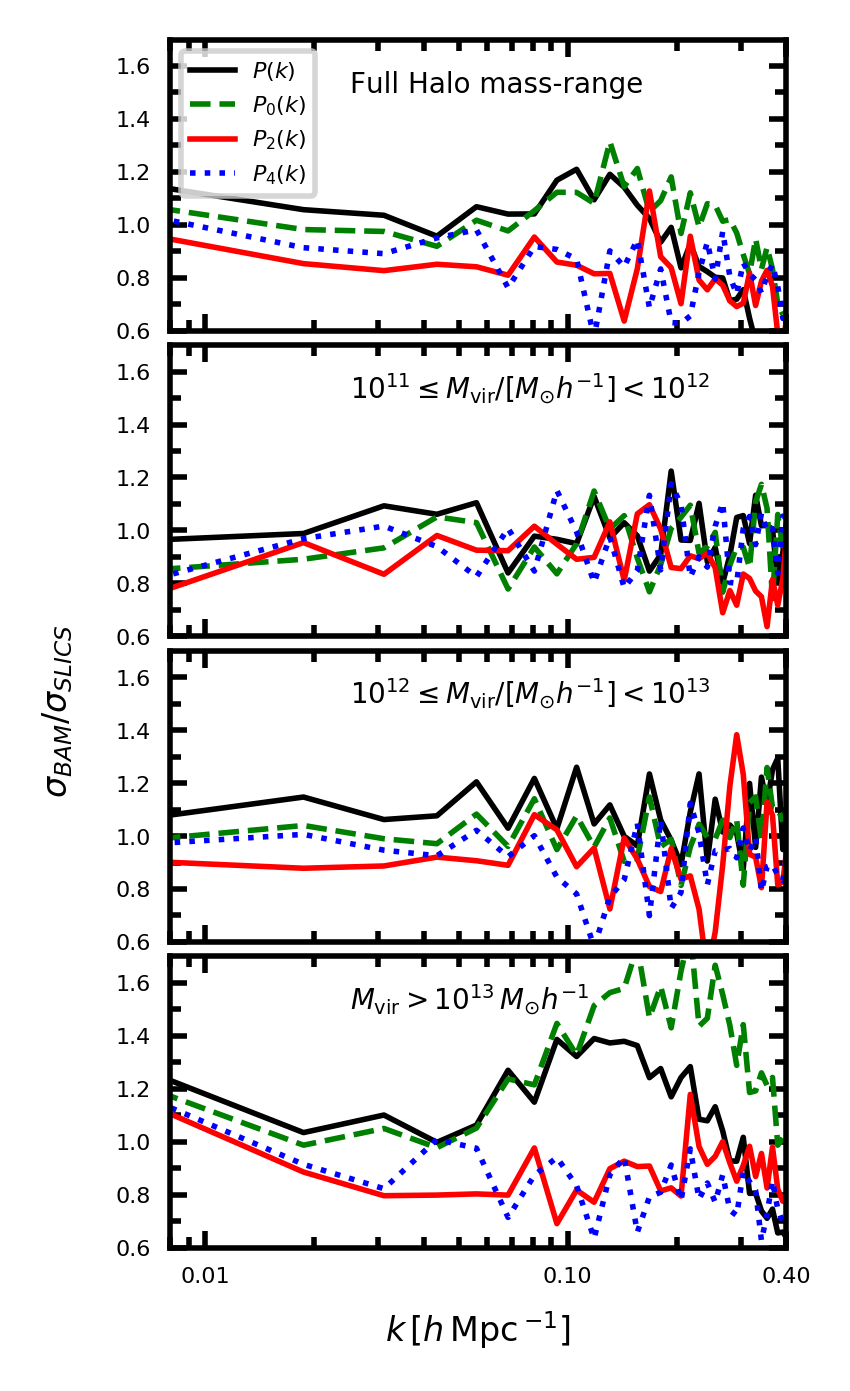}
\includegraphics[trim = 0cm 0cm 0cm 0cm ,clip=true, width=0.62\textwidth]{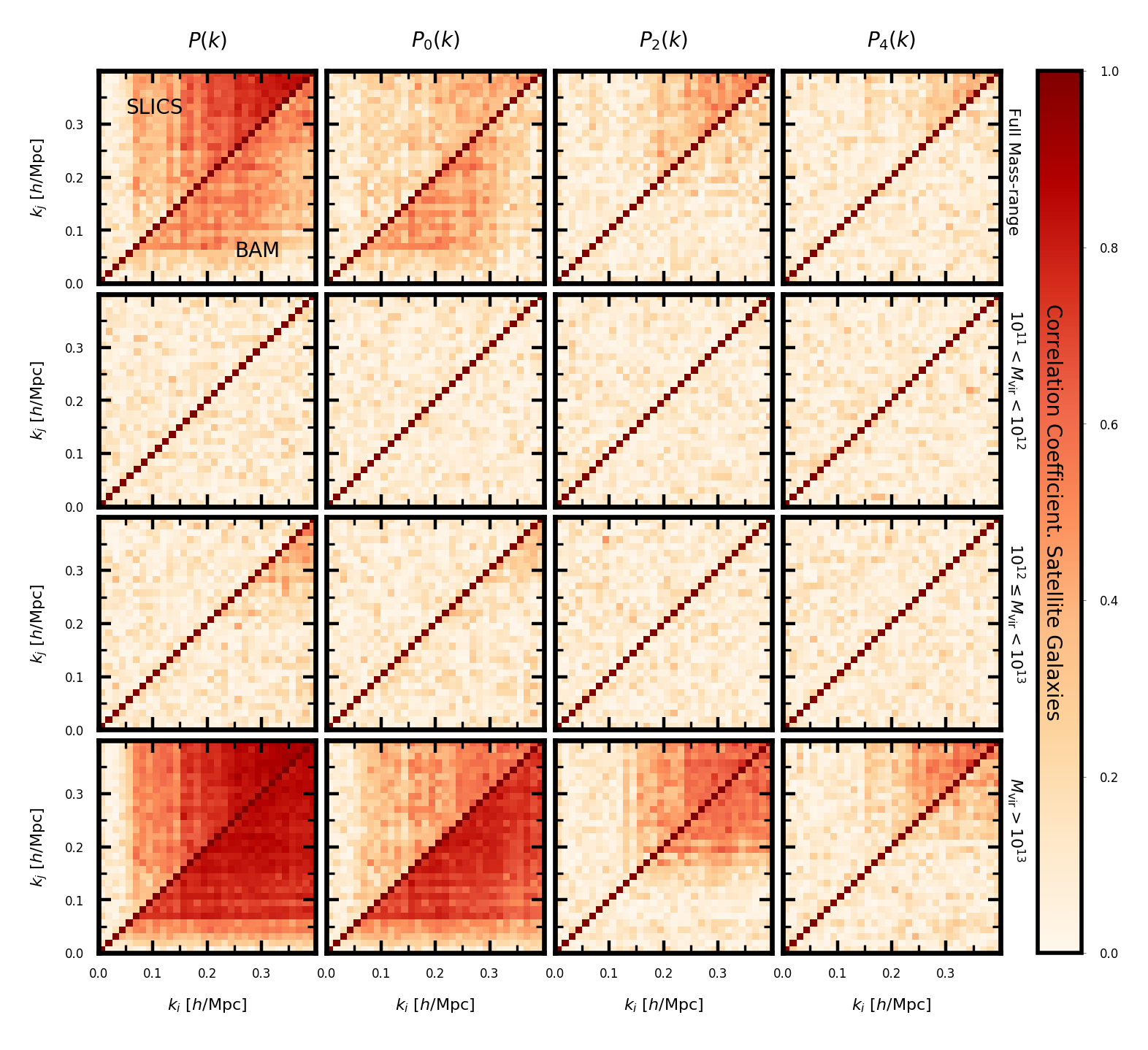}
\caption{Same as Fig.~\ref{fig:cova_power_rss} but for the satellite galaxy population.}
    \label{fig:cova_power_rss_sat}
\end{figure*}

Figures \ref{fig:cova_power_rss_gals}, \ref{fig:cova_power_rss_bcg}, and \ref{fig:cova_power_rss_sat} show the variance and the correlation matrix of the mock galaxy catalogues, split into the two types of population and into different bins of (host) halo mass. In general, the covariance matrices show good agreement with the reference. The extra correlation in the high-halo-mass bins is inherited from the parent halo distribution but is only embodied within the statistics of the satellite population. The behaviour of the correlation matrices for central galaxies (in all host-mass bins) contrasts with that from the parent halo distribution, as it lacks the extra mode coupling presented in Fig.~\ref{fig:cova_power_rss_bcg}. The reason for this is that, as pointed out above, the HOD model applied here suppresses the abundance of central ELGs in high-mass halos, which negates the possible deviations coming from the combination of cosmic variance and inaccuracies in the kernel--bias connection (see \S\ref{sec:meth}).

\begin{figure}
\includegraphics[trim = 0.2cm 0cm 0cm 0cm ,clip=true, width=0.50\textwidth]{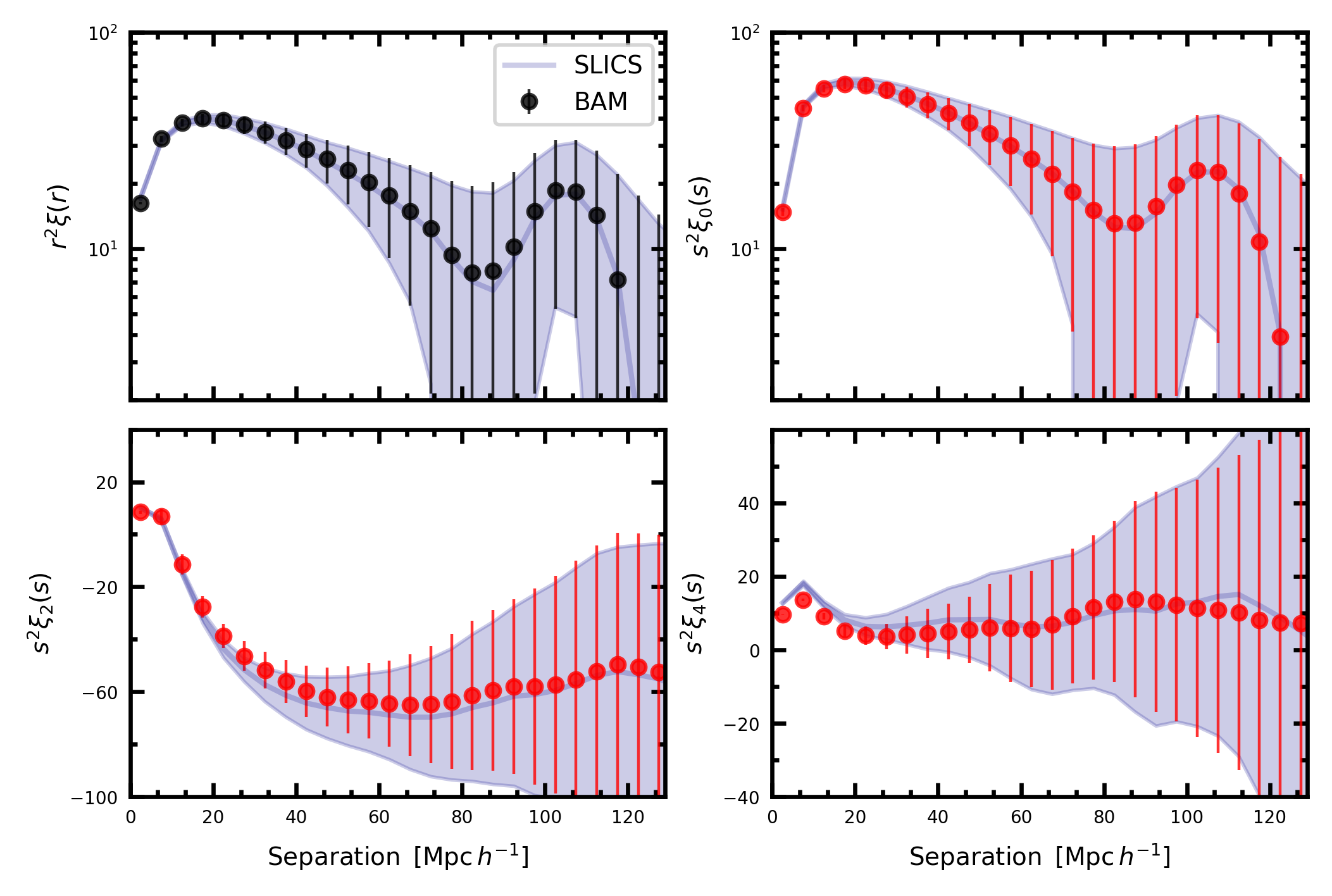}
\caption{Galaxy correlation function in real space $\xi(r)$ (top panel) and redshift space in the form of monopole $\xi_{0}(s)$, quadrupole $\xi_{2}(s),$ and hexadecapole $\xi_{4}(s)$. The solid line and shaded area denote  the mean and sample variance from the \texttt{SLICS} simulation, respectively.}
    \label{fig:cf_gal}
\end{figure}

\emph{Configuration space}: We measure the standard correlation function \citep[][]{1980lssu.book.....P} in real and redshift space (the latter similarly represented by the multipole decomposition) based on the natural estimator \citep[e.g.][]{2000ApJ...535L..13K} $\xi(s,\mu)+1=DD(s,\mu)/RR(s,\mu)
$\footnote{We measure the correlation function using the publicly available code at \url{https://github.com/cheng-zhao/FCFC}, \citet[][]{2021MNRAS.503.1149Z}}, where $DD(s)$ ($RR(s)$) is the number of (random) galaxy  pairs at a separation $s$ in redshift space, and $\mu=\hat{z}$ (according to the distant observer approximation; see \S~\ref{sec:vels}). The multipoles of the two-point correlation function are obtained as a Legendre decomposition, in analogy to Eq.~(\ref{pdec}). Figure~\ref{fig:cf_gal} shows the comparison of the two-point correlation function of the two data sets \citep[see e.g.][for a clustering analysis of ELGs]{2017MNRAS.472..550F}. The picture observed in Fourier space (see Fig.\ref{fig:power_rss_gal}) is replicated here, in which a systematic bias ($\sim 5\%$) can be seen in the quadrupole, albeit not statistically significant. The real space correlation function and its monopole in redshift space agrees very well with that of the reference. It is of paramount relevance for the quality validation of this suite to show how the position and the amplitude of the baryonic acoustic peak is well preserved in the \texttt{BAM} mocks.

\begin{figure}
\includegraphics[trim = .2cm 0cm 0cm 0cm ,clip=true, width=0.5\textwidth]{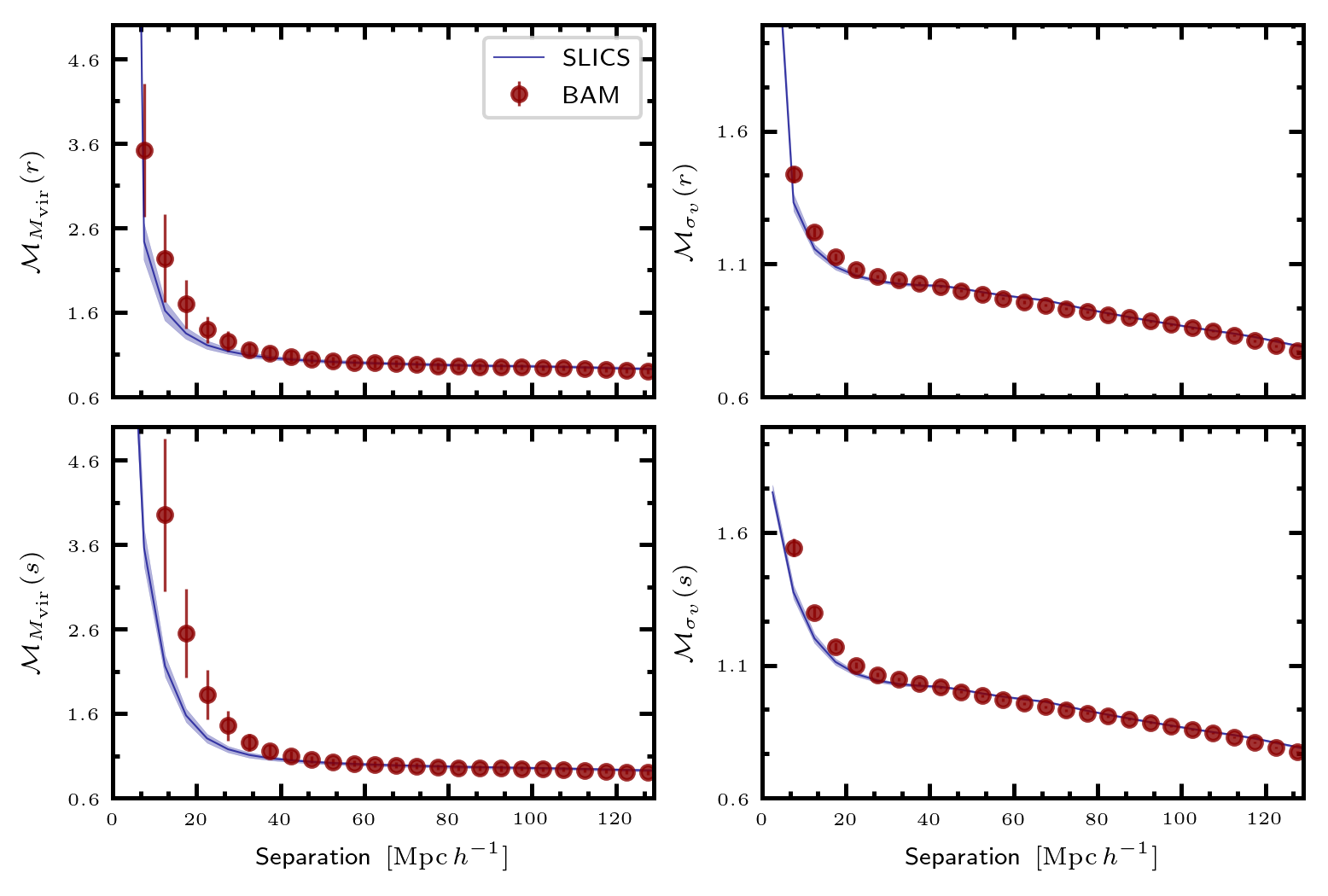}
\caption{Galaxy marked correlation in real $\mathcal{M}(r)$ and redshift space $\mathcal{M}(s)$. The top panel show  the result of using the halo virial mass as the mark. The middle panel uses the velocity dispersion and the third shows the results of using the cross-marked correlation function.}
\label{fig:marked_gal}
\end{figure}

\emph{Marked statistics}: We can jointly assess the clustering properties and the quality of the assignment of halo properties using marked statistics in Fourier space \cite[see e.g.][]{2014A&A...563A.141B} or in configuration space \cite[see e.g.][]{2005MNRAS.364..796S,2005astro.ph.11773S,2006MNRAS.369...68S,2019MNRAS.484.2148S}. The marked correlation function can be regarded as a measure of the clustering of a given property (mark), defined as $\mathcal{M}_{\alpha}(s)=WW_{\alpha}(s)/DD(s)$ where $WW_{\alpha}(s)$ represents the count pair of galaxies weighted with a given property $\alpha$ at separation $s$. Figure \ref{fig:marked_gal} shows the galaxy marked correlation function in real and redshift space, using the halo mass and velocity dispersion as marks, as obtained from the two suites of mocks. In general, the trend followed by the measurements from the two ensembles is consistent, showing how the \texttt{BAM} mocks can properly encode the information of galaxy bias as a function of different properties. Statistically significant differences are sizable when the halo mass is marked under inspection, especially on small separations, evidencing the trends observed in terms of the power spectrum of Fig.~\ref{fig:power_rss_gal}.

\emph{Statistical compatibility with the reference suite}: We verified the statistical compatibility between the two sets of mock catalogues by means of a Kolmogorov-Smirnov test \citep[e.g.][]{2002nrca.book.....P} based on the $\chi^{2}$ distributions of power spectra. The test, displaying $p$-values of $\sim 0.3$, forecasts precise and accurate results (compared to those obtained from the reference $N$-body simulation) when the set of \texttt{BAM} mocks are subject to likelihood analysis \citep[][]{Chuang2023}.

The main body of this paper is devoted to the construction of galaxy catalogues based on the generation of halo catalogues endowed with intrinsic properties (such as virial mass and velocity dispersion) to which an HOD prescription is applied. This procedure allows us to discriminate between central and satellite galaxies, and paves the way towards the assignment of galaxy properties by linking them with the properties of their host halos \cite[see e.g.][]{2022MNRAS.514.2463D}. Nevertheless, if the main goal is to generate galaxy catalogues without further properties, we could similarly have applied the \texttt{BAM} machinery directly using a set of reference galaxy catalogues as the training set. In Appendix \ref{ap:gala}, we discuss this option and present the results of such a direct approach.

%------------------------------------------------------------------------------
%------------------------------------------------------------------------------

%==============================================================================================================
%==============================================================================================================
%==============================================================================================================

\section{Discussions and conclusions}

In this paper, we present the construction of mock galaxy catalogues based on the bias assignment method (\texttt{BAM}) \citep[][]{2019MNRAS.483L..58B}. We used the initial conditions of the reference $N$-body simulation (the \texttt{SLICS} simulation) \citet[][]{2018MNRAS.481.1337H}, down-sampled, and evolved it using augmented Lagrangian perturbation theory \citep[][]{doi:10.1093/mnrasl/slt101}.
This approximated density field is accurate enough at scales of $\sim 0.4\,h^{-1}$Mpc to robustly study the bias relation between the cosmic web and the halo number counts. The remaining differences between the approximate gravity solver and the exact solution from an $N$-body simulation (on scales above the mesh resolution) are automatically taken into account in the effective halo bias extracted from the reference simulation. 
In particular, the approximated density field is used along with the corresponding halo catalogue reference $N$-body simulation to iteratively learn the halo bias and a kernel, which are the main outputs of the learning phase of the method. We show how the characterisation of the halo bias as a function of cosmic-web type and the implementation of a number of realisations as a training data set (so as to increase the effective volume of the reference simulation) can generate ensembles of independent realisations of halo number counts with $\sim 2\%-5\%$ precision in the two- and three-point statistics, as well as in the variance obtained from the corresponding covariance matrices. 

We describe a procedure to assign halo coordinates, velocities, and intrinsic properties, with the aim being to generate a set of $770$ dark matter halos with the same number of properties as in the reference set. We assigned velocity dispersion and virial mass following a hierarchical and multi-scaling approach (see \S~\ref{sec:hprops}) designed to replicate the abundance as a function of these properties. The halo two-point statistics replicates that of the reference with $5\%$ precision at $k\sim 0.4 h\, {\rm Mpc}^{-1}$, which is the maximum wave number adopted by the DESI mock challenge. We verified that covariance matrices of the two- and three-point statistics measured from the mock catalogues generated by \texttt{BAM} are in good qualitative agreement with those obtained from the reference $N$-body simulation. A thorough likelihood analysis using these covariance matrices will be performed by \cite[][]{Chuang2023} as part of the DESI mock comparison project.

Based on this set of halo catalogues, we  generated the same number of galaxy catalogues using an HOD prescription designed to replicate the abundance of emission-line galaxies. It is important to stress that the same HOD parameters were applied to both \texttt{BAM} and the $N$-body-based halo catalogues. The goodness of the suite of mocks is assessed not only in Fourier space but also in configuration space through the correlation function and the marked correlation function.

Despite the good performance of the method in terms of the different statistical probes explored, we identified a number of items in which \texttt{BAM} has to be improved to reduce the deviations observed with respect to a reference simulation. These are
\begin{itemize}
    \item \emph{Peculiar velocities}. Along with the density-dependent bias correction (described in \S\ref{sec:vels}), the small-scale clustering signal in redshift space demands the further treatment of the peculiar velocities. This treatment starts with a thorough analysis of the velocity assignment, especially for random tracers, as shown in Fig.~\ref{fig:col_scheme}. Generalisations of such an approach to taking into account halo properties and/or modification of the full population (dark matter and tracers) are part of this task. A deeper understanding of the origin of the different corrections in the velocities applied in this work is to be addressed in forthcoming publications.
    \item \emph{Assignment of halo properties}. 
    A thorough approach to this task is to impose the pair distribution of tracers as a function of the different halo properties. However, this is a highly expensive task. Although the multi-scale algorithm described in \S\ref{sec:multi}
    is an improvement with respect to previous algorithms, further developments need to be investigated and implemented.
    To that end, we are currently including a second learning phase and using marked statistics as the main diagnosis. The goal is to replicate the clustering pattern observed in the reference as a function of all halo properties.
\end{itemize}

In general, the accurate performance of \texttt{BAM} does not only depend on these planned improvements. It is also complemented by the characteristics of the reference simulation used as the training data set. The \texttt{SLICS}, with a relatively small cosmological volume, and initial conditions generated from Gaussian random fields, is highly prone to cosmic variance. This is reflected in the limited statistical information contained in a halo bias obtained from only one reference, which in turn can lead to inaccuracies in the generation of mock  halo catalogues with different seeds. To take this into account, we pushed the method to the implementation of more than one reference simulation (see Eq.~\ref{eq:btot}) as the training data set.

In Appendix \ref{sec:unit}, we describe our motivation to implement a reference simulation based on initial conditions with variance suppression \citep[fixed amplitude initial conditions; see e.g.][]{2016MNRAS.462L...1A,2019MNRAS.487...48C,2022arXiv220403868M}
covering larger cosmological volumes. This scenario can substantially improve the accuracy in the two- and three-point statistics, as well as the procedure to assign halo properties. Such a setup will also allow the method to extrapolate the generation of mock catalogues to volumes larger than that of the reference.

The present paper demonstrates the potential of \texttt{BAM} to speedily deliver mock halo catalogues ---with a number of properties--- that are both flexible and accurate enough to implement any mechanism to generate galaxy samples within the context of the halo model. The next step is the generation of larger sets of mock halo catalogues (larger cosmological volumes and light cones) with more halo properties (e.g. spin, concentration, maximum circular velocity) on which different methods for galaxy occupation and selection functions can be imposed to replicate the sky observed by different experiments.

\begin{acknowledgements}
We would like to thank Joachim Harnois-Deraps for making public the \texttt{SLICS} data,
available at \url{http://slics.roe.ac.uk/}.  ABA and FSK acknowledge the IAC facilities and  the Spanish Ministry of Economy and Competitiveness (MINECO) under the Severo Ochoa program SEV-2015-0548, PID2020-120612GB-I00 and CEX2019-000920-S grants. FSK also thanks the  RYC2015-18693 grant. ABA wishes to give deep thanks to Olga Alemán for her continuous support during this work.

This research is supported by the Director, Office of Science, Office of High Energy Physics of the U.S. Department of Energy under Contract No. DE–AC02–05CH11231, and by the National Energy Research Scientific Computing Center, a DOE Office of Science User Facility under the same contract; additional support for DESI is provided by the U.S. National Science Foundation, Division of Astronomical Sciences under Contract No. AST-0950945 to the NSF’s National Optical-Infrared Astronomy Research Laboratory; the Science and Technologies Facilities Council of the United Kingdom; the Gordon and Betty Moore Foundation; the Heising-Simons Foundation; the French Alternative Energies and Atomic Energy Commission (CEA); the National Council of Science and Technology of Mexico (CONACYT); the Ministry of Science and Innovation of Spain (MICINN), and by the DESI Member Institutions: \url{https://www.desi.lbl.gov/collaborating-institutions}.

The authors are honored to be permitted to conduct scientific research on Iolkam Du’ag (Kitt Peak), a mountain with particular significance to the Tohono O’odham Nation.
\end{acknowledgements}
%------------------------------------------------------------------------------
%------------------------------------------------------------------------------
%------------------------------------------------------------------------------

% 00000000000000000000000000000000000000000000000000000000000000000000000000000000000000000000000000000000000000000000000000000000000000000000000000000000000000000000000000000
% 0000000000000000000000000000000000000000000000000000000000000000000000000000000000000000000000000000000000000000000000000000000000000000000000000000000000000000000000000000
\bibliographystyle{aa}
\bibliography{refs}  

\begin{thebibliography}{222}
\expandafter\ifx\csname natexlab\endcsname\relax\def\natexlab#1{#1}\fi

\bibitem[{{Abel} {et~al.}(2012){Abel}, {Hahn}, \&
  {Kaehler}}]{2012MNRAS.427...61A}
{Abel}, T., {Hahn}, O., \& {Kaehler}, R. 2012, \mnras, 427, 61

\bibitem[{{Abramo} \& {Leonard}(2013)}]{2013MNRAS.432..318A}
{Abramo}, L.~R. \& {Leonard}, K.~E. 2013, \mnras, 432, 318

\bibitem[{{Abramo} {et~al.}(2016){Abramo}, {Secco}, \&
  {Loureiro}}]{2016MNRAS.455.3871A}
{Abramo}, L.~R., {Secco}, L.~F., \& {Loureiro}, A. 2016, \mnras, 455, 3871

\bibitem[{{Agrawal} {et~al.}(2017){Agrawal}, {Makiya}, {Chiang}, {Jeong},
  {Saito}, \& {Komatsu}}]{2017JCAP...10..003A}
{Agrawal}, A., {Makiya}, R., {Chiang}, C.-T., {et~al.} 2017, \jcap, 2017, 003

\bibitem[{{Ahn} {et~al.}(2015){Ahn}, {Iliev}, {Shapiro}, \&
  {Srisawat}}]{2015MNRAS.450.1486A}
{Ahn}, K., {Iliev}, I.~T., {Shapiro}, P.~R., \& {Srisawat}, C. 2015, \mnras,
  450, 1486

\bibitem[{Alam {et~al.}(2021)Alam, Arnold, Aviles, Bean, Cai, Cautun,
  Cervantes-Cota, Cuesta-Lazaro, Devi, Eggemeier, Fromenteau, Gonzalez-Morales,
  Halenka, hua He, Hellwing, Hern{\'{a}}ndez-Aguayo, Ishak, Koyama, Li, de~la
  Macorra, Rizo, Miller, Mueller, Niz, Ntelis, Otero, Sabiu, Slepian, Stark,
  Valenzuela, Valogiannis, Vargas-Maga{\~{n}}a, Winther, Zarrouk, Zhao, \&
  Zheng}]{Alam_2021}
Alam, S., Arnold, C., Aviles, A., {et~al.} 2021, Journal of Cosmology and
  Astroparticle Physics, 2021, 050

\bibitem[{{Alam} {et~al.}(2020){Alam}, {Peacock}, {Kraljic}, {Ross}, \&
  {Comparat}}]{2020MNRAS.497..581A}
{Alam}, S., {Peacock}, J.~A., {Kraljic}, K., {Ross}, A.~J., \& {Comparat}, J.
  2020, \mnras, 497, 581

\bibitem[{{Alexander} {et~al.}(2022){Alexander}, {Davis}, {Chaussidon},
  {Fawcett}, {Gonzalez-Morales}, {Lan}, {Yeche}, {Ahlen}, {Aguilar},
  {Armengaud}, {Bailey}, {Brooks}, {Cai}, {Canning}, {Carr}, {Chabanier},
  {Cousinou}, {Dawson}, {de la Macorra}, {Dey}, {Dey}, {Dhungana}, {Edge},
  {Eftekharzadeh}, {Fanning}, {Farr}, {Font-Ribera}, {Garcia-Bellido},
  {Garrison}, {Gaztanaga}, {Gontcho}, {Gordon}, {Guadalupe Medellin Gonzalez},
  {Guy}, {Herrera-Alcantar}, {Jiang}, {Juneau}, {Karacayli}, {Kehoe}, {Kisner},
  {Kovacs}, {Landriau}, {Levi}, {Magneville}, {Martini}, {Meisner}, {Mezcua},
  {Miquel}, {Montero Camacho}, {Moustakas}, {Munoz-Gutierrez}, {Myers},
  {Nadathur}, {Napolitano}, {Nie}, {Palanque-Delabrouille}, {Pan}, {Percival},
  {Perez-Rafols}, {Poppett}, {Prada}, {Ramirez-Perez}, {Ravoux}, {Rosario},
  {Schubnell}, {Tarle}, {Walther}, {Weiner}, {Youles}, {Zhou}, {Zou}, \&
  {Zou}}]{2022arXiv220808517A}
{Alexander}, D.~M., {Davis}, T.~M., {Chaussidon}, E., {et~al.} 2022, arXiv
  e-prints, arXiv:2208.08517

\bibitem[{{Allende Prieto} {et~al.}(2020){Allende Prieto}, {Cooper}, {Dey},
  {G{\"a}nsicke}, {Koposov}, {Li}, {Manser}, {Nidever}, {Rockosi}, {Wang},
  {Aguado}, {Blum}, {Brooks}, {Eisenstein}, {Duan}, {Eftekharzadeh},
  {Gazta{\~n}aga}, {Kehoe}, {Landriau}, {Lee}, {Levi}, {Meisner}, {Myers},
  {Najita}, {Olsen}, {Palanque-Delabrouille}, {Poppett}, {Prada}, {Schlegel},
  {Schubnell}, {Tarl{\'e}}, {Valluri}, {Wechsler}, \&
  {Y{\`e}che}}]{2020RNAAS...4..188A}
{Allende Prieto}, C., {Cooper}, A.~P., {Dey}, A., {et~al.} 2020, Research Notes
  of the American Astronomical Society, 4, 188

\bibitem[{{Amendola} {et~al.}(2016){Amendola}, {Appleby}, {Avgoustidis},
  {Bacon}, {Baker}, {Baldi}, {Bartolo}, {Blanchard}, {Bonvin}, {Borgani},
  {Branchini}, {Burrage}, {Camera}, {Carbone}, {Casarini}, {Cropper}, {de
  Rham}, {Dietrich}, {Di Porto}, {Durrer}, {Ealet}, {Ferreira}, {Finelli},
  {Garcia-Bellido}, {Giannantonio}, {Guzzo}, {Heavens}, {Heisenberg},
  {Heymans}, {Hoekstra}, {Hollenstein}, {Holmes}, {Horst}, {Hwang}, {Jahnke},
  {Kitching}, {Koivisto}, {Kunz}, {La Vacca}, {Linder}, {March}, {Marra},
  {Martins}, {Majerotto}, {Markovic}, {Marsh}, {Marulli}, {Massey}, {Mellier},
  {Montanari}, {Mota}, {Nunes}, {Percival}, {Pettorino}, {Porciani},
  {Quercellini}, {Read}, {Rinaldi}, {Sapone}, {Sawicki}, {Scaramella},
  {Skordis}, {Simpson}, {Taylor}, {Thomas}, {Trotta}, {Verde}, {Vernizzi},
  {Vollmer}, {Wang}, {Weller}, \& {Zlosnik}}]{Euclid}
{Amendola}, L., {Appleby}, S., {Avgoustidis}, A., {et~al.} 2016, ArXiv e-prints

\bibitem[{{Angulo} {et~al.}(2008){Angulo}, {Baugh}, \&
  {Lacey}}]{2008MNRAS.387..921A}
{Angulo}, R.~E., {Baugh}, C.~M., \& {Lacey}, C.~G. 2008, \mnras, 387, 921

\bibitem[{{Angulo} \& {Pontzen}(2016)}]{2016MNRAS.462L...1A}
{Angulo}, R.~E. \& {Pontzen}, A. 2016, \mnras, 462, L1

\bibitem[{{Aragon-Calvo}(2016)}]{2016MNRAS.455..438A}
{Aragon-Calvo}, M.~A. 2016, \mnras, 455, 438

\bibitem[{{Avila} {et~al.}(2015){Avila}, {Murray}, {Knebe}, {Power},
  {Robotham}, \& {Garcia-Bellido}}]{2015MNRAS.450.1856A}
{Avila}, S., {Murray}, S.~G., {Knebe}, A., {et~al.} 2015, \mnras, 450, 1856

\bibitem[{{Bailey et al.}(2022)}]{redrock2022}
{Bailey et al.} 2022, in preparation

\bibitem[{{Balaguera-Antol{\'\i}nez}(2014)}]{2014A&A...563A.141B}
{Balaguera-Antol{\'\i}nez}, A. 2014, \aap, 563, A141

\bibitem[{{Balaguera-Antol{\'\i}nez} {et~al.}(2020){Balaguera-Antol{\'\i}nez},
  {Kitaura}, {Pellejero-Ib{\'a}{\~n}ez}, {Lippich}, {Zhao}, {S{\'a}nchez},
  {Dalla Vecchia}, {Angulo}, \& {Crocce}}]{2020MNRAS.491.2565B}
{Balaguera-Antol{\'\i}nez}, A., {Kitaura}, F.-S., {Pellejero-Ib{\'a}{\~n}ez},
  M., {et~al.} 2020, \mnras, 491, 2565

\bibitem[{{Balaguera-Antol{\'{\i}}nez}
  {et~al.}(2019){Balaguera-Antol{\'{\i}}nez}, {Kitaura},
  {Pellejero-Ib{\'a}{\~n}ez}, {Zhao}, \& {Abel}}]{2019MNRAS.483L..58B}
{Balaguera-Antol{\'{\i}}nez}, A., {Kitaura}, F.-S., {Pellejero-Ib{\'a}{\~n}ez},
  M., {Zhao}, C., \& {Abel}, T. 2019, \mnras, 483, L58

\bibitem[{{Balaguera-Antol{\'\i}nez} {et~al.}(2012){Balaguera-Antol{\'\i}nez},
  {S{\'a}nchez}, {B{\"o}hringer}, \& {Collins}}]{2012MNRAS.425.2244B}
{Balaguera-Antol{\'\i}nez}, A., {S{\'a}nchez}, A.~G., {B{\"o}hringer}, H., \&
  {Collins}, C. 2012, \mnras, 425, 2244

\bibitem[{{Balaguera-Antolínez et al.}(2023)}]{Bala2023}
{Balaguera-Antolínez et al.} 2023, in preparation

\bibitem[{{Baldauf} {et~al.}(2013){Baldauf}, {Seljak}, {Smith}, {Hamaus}, \&
  {Desjacques}}]{2013PhRvD..88h3507B}
{Baldauf}, T., {Seljak}, U., {Smith}, R.~E., {Hamaus}, N., \& {Desjacques}, V.
  2013, \prd, 88, 083507

\bibitem[{{Baratta} {et~al.}(2022){Baratta}, {Bel}, {Gouyou Beauchamps}, \&
  {Carbone}}]{2022arXiv221113590B}
{Baratta}, P., {Bel}, J., {Gouyou Beauchamps}, S., \& {Carbone}, C. 2022, arXiv
  e-prints, arXiv:2211.13590

\bibitem[{{Baratta} {et~al.}(2020){Baratta}, {Bel}, {Plaszczynski}, \&
  {Ealet}}]{2020A&A...633A..26B}
{Baratta}, P., {Bel}, J., {Plaszczynski}, S., \& {Ealet}, A. 2020, \aap, 633,
  A26

\bibitem[{{Bardeen} {et~al.}(1986){Bardeen}, {Bond}, {Kaiser}, \&
  {Szalay}}]{1986ApJ...304...15B}
{Bardeen}, J.~M., {Bond}, J.~R., {Kaiser}, N., \& {Szalay}, A.~S. 1986, \apj,
  304, 15

\bibitem[{{Benitez} {et~al.}(2014){Benitez}, {Dupke}, {Moles}, {Sodre},
  {Cenarro}, {Marin-Franch}, {Taylor}, {Cristobal}, {Fernandez-Soto}, {Mendes
  de Oliveira}, {Cepa-Nogue}, {Abramo}, {Alcaniz}, {Overzier},
  {Hernandez-Monteagudo}, {Alfaro}, {Kanaan}, {Carvano}, {Reis}, {Martinez
  Gonzalez}, {Ascaso}, {Ballesteros}, {Xavier}, {Varela}, {Ederoclite},
  {Vazquez Ramio}, {Broadhurst}, {Cypriano}, {Angulo}, {Diego}, {Zandivarez},
  {Diaz}, {Melchior}, {Umetsu}, {Spinelli}, {Zitrin}, {Coe}, {Yepes}, {Vielva},
  {Sahni}, {Marcos-Caballero}, {Shu Kitaura}, {Maroto}, {Masip}, {Tsujikawa},
  {Carneiro}, {Gonzalez Nuevo}, {Bongiovanni}, {Bonoli}, {Bruzual}, {Cardiel},
  {Cava}, {Cid Fernandes}, {Coelho}, {Cortesi}, {Delgado}, {Diaz Garcia},
  {Espinosa}, {Galliano}, {Gonzalez-Serrano}, {Falcon-Barroso}, {Fritz},
  {Fernandes}, {Gorgas}, {Hoyos}, {Jimenez-Teja}, {Lopez-Aguerri}, {Lopez-San
  Juan}, {Mateus}, {Molino}, {Novais}, {OMill}, {Oteo}, {Perez-Gonzalez},
  {Poggianti}, {Proctor}, {Ricciardelli}, {Sanchez-Blazquez},
  {Storchi-Bergmann}, {Telles}, {Schoennell}, {Trujillo}, {Vazdekis},
  {Viironen}, {Daflon}, {Aparicio-Villegas}, {Rocha}, {Ribeiro}, {Borges},
  {Martins}, {Marcolino}, {Martinez-Delgado}, {Perez-Torres}, {Siffert},
  {Calvao}, {Sako}, {Kessler}, {Alvarez-Cand al}, {De Pra}, {Roig}, {Lazzaro},
  {Gorosabel}, {Lopes de Oliveira}, {Lima-Neto}, {Irwin}, {Liu}, {Alvarez},
  {Balmes}, {Chueca}, {Costa-Duarte}, {da Costa}, {Dantas}, {Diaz}, {Fabregat},
  {Ferrari}, {Gavela}, {Gracia}, {Gruel}, {Gutierrez}, {Guzman},
  {Hernandez-Fernand ez}, {Herranz}, {Hurtado-Gil}, {Jablonsky}, {Laporte}, {Le
  Tiran}, {Licandro}, {Lima}, {Martin}, {Martinez}, {Montero}, {Penteado},
  {Pereira}, {Peris}, {Quilis}, {Sanchez-Portal}, {Soja}, {Solano}, {Torra}, \&
  {Valdivielso}}]{2014arXiv1403.5237B}
{Benitez}, N., {Dupke}, R., {Moles}, M., {et~al.} 2014, arXiv e-prints,
  arXiv:1403.5237

\bibitem[{{Berlind} \& {Weinberg}(2002)}]{2002ApJ...575..587B}
{Berlind}, A.~A. \& {Weinberg}, D.~H. 2002, \apj, 575, 587

\bibitem[{{Bernardeau}(1994)}]{1994ApJ...427...51B}
{Bernardeau}, F. 1994, \apj, 427, 51

\bibitem[{{Bernardeau} {et~al.}(2002){Bernardeau}, {Colombi}, {Gazta{\~n}aga},
  \& {Scoccimarro}}]{2002PhR...367....1B}
{Bernardeau}, F., {Colombi}, S., {Gazta{\~n}aga}, E., \& {Scoccimarro}, R.
  2002, \physrep, 367, 1

\bibitem[{{Berner} {et~al.}(2022){Berner}, {Refregier}, {Sgier}, {Kacprzak},
  {Tortorelli}, \& {Monaco}}]{2022JCAP...11..002B}
{Berner}, P., {Refregier}, A., {Sgier}, R., {et~al.} 2022, \jcap, 2022, 002

\bibitem[{{Blanton}(2000)}]{2000ApJ...544...63B}
{Blanton}, M. 2000, \apj, 544, 63

\bibitem[{{Blot} {et~al.}(2019){Blot}, {Crocce}, {Sefusatti}, {Lippich},
  {S{\'a}nchez}, {Colavincenzo}, {Monaco}, {Alvarez}, {Agrawal}, {Avila},
  {Balaguera-Antol{\'\i}nez}, {Bond}, {Codis}, {Vecchia}, {Dorta}, {Fosalba},
  {Izard}, {Kitaura}, {Pellejero-Ibanez}, {Stein}, {Vakili}, \&
  {Yepes}}]{2018arXiv180609497B}
{Blot}, L., {Crocce}, M., {Sefusatti}, E., {et~al.} 2019, \mnras, 500

\bibitem[{{Bond} {et~al.}(1996){Bond}, {Kofman}, \&
  {Pogosyan}}]{1996Natur.380..603B}
{Bond}, J.~R., {Kofman}, L., \& {Pogosyan}, D. 1996, \nat, 380, 603

\bibitem[{{Bond} \& {Myers}(1996{\natexlab{a}})}]{1996ApJS..103....1B}
{Bond}, J.~R. \& {Myers}, S.~T. 1996{\natexlab{a}}, \apjs, 103, 1

\bibitem[{{Bond} \& {Myers}(1996{\natexlab{b}})}]{1996ApJS..103...41B}
{Bond}, J.~R. \& {Myers}, S.~T. 1996{\natexlab{b}}, \apjs, 103, 41

\bibitem[{{Bond} \& {Myers}(1996{\natexlab{c}})}]{1996ApJS..103...63B}
{Bond}, J.~R. \& {Myers}, S.~T. 1996{\natexlab{c}}, \apjs, 103, 63

\bibitem[{{Bouchet} {et~al.}(1995){Bouchet}, {Colombi}, {Hivon}, \&
  {Juszkiewicz}}]{1995A&A...296..575B}
{Bouchet}, F.~R., {Colombi}, S., {Hivon}, E., \& {Juszkiewicz}, R. 1995, \aap,
  296, 575

\bibitem[{Buchert \& Ehlers(1993)}]{10.1093/mnras/264.2.375}
Buchert, T. \& Ehlers, J. 1993, Monthly Notices of the Royal Astronomical
  Society, 264, 375

\bibitem[{{Cai} {et~al.}(2009){Cai}, {Angulo}, {Baugh}, {Cole}, {Frenk}, \&
  {Jenkins}}]{2009MNRAS.395.1185C}
{Cai}, Y.-C., {Angulo}, R.~E., {Baugh}, C.~M., {et~al.} 2009, \mnras, 395, 1185

\bibitem[{{Chartier} {et~al.}(2021){Chartier}, {Wandelt}, {Akrami}, \&
  {Villaescusa-Navarro}}]{2021MNRAS.503.1897C}
{Chartier}, N., {Wandelt}, B., {Akrami}, Y., \& {Villaescusa-Navarro}, F. 2021,
  \mnras, 503, 1897

\bibitem[{{Chaussidon} {et~al.}(2022){Chaussidon}, {Y{\`e}che},
  {Palanque-Delabrouille}, {Alexander}, {Yang}, {Ahlen}, {Bailey}, {Brooks},
  {Cai}, { Chabanier}, {Davis}, {Dawson}, {de la Macorra}, {Dey}, {Dey},
  {Eftekharzadeh}, {Eisenstein}, {Fanning}, {Font-Ribera}, {Gazta{\~n}aga},
  {Gontcho}, {Gonzalez-Morales}, {Guy}, {Herrera-Alcantar}, {Honscheid},
  {Ishak}, {Jiang}, {Juneau}, {Kehoe}, {Kisner}, {Kov{\'a}cs}, {Kremin}, {Lan},
  {Landriau}, {Le Guillou}, {Levi}, {Magneville}, {Martini}, {Meisner},
  {Moustakas}, {Mu{\~n}oz-Guti{\'e}rrez}, {Myers}, {Newman}, {Nie}, {Percival},
  {Poppett}, {Prada}, {Raichoor}, {Ravoux}, {Ross}, {Schlafly}, {Schlegel},
  {Tan}, {Tarl{\'e}}, {Zhou}, {Zhou}, \& {Zou}}]{2022arXiv220808511C}
{Chaussidon}, E., {Y{\`e}che}, C., {Palanque-Delabrouille}, N., {et~al.} 2022,
  arXiv e-prints, arXiv:2208.08511

\bibitem[{{Chen} {et~al.}(2018){Chen}, {Zhang}, {Zheng}, {Yu}, \&
  {Jing}}]{2018ApJ...861...58C}
{Chen}, J., {Zhang}, P., {Zheng}, Y., {Yu}, Y., \& {Jing}, Y. 2018, \apj, 861,
  58

\bibitem[{{Chuang} {et~al.}(2015{\natexlab{a}}){Chuang}, {Kitaura}, {Prada},
  {Zhao}, \& {Yepes}}]{2015MNRAS.446.2621C}
{Chuang}, C.-H., {Kitaura}, F.-S., {Prada}, F., {Zhao}, C., \& {Yepes}, G.
  2015{\natexlab{a}}, \mnras, 446, 2621

\bibitem[{{Chuang} {et~al.}(2019){Chuang}, {Yepes}, {Kitaura},
  {Pellejero-Ibanez}, {Rodr{\'\i}guez-Torres}, {Feng}, {Metcalf}, {Wechsler},
  {Zhao}, {To}, {Alam}, {Banerjee}, {DeRose}, {Giocoli}, {Knebe}, \&
  {Reyes}}]{2019MNRAS.487...48C}
{Chuang}, C.-H., {Yepes}, G., {Kitaura}, F.-S., {et~al.} 2019, \mnras, 487, 48

\bibitem[{{Chuang} {et~al.}(2015{\natexlab{b}}){Chuang}, {Zhao}, {Prada},
  {Munari}, {Avila}, {Izard}, {Kitaura}, {Manera}, {Monaco}, {Murray}, {Knebe},
  {Sc{\'o}ccola}, {Yepes}, {Garcia-Bellido}, {Mar{\'\i}n}, {M{\"u}ller},
  {Skibba}, {Crocce}, {Fosalba}, {Gottl{\"o}ber}, {Klypin}, {Power}, {Tao}, \&
  {Turchaninov}}]{2015MNRAS.452..686C}
{Chuang}, C.-H., {Zhao}, C., {Prada}, F., {et~al.} 2015{\natexlab{b}}, \mnras,
  452, 686

\bibitem[{{Chuang et al.}(2023)}]{Chuang2023}
{Chuang et al.} 2023, in preparation

\bibitem[{{Colavincenzo} {et~al.}(2018){Colavincenzo}, {Sefusatti}, {Monaco},
  {Blot}, {Crocce}, {Lippich}, {S{\'a}nchez}, {Alvarez}, {Agrawal}, {Avila},
  {Balaguera-Antol{\'{\i}}nez}, {Bond}, {Codis}, {Dalla Vecchia}, {Dorta},
  {Fosalba}, {Izard}, {Kitaura}, {Pellejero-Ibanez}, {Stein}, {Vakili}, \&
  {Yepes}}]{2018MNRAS.tmp.2818C}
{Colavincenzo}, M., {Sefusatti}, E., {Monaco}, P., {et~al.} 2018, \mnras

\bibitem[{{Coles} \& {Jones}(1991)}]{1991MNRAS.248....1C}
{Coles}, P. \& {Jones}, B. 1991, \mnras, 248, 1

\bibitem[{{Conroy} {et~al.}(2006){Conroy}, {Wechsler}, \&
  {Kravtsov}}]{2006ApJ...647..201C}
{Conroy}, C., {Wechsler}, R.~H., \& {Kravtsov}, A.~V. 2006, \apj, 647, 201

\bibitem[{{Contreras} {et~al.}(2019){Contreras}, {Zehavi}, {Padilla}, {Baugh},
  {Jim{\'e}nez}, \& {Lacerna}}]{2019MNRAS.484.1133C}
{Contreras}, S., {Zehavi}, I., {Padilla}, N., {et~al.} 2019, \mnras, 484, 1133

\bibitem[{{Cooper} {et~al.}(2022){Cooper}, {Koposov}, {Allende Prieto},
  {Manser}, {Kizhuprakkat}, {Myers}, {Dey}, {Gaensicke}, {Li}, {Rockosi},
  {Valluri}, {Najita}, {Deason}, {Raichoor}, {Wang}, {Ting}, {Kim}, {Carrillo},
  {Wang}, {Beraldo e Silva}, {Han}, {Ding}, {Sanchez-Conde}, {Aguilar},
  {Ahlen}, {Bailey}, {Belokurov}, {Brooks}, {Cunha}, {Dawson}, {de la Macorra},
  {Doel}, {Eisenstein}, {Fagrelius}, {Fanning}, {Font-Ribera}, {Forero-Romero},
  {Gaztanaga}, {Gontcho}, {Guy}, {Honscheid}, {Kehoe}, {Kisner}, {Kremin},
  {Landriau}, {Levi}, {Martini}, {Meisner}, {Miquel}, {Moustakas}, {Nie},
  {Palanque-Delabrouille}, {Percival}, {Poppett}, {Prada}, {Rehemtulla},
  {Schlafly}, {Schlegel}, {Schubnell}, {Sharples}, {Tarle}, {Wechsler},
  {Weinberg}, {Zhou}, \& {Zou}}]{2022arXiv220808514C}
{Cooper}, A.~P., {Koposov}, S.~E., {Allende Prieto}, C., {et~al.} 2022, arXiv
  e-prints, arXiv:2208.08514

\bibitem[{{Cooray}(2002)}]{2002ApJ...576L.105C}
{Cooray}, A. 2002, \apjl, 576, L105

\bibitem[{{Cooray} \& {Sheth}(2002)}]{2002PhR...372....1C}
{Cooray}, A. \& {Sheth}, R. 2002, Phys. Rep., 372, 1

\bibitem[{{Crocce} \& {Scoccimarro}(2006)}]{2006PhRvD..73f3520C}
{Crocce}, M. \& {Scoccimarro}, R. 2006, \prd, 73, 063520

\bibitem[{{Croton} {et~al.}(2007){Croton}, {Gao}, \&
  {White}}]{2007MNRAS.374.1303C}
{Croton}, D.~J., {Gao}, L., \& {White}, S. D.~M. 2007, \mnras, 374, 1303

\bibitem[{{Dalal} {et~al.}(2008){Dalal}, {White}, {Bond}, \&
  {Shirokov}}]{2008ApJ...687...12D}
{Dalal}, N., {White}, M., {Bond}, J.~R., \& {Shirokov}, A. 2008, \apj, 687, 12

\bibitem[{{de Santi} \& {Abramo}(2022)}]{2022arXiv220510881D}
{de Santi}, N. S.~M. \& {Abramo}, L.~R. 2022, arXiv e-prints, arXiv:2205.10881

\bibitem[{{de Santi} {et~al.}(2022){de Santi}, {Rodrigues}, {Montero-Dorta},
  {Abramo}, {Tucci}, \& {Artale}}]{2022MNRAS.514.2463D}
{de Santi}, N. S.~M., {Rodrigues}, N. V.~N., {Montero-Dorta}, A.~D., {et~al.}
  2022, \mnras, 514, 2463

\bibitem[{{Dekel} \& {Lahav}(1999)}]{1999ApJ...520...24D}
{Dekel}, A. \& {Lahav}, O. 1999, ApJ, 520, 24

\bibitem[{{DESI Collaboration} {et~al.}(2022){DESI Collaboration}, {Abareshi},
  {Aguilar}, {Ahlen}, {Alam}, {Alexander}, {Alfarsy}, {Allen}, {Allende
  Prieto}, {Alves}, {Ameel}, {Armengaud}, {Asorey}, {Aviles}, {Bailey},
  {Balaguera-Antol{\'\i}nez}, {Ballester}, {Baltay}, {Bault}, {Beltran},
  {Benavides}, {BenZvi}, {Berti}, {Besuner}, {Beutler}, {Bianchi}, {Blake},
  {Blanc}, {Blum}, {Bolton}, {Bose}, {Bramall}, {Brieden}, {Brodzeller},
  {Brooks}, {Brownewell}, {Buckley-Geer}, {Cahn}, {Cai}, {Canning}, {Carnero
  Rosell}, {Carton}, {Casas}, {Castander}, {Cervantes-Cota}, {Chabanier},
  {Chaussidon}, {Chuang}, {Circosta}, {Cole}, {Cooper}, {da Costa}, {Cousinou},
  {Cuceu}, {Davis}, {Dawson}, {de la Cruz-Noriega}, {de la Macorra}, {de
  Mattia}, {Della Costa}, {Demmer}, {Derwent}, {Dey}, {Dey}, {Dhungana},
  {Ding}, {Dobson}, {Doel}, {Donald-McCann}, {Donaldson}, {Douglass}, {Duan},
  {Dunlop}, {Edelstein}, {Eftekharzadeh}, {Eisenstein}, {Enriquez-Vargas},
  {Escoffier}, {Evatt}, {Fagrelius}, {Fan}, {Fanning}, {Fawcett}, {Ferraro},
  {Ereza}, {Flaugher}, {Font-Ribera}, {Forero-Romero}, {Frenk}, {Fromenteau},
  {G{\"a}nsicke}, {Garcia-Quintero}, {Garrison}, {Gazta{\~n}aga}, {Gerardi},
  {Gil-Mar{\'\i}n}, {Gontcho}, {Gonzalez-Morales}, {Gonzalez-de-Rivera},
  {Gonzalez-Perez}, {Gordon}, {Graur}, {Green}, {Grove}, {Gruen}, {Gutierrez},
  {Guy}, {Hahn}, {Harris}, {Herrera}, {Herrera-Alcantar}, {Honscheid},
  {Howlett}, {Huterer}, {Ir{\v{s}}i{\v{c}}}, {Ishak}, {Jelinsky}, {Jiang},
  {Jimenez}, {Jing}, {Joyce}, {Jullo}, {Juneau}, {Kara{\c{c}}ayl{\i}},
  {Karamanis}, {Karcher}, {Karim}, {Kehoe}, {Kent}, {Kirkby}, {Kisner},
  {Kitaura}, {Koposov}, {Kov{\'a}cs}, {Kremin}, {Krolewski}, {L'Huillier},
  {Lahav}, {Lambert}, {Lamman}, {Lan}, {Landriau}, {Lane}, {Lang}, {Lange},
  {Lasker}, {Le Guillou}, {Leauthaud}, {Le Van Suu}, {Levi}, {Li},
  {Magneville}, {Manera}, {Manser}, {Marshall}, {McCollam}, {McDonald},
  {Meisner}, {Mezcua}, {Miller}, {Miquel}, {Montero-Camacho}, {Moon},
  {Martini}, {Meneses-Rizo}, {Moustakas}, {Mueller}, {Mu{\~n}oz-Guti{\'e}rrez},
  {Myers}, {Nadathur}, {Najita}, {Napolitano}, {Neilsen}, {Newman}, {Nie},
  {Ning}, {Niz}, {Norberg}, {Noriega}, {O'Brien}, {Obuljen},
  {Palanque-Delabrouille}, {Palmese}, {Zhiwei}, {Pappalardo}, {Peng},
  {Percival}, {Perruchot}, {Pogge}, {Poppett}, {Porredon}, {Prada},
  {Prochaska}, {Pucha}, {P{\'e}rez-Fern{\'a}ndez}, {P{\'e}rez-R{\'a}fols},
  {Rabinowitz}, {Raichoor}, {Ramirez-Solano}, {Ram{\'\i}rez-P{\'e}rez},
  {Ravoux}, {Reil}, {Rezaie}, {Rocher}, {Rockosi}, {Roe}, {Roodman}, {Ross},
  {Rossi}, {Ruggeri}, {Ruhlmann-Kleider}, {Sabiu}, {Safonova}, {Said},
  {Saintonge}, {Salas Catonga}, {Samushia}, {Sanchez}, {Saulder}, {Schaan},
  {Schlafly}, {Schlegel}, {Schmoll}, {Scholte}, {Schubnell}, {Secroun}, {Seo},
  {Serrano}, {Sharples}, {Sholl}, {Silber}, {Silva}, {Sirk}, {Siudek}, {Smith},
  {Sprayberry}, {Staten}, {Stupak}, {Tan}, {Tarl{\'e}}, {Sien Tie}, {Tojeiro},
  {Ure{\~n}a-L{\'o}pez}, {Valdes}, {Valenzuela}, {Valluri},
  {Vargas-Maga{\~n}a}, {Verde}, {Walther}, {Wang}, {Wang}, {Weaver},
  {Weaverdyck}, {Wechsler}, {Wilson}, {Yang}, {Yu}, {Yuan}, {Y{\`e}che},
  {Zhang}, {Zhang}, {Zhao}, {Zhou}, {Zhou}, {Zou}, {Zou}, {Zou}, \&
  {Zu}}]{2022arXiv220510939A}
{DESI Collaboration}, {Abareshi}, B., {Aguilar}, J., {et~al.} 2022, arXiv
  e-prints, arXiv:2205.10939

\bibitem[{{DESI Collaboration} {et~al.}(2016{\natexlab{a}}){DESI
  Collaboration}, {Aghamousa}, {Aguilar}, {Ahlen}, {Alam}, {Allen}, {Allende
  Prieto}, {Annis}, {Bailey}, {Balland}, {Ballester}, {Baltay}, {Beaufore},
  {Bebek}, {Beers}, {Bell}, {Bernal}, {Besuner}, {Beutler}, {Blake}, {Bleuler},
  {Blomqvist}, {Blum}, {Bolton}, {Briceno}, {Brooks}, {Brownstein},
  {Buckley-Geer}, {Burden}, {Burtin}, {Busca}, {Cahn}, {Cai}, {Cardiel-Sas},
  {Carlberg}, {Carton}, {Casas}, {Castander}, {Cervantes-Cota}, {Claybaugh},
  {Close}, {Coker}, {Cole}, {Comparat}, {Cooper}, {Cousinou}, {Crocce}, {Cuby},
  {Cunningham}, {Davis}, {Dawson}, {de la Macorra}, {De Vicente}, {Delubac},
  {Derwent}, {Dey}, {Dhungana}, {Ding}, {Doel}, {Duan}, {Ealet}, {Edelstein},
  {Eftekharzadeh}, {Eisenstein}, {Elliott}, {Escoffier}, {Evatt}, {Fagrelius},
  {Fan}, {Fanning}, {Farahi}, {Farihi}, {Favole}, {Feng}, {Fernandez},
  {Findlay}, {Finkbeiner}, {Fitzpatrick}, {Flaugher}, {Flender}, {Font-Ribera},
  {Forero-Romero}, {Fosalba}, {Frenk}, {Fumagalli}, {Gaensicke}, {Gallo},
  {Garcia-Bellido}, {Gaztanaga}, {Pietro Gentile Fusillo}, {Gerard},
  {Gershkovich}, {Giannantonio}, {Gillet}, {Gonzalez-de-Rivera},
  {Gonzalez-Perez}, {Gott}, {Graur}, {Gutierrez}, {Guy}, {Habib}, {Heetderks},
  {Heetderks}, {Heitmann}, {Hellwing}, {Herrera}, {Ho}, {Holland}, {Honscheid},
  {Huff}, {Hutchinson}, {Huterer}, {Hwang}, {Illa Laguna}, {Ishikawa},
  {Jacobs}, {Jeffrey}, {Jelinsky}, {Jennings}, {Jiang}, {Jimenez}, {Johnson},
  {Joyce}, {Jullo}, {Juneau}, {Kama}, {Karcher}, {Karkar}, {Kehoe}, {Kennamer},
  {Kent}, {Kilbinger}, {Kim}, {Kirkby}, {Kisner}, {Kitanidis}, {Kneib},
  {Koposov}, {Kovacs}, {Koyama}, {Kremin}, {Kron}, {Kronig}, {Kueter-Young},
  {Lacey}, {Lafever}, {Lahav}, {Lambert}, {Lampton}, {Landriau}, {Lang},
  {Lauer}, {Le Goff}, {Le Guillou}, {Le Van Suu}, {Lee}, {Lee}, {Leitner},
  {Lesser}, {Levi}, {L'Huillier}, {Li}, {Liang}, {Lin}, {Linder}, {Loebman},
  {Luki{\'c}}, {Ma}, {MacCrann}, {Magneville}, {Makarem}, {Manera}, {Manser},
  {Marshall}, {Martini}, {Massey}, {Matheson}, {McCauley}, {McDonald},
  {McGreer}, {Meisner}, {Metcalfe}, {Miller}, {Miquel}, {Moustakas}, {Myers},
  {Naik}, {Newman}, {Nichol}, {Nicola}, {Nicolati da Costa}, {Nie}, {Niz},
  {Norberg}, {Nord}, {Norman}, {Nugent}, {O'Brien}, {Oh}, {Olsen}, {Padilla},
  {Padmanabhan}, {Padmanabhan}, {Palanque-Delabrouille}, {Palmese},
  {Pappalardo}, {P{\^a}ris}, {Park}, {Patej}, {Peacock}, {Peiris}, {Peng},
  {Percival}, {Perruchot}, {Pieri}, {Pogge}, {Pollack}, {Poppett}, {Prada},
  {Prakash}, {Probst}, {Rabinowitz}, {Raichoor}, {Ree}, {Refregier}, {Regal},
  {Reid}, {Reil}, {Rezaie}, {Rockosi}, {Roe}, {Ronayette}, {Roodman}, {Ross},
  {Ross}, {Rossi}, {Rozo}, {Ruhlmann-Kleider}, {Rykoff}, {Sabiu}, {Samushia},
  {Sanchez}, {Sanchez}, {Schlegel}, {Schneider}, {Schubnell}, {Secroun},
  {Seljak}, {Seo}, {Serrano}, {Shafieloo}, {Shan}, {Sharples}, {Sholl},
  {Shourt}, {Silber}, {Silva}, {Sirk}, {Slosar}, {Smith}, {Smoot}, {Som},
  {Song}, {Sprayberry}, {Staten}, {Stefanik}, {Tarle}, {Sien Tie}, {Tinker},
  {Tojeiro}, {Valdes}, {Valenzuela}, {Valluri}, {Vargas-Magana}, {Verde},
  {Walker}, {Wang}, {Wang}, {Weaver}, {Weaverdyck}, {Wechsler}, {Weinberg},
  {White}, {Yang}, {Yeche}, {Zhang}, {Zhao}, {Zheng}, {Zhou}, {Zhou}, {Zhu},
  {Zou}, \& {Zu}}]{2016arXiv161100036D}
{DESI Collaboration}, {Aghamousa}, A., {Aguilar}, J., {et~al.}
  2016{\natexlab{a}}, arXiv e-prints, arXiv:1611.00036

\bibitem[{{DESI Collaboration} {et~al.}(2016{\natexlab{b}}){DESI
  Collaboration}, {Aghamousa}, {Aguilar}, {Ahlen}, {Alam}, {Allen}, {Allende
  Prieto}, {Annis}, {Bailey}, {Balland}, {Ballester}, {Baltay}, {Beaufore},
  {Bebek}, {Beers}, {Bell}, {Bernal}, {Besuner}, {Beutler}, {Blake}, {Bleuler},
  {Blomqvist}, {Blum}, {Bolton}, {Briceno}, {Brooks}, {Brownstein},
  {Buckley-Geer}, {Burden}, {Burtin}, {Busca}, {Cahn}, {Cai}, {Cardiel-Sas},
  {Carlberg}, {Carton}, {Casas}, {Castander}, {Cervantes-Cota}, {Claybaugh},
  {Close}, {Coker}, {Cole}, {Comparat}, {Cooper}, {Cousinou}, {Crocce}, {Cuby},
  {Cunningham}, {Davis}, {Dawson}, {de la Macorra}, {De Vicente}, {Delubac},
  {Derwent}, {Dey}, {Dhungana}, {Ding}, {Doel}, {Duan}, {Ealet}, {Edelstein},
  {Eftekharzadeh}, {Eisenstein}, {Elliott}, {Escoffier}, {Evatt}, {Fagrelius},
  {Fan}, {Fanning}, {Farahi}, {Farihi}, {Favole}, {Feng}, {Fernandez},
  {Findlay}, {Finkbeiner}, {Fitzpatrick}, {Flaugher}, {Flender}, {Font-Ribera},
  {Forero-Romero}, {Fosalba}, {Frenk}, {Fumagalli}, {Gaensicke}, {Gallo},
  {Garcia-Bellido}, {Gaztanaga}, {Pietro Gentile Fusillo}, {Gerard},
  {Gershkovich}, {Giannantonio}, {Gillet}, {Gonzalez-de-Rivera},
  {Gonzalez-Perez}, {Gott}, {Graur}, {Gutierrez}, {Guy}, {Habib}, {Heetderks},
  {Heetderks}, {Heitmann}, {Hellwing}, {Herrera}, {Ho}, {Holland}, {Honscheid},
  {Huff}, {Hutchinson}, {Huterer}, {Hwang}, {Illa Laguna}, {Ishikawa},
  {Jacobs}, {Jeffrey}, {Jelinsky}, {Jennings}, {Jiang}, {Jimenez}, {Johnson},
  {Joyce}, {Jullo}, {Juneau}, {Kama}, {Karcher}, {Karkar}, {Kehoe}, {Kennamer},
  {Kent}, {Kilbinger}, {Kim}, {Kirkby}, {Kisner}, {Kitanidis}, {Kneib},
  {Koposov}, {Kovacs}, {Koyama}, {Kremin}, {Kron}, {Kronig}, {Kueter-Young},
  {Lacey}, {Lafever}, {Lahav}, {Lambert}, {Lampton}, {Landriau}, {Lang},
  {Lauer}, {Le Goff}, {Le Guillou}, {Le Van Suu}, {Lee}, {Lee}, {Leitner},
  {Lesser}, {Levi}, {L'Huillier}, {Li}, {Liang}, {Lin}, {Linder}, {Loebman},
  {Luki{\'c}}, {Ma}, {MacCrann}, {Magneville}, {Makarem}, {Manera}, {Manser},
  {Marshall}, {Martini}, {Massey}, {Matheson}, {McCauley}, {McDonald},
  {McGreer}, {Meisner}, {Metcalfe}, {Miller}, {Miquel}, {Moustakas}, {Myers},
  {Naik}, {Newman}, {Nichol}, {Nicola}, {Nicolati da Costa}, {Nie}, {Niz},
  {Norberg}, {Nord}, {Norman}, {Nugent}, {O'Brien}, {Oh}, {Olsen}, {Padilla},
  {Padmanabhan}, {Padmanabhan}, {Palanque-Delabrouille}, {Palmese},
  {Pappalardo}, {P{\^a}ris}, {Park}, {Patej}, {Peacock}, {Peiris}, {Peng},
  {Percival}, {Perruchot}, {Pieri}, {Pogge}, {Pollack}, {Poppett}, {Prada},
  {Prakash}, {Probst}, {Rabinowitz}, {Raichoor}, {Ree}, {Refregier}, {Regal},
  {Reid}, {Reil}, {Rezaie}, {Rockosi}, {Roe}, {Ronayette}, {Roodman}, {Ross},
  {Ross}, {Rossi}, {Rozo}, {Ruhlmann-Kleider}, {Rykoff}, {Sabiu}, {Samushia},
  {Sanchez}, {Sanchez}, {Schlegel}, {Schneider}, {Schubnell}, {Secroun},
  {Seljak}, {Seo}, {Serrano}, {Shafieloo}, {Shan}, {Sharples}, {Sholl},
  {Shourt}, {Silber}, {Silva}, {Sirk}, {Slosar}, {Smith}, {Smoot}, {Som},
  {Song}, {Sprayberry}, {Staten}, {Stefanik}, {Tarle}, {Sien Tie}, {Tinker},
  {Tojeiro}, {Valdes}, {Valenzuela}, {Valluri}, {Vargas-Magana}, {Verde},
  {Walker}, {Wang}, {Wang}, {Weaver}, {Weaverdyck}, {Wechsler}, {Weinberg},
  {White}, {Yang}, {Yeche}, {Zhang}, {Zhao}, {Zheng}, {Zhou}, {Zhou}, {Zhu},
  {Zou}, \& {Zu}}]{2016arXiv161100037D}
{DESI Collaboration}, {Aghamousa}, A., {Aguilar}, J., {et~al.}
  2016{\natexlab{b}}, arXiv e-prints, arXiv:1611.00037

\bibitem[{{DESI collaboration et al.}(2022)}]{sv}
{DESI collaboration et al.} 2022, in preparation

\bibitem[{{DESI collaboration et al.}(2023)}]{dr}
{DESI collaboration et al.} 2023, in preparation

\bibitem[{{Desjacques} {et~al.}(2018){Desjacques}, {Jeong}, \&
  {Schmidt}}]{2018PhR...733....1D}
{Desjacques}, V., {Jeong}, D., \& {Schmidt}, F. 2018, \physrep, 733, 1

\bibitem[{{Dey} {et~al.}(2019){Dey}, {Schlegel}, {Lang}, {Blum}, {Burleigh},
  {Fan}, {Findlay}, {Finkbeiner}, {Herrera}, {Juneau}, {Landriau}, {Levi},
  {McGreer}, {Meisner}, {Myers}, {Moustakas}, {Nugent}, {Patej}, {Schlafly},
  {Walker}, {Valdes}, {Weaver}, {Y{\`e}che}, {Zou}, {Zhou}, {Abareshi},
  {Abbott}, {Abolfathi}, {Aguilera}, {Alam}, {Allen}, {Alvarez}, {Annis},
  {Ansarinejad}, {Aubert}, {Beechert}, {Bell}, {BenZvi}, {Beutler}, {Bielby},
  {Bolton}, {Brice{\~n}o}, {Buckley-Geer}, {Butler}, {Calamida}, {Carlberg},
  {Carter}, {Casas}, {Castander}, {Choi}, {Comparat}, {Cukanovaite}, {Delubac},
  {DeVries}, {Dey}, {Dhungana}, {Dickinson}, {Ding}, {Donaldson}, {Duan},
  {Duckworth}, {Eftekharzadeh}, {Eisenstein}, {Etourneau}, {Fagrelius},
  {Farihi}, {Fitzpatrick}, {Font-Ribera}, {Fulmer}, {G{\"a}nsicke},
  {Gaztanaga}, {George}, {Gerdes}, {Gontcho}, {Gorgoni}, {Green}, {Guy},
  {Harmer}, {Hernandez}, {Honscheid}, {Huang}, {James}, {Jannuzi}, {Jiang},
  {Joyce}, {Karcher}, {Karkar}, {Kehoe}, {Kneib}, {Kueter-Young}, {Lan},
  {Lauer}, {Le Guillou}, {Le Van Suu}, {Lee}, {Lesser}, {Perreault Levasseur},
  {Li}, {Mann}, {Marshall}, {Mart{\'\i}nez-V{\'a}zquez}, {Martini}, {du Mas des
  Bourboux}, {McManus}, {Meier}, {M{\'e}nard}, {Metcalfe},
  {Mu{\~n}oz-Guti{\'e}rrez}, {Najita}, {Napier}, {Narayan}, {Newman}, {Nie},
  {Nord}, {Norman}, {Olsen}, {Paat}, {Palanque-Delabrouille}, {Peng},
  {Poppett}, {Poremba}, {Prakash}, {Rabinowitz}, {Raichoor}, {Rezaie},
  {Robertson}, {Roe}, {Ross}, {Ross}, {Rudnick}, {Safonova}, {Saha},
  {S{\'a}nchez}, {Savary}, {Schweiker}, {Scott}, {Seo}, {Shan}, {Silva},
  {Slepian}, {Soto}, {Sprayberry}, {Staten}, {Stillman}, {Stupak}, {Summers},
  {Sien Tie}, {Tirado}, {Vargas-Maga{\~n}a}, {Vivas}, {Wechsler}, {Williams},
  {Yang}, {Yang}, {Yapici}, {Zaritsky}, {Zenteno}, {Zhang}, {Zhang}, {Zhou}, \&
  {Zhou}}]{2019AJ....157..168D}
{Dey}, A., {Schlegel}, D.~J., {Lang}, D., {et~al.} 2019, \aj, 157, 168

\bibitem[{{Ding} {et~al.}(2022){Ding}, {Chuang}, {Yu}, {Garrison}, {Bayer},
  {Feng}, {Modi}, {Eisenstein}, {White}, {Variu}, {Zhao}, {Zhang}, {Meneses
  Rizo}, {Brooks}, {Dawson}, {Doel}, {Gaztanaga}, {Kehoe}, {Krolewski},
  {Landriau}, {Palanque-Delabrouille}, \& {Poppett}}]{2022arXiv220206074D}
{Ding}, Z., {Chuang}, C.-H., {Yu}, Y., {et~al.} 2022, arXiv e-prints,
  arXiv:2202.06074

\bibitem[{Dodelson(2003)}]{dodelson:2003}
Dodelson, S. 2003, {Modern Cosmology} (Academic Press, Elsevier Science)

\bibitem[{{Dodelson} \& {Schneider}(2013)}]{2013PhRvD..88f3537D}
{Dodelson}, S. \& {Schneider}, M.~D. 2013, \prd, 88, 063537

\bibitem[{{Dvorkin} {et~al.}(2022){Dvorkin}, {Mishra-Sharma}, {Nord}, {Villar},
  {Avestruz}, {Bechtol}, {{\'C}iprijanovi{\'c}}, {Connolly}, {Garrison},
  {Narayan}, \& {Villaescusa-Navarro}}]{2022arXiv220308056D}
{Dvorkin}, C., {Mishra-Sharma}, S., {Nord}, B., {et~al.} 2022, arXiv e-prints,
  arXiv:2203.08056

\bibitem[{{Faltenbacher} \& {White}(2010)}]{2010ApJ...708..469F}
{Faltenbacher}, A. \& {White}, S. D.~M. 2010, \apj, 708, 469

\bibitem[{{Favole} {et~al.}(2016){Favole}, {Comparat}, {Prada}, {Yepes},
  {Jullo}, {Niemiec}, {Kneib}, {Rodr{\'\i}guez-Torres}, {Klypin}, {Skibba},
  {McBride}, {Eisenstein}, {Schlegel}, {Nuza}, {Chuang}, {Delubac},
  {Y{\`e}che}, \& {Schneider}}]{2016MNRAS.461.3421F}
{Favole}, G., {Comparat}, J., {Prada}, F., {et~al.} 2016, \mnras, 461, 3421

\bibitem[{{Favole} {et~al.}(2017){Favole}, {Rodr{\'\i}guez-Torres}, {Comparat},
  {Prada}, {Guo}, {Klypin}, \& {Montero-Dorta}}]{2017MNRAS.472..550F}
{Favole}, G., {Rodr{\'\i}guez-Torres}, S.~A., {Comparat}, J., {et~al.} 2017,
  \mnras, 472, 550

\bibitem[{{Feng} {et~al.}(2016){Feng}, {Chu}, {Seljak}, \&
  {McDonald}}]{2016MNRAS.463.2273F}
{Feng}, Y., {Chu}, M.-Y., {Seljak}, U., \& {McDonald}, P. 2016, \mnras, 463,
  2273

\bibitem[{{Fisher} \& {Faltenbacher}(2018)}]{2018MNRAS.473.3941F}
{Fisher}, J.~D. \& {Faltenbacher}, A. 2018, \mnras, 473, 3941

\bibitem[{{Forero-Romero} {et~al.}(2009){Forero-Romero}, {Hoffman},
  {Gottl{\"o}ber}, {Klypin}, \& {Yepes}}]{2009MNRAS.396.1815F}
{Forero-Romero}, J.~E., {Hoffman}, Y., {Gottl{\"o}ber}, S., {Klypin}, A., \&
  {Yepes}, G. 2009, \mnras, 396, 1815

\bibitem[{{Forero-S{\'a}nchez} {et~al.}(2022){Forero-S{\'a}nchez}, {Chuang},
  {Rodr{\'\i}guez-Torres}, {Yepes}, {Gottl{\"o}ber}, \&
  {Zhao}}]{2022arXiv220312669F}
{Forero-S{\'a}nchez}, D., {Chuang}, C.-H., {Rodr{\'\i}guez-Torres}, S.,
  {et~al.} 2022, \mnras, 513, 4318

\bibitem[{{Fry} \& {Gaztanaga}(1993)}]{1993ApJ...413..447F}
{Fry}, J.~N. \& {Gaztanaga}, E. 1993, \apj, 413, 447

\bibitem[{{Gao} {et~al.}(2005){Gao}, {Springel}, \&
  {White}}]{2005MNRAS.363L..66G}
{Gao}, L., {Springel}, V., \& {White}, S. D.~M. 2005, \mnras, 363, L66

\bibitem[{{Gao} \& {White}(2007)}]{2007MNRAS.377L...5G}
{Gao}, L. \& {White}, S. D.~M. 2007, \mnras, 377, L5

\bibitem[{{Garc{\'\i}a} \& {Rozo}(2019)}]{2019MNRAS.489.4170G}
{Garc{\'\i}a}, R. \& {Rozo}, E. 2019, \mnras, 489, 4170

\bibitem[{{Garrison} {et~al.}(2018){Garrison}, {Eisenstein}, {Ferrer},
  {Tinker}, {Pinto}, \& {Weinberg}}]{2018ApJS..236...43G}
{Garrison}, L.~H., {Eisenstein}, D.~J., {Ferrer}, D., {et~al.} 2018, \apjs,
  236, 43

\bibitem[{{Gil-Mar{\'\i}n} {et~al.}(2012){Gil-Mar{\'\i}n}, {Wagner},
  {Fragkoudi}, {Jimenez}, \& {Verde}}]{2012JCAP...02..047G}
{Gil-Mar{\'\i}n}, H., {Wagner}, C., {Fragkoudi}, F., {Jimenez}, R., \& {Verde},
  L. 2012, \jcap, 2012, 047

\bibitem[{{Goldstein} {et~al.}(2022){Goldstein}, {Pandey}, {Slosar}, {Blazek},
  {Jain}, \& {LSST Dark Energy Science Collaboration}}]{2022PhRvD.105l3518G}
{Goldstein}, S., {Pandey}, S., {Slosar}, A., {et~al.} 2022, \prd, 105, 123518

\bibitem[{{Gonzalez-Perez} {et~al.}(2018){Gonzalez-Perez}, {Comparat},
  {Norberg}, {Baugh}, {Contreras}, {Lacey}, {McCullagh}, {Orsi}, {Helly}, \&
  {Humphries}}]{2018MNRAS.474.4024G}
{Gonzalez-Perez}, V., {Comparat}, J., {Norberg}, P., {et~al.} 2018, \mnras,
  474, 4024

\bibitem[{{Grove} {et~al.}(2022){Grove}, {Chuang}, {Devi}, {Garrison},
  {L'Huillier}, {Feng}, {Helly}, {Hern{\'a}ndez-Aguayo}, {Alam}, {Zhang}, {Yu},
  {Cole}, {Eisenstein}, {Norberg}, {Wechsler}, {Brooks}, {Dawson}, {Landriau},
  {Meisner}, {Poppett}, {Tarl{\'e}}, \& {Valenzuela}}]{2021arXiv211209138G}
{Grove}, C., {Chuang}, C.-H., {Devi}, N.~C., {et~al.} 2022, \mnras, 515, 1854

\bibitem[{{Guy et al.}(2022)}]{spec2022}
{Guy et al.} 2022, in preparation

\bibitem[{{Hahn} {et~al.}(2022){Hahn}, {Wilson}, {Ruiz-Macias}, {Cole},
  {Weinberg}, {Moustakas}, {Kremin}, {Tinker}, {Smith}, {Wechsler}, {Ahlen},
  {Alam}, {Bailey}, {Brooks}, {Cooper}, {Davis}, {Dawson}, {Dey}, {Dey},
  {Eftekharzadeh}, {Eisenstein}, {Fanning}, {Forero-Romero}, {Frenk},
  {Gazta{\~n}aga}, {Gontcho}, {Guy}, {Honscheid}, {Ishak}, {Juneau}, {Kehoe},
  {Kisner}, {Lan}, {Landriau}, {Le Guillou}, {Levi}, {Magneville}, {Martini},
  {Meisner}, {Myers}, {Nie}, {Norberg}, {Palanque-Delabrouille}, {Percival},
  {Poppett}, {Prada}, {Raichoor}, {Ross}, {Safonova}, {Saulder}, {Schlafly},
  {Schlegel}, {Sierra-Porta}, {Tarle}, {Weaver}, {Y{\`e}che}, {Zarrouk},
  {Zhou}, {Zhou}, \& {Zou}}]{2022arXiv220808512H}
{Hahn}, C., {Wilson}, M.~J., {Ruiz-Macias}, O., {et~al.} 2022, arXiv e-prints,
  arXiv:2208.08512

\bibitem[{{Hahn} {et~al.}(2013){Hahn}, {Abel}, \& {Kaehler}}]{Hahn:2013aa}
{Hahn}, O., {Abel}, T., \& {Kaehler}, R. 2013, \mnras, 434, 1171

\bibitem[{{Hahn} {et~al.}(2007){Hahn}, {Porciani}, {Carollo}, \&
  {Dekel}}]{2007MNRAS.375..489H}
{Hahn}, O., {Porciani}, C., {Carollo}, C.~M., \& {Dekel}, A. 2007, \mnras, 375,
  489

\bibitem[{{Hamaus} {et~al.}(2012){Hamaus}, {Seljak}, \&
  {Desjacques}}]{2012PhRvD..86j3513H}
{Hamaus}, N., {Seljak}, U., \& {Desjacques}, V. 2012, \prd, 86, 103513

\bibitem[{{Hamilton}(1998)}]{1998ASSL..231..185H}
{Hamilton}, A.~J.~S. 1998, in Astrophysics and Space Science Library, Vol. 231,
  The Evolving Universe, ed. D.~{Hamilton}, 185

\bibitem[{{Han} {et~al.}(2019){Han}, {Li}, {Jing}, {Nishimichi}, {Wang}, \&
  {Jiang}}]{2019MNRAS.482.1900H}
{Han}, J., {Li}, Y., {Jing}, Y., {et~al.} 2019, \mnras, 482, 1900

\bibitem[{{Harnois-D{\'e}raps} {et~al.}(2018){Harnois-D{\'e}raps}, {Amon},
  {Choi}, {Demchenko}, {Heymans}, {Kannawadi}, {Nakajima}, {Sirks}, {van
  Waerbeke}, {Cai}, {Giblin}, {Hildebrandt}, {Hoekstra}, {Miller}, \&
  {Tr{\"o}ster}}]{2018MNRAS.481.1337H}
{Harnois-D{\'e}raps}, J., {Amon}, A., {Choi}, A., {et~al.} 2018, \mnras, 481,
  1337

\bibitem[{Harnois-Déraps {et~al.}(2013)Harnois-Déraps, Pen, Iliev, Merz,
  Emberson, \& Desjacques}]{10.1093/mnras/stt1591}
Harnois-Déraps, J., Pen, U.-L., Iliev, I.~T., {et~al.} 2013, Monthly Notices
  of the Royal Astronomical Society, 436, 540

\bibitem[{{He} {et~al.}(2019){He}, {Li}, {Feng}, {Ho}, {Ravanbakhsh}, {Chen},
  \& {P{\'o}czos}}]{2019PNAS..11613825H}
{He}, S., {Li}, Y., {Feng}, Y., {et~al.} 2019, Proceedings of the National
  Academy of Science, 116, 13825

\bibitem[{{Hearin} {et~al.}(2016){Hearin}, {Zentner}, {van den Bosch},
  {Campbell}, \& {Tollerud}}]{2016MNRAS.460.2552H}
{Hearin}, A.~P., {Zentner}, A.~R., {van den Bosch}, F.~C., {Campbell}, D., \&
  {Tollerud}, E. 2016, \mnras, 460, 2552

\bibitem[{{Heath}(1977)}]{1977MNRAS.179..351H}
{Heath}, D.~J. 1977, \mnras, 179, 351

\bibitem[{{Heavens}(2009)}]{2009arXiv0906.0664H}
{Heavens}, A. 2009, arXiv e-prints, arXiv:0906.0664

\bibitem[{{He{\ss}} {et~al.}(2013){He{\ss}}, {Kitaura}, \&
  {Gottl{\"o}ber}}]{2013MNRAS.435.2065H}
{He{\ss}}, S., {Kitaura}, F.-S., \& {Gottl{\"o}ber}, S. 2013, \mnras, 435, 2065

\bibitem[{{Hikage} \& {Yamamoto}(2016)}]{2016MNRAS.455L..77H}
{Hikage}, C. \& {Yamamoto}, K. 2016, \mnras, 455, L77

\bibitem[{{Hockney} \& {Eastwood}(1988)}]{1988csup.bookH}
{Hockney}, R.~W. \& {Eastwood}, J.~W. 1988, {Computer simulation using
  particles}

\bibitem[{{Howlett} {et~al.}(2015){Howlett}, {Manera}, \&
  {Percival}}]{2015A&C....12..109H}
{Howlett}, C., {Manera}, M., \& {Percival}, W.~J. 2015, Astronomy and
  Computing, 12, 109

\bibitem[{{Howlett} \& {Percival}(2017)}]{2017MNRAS.472.4935H}
{Howlett}, C. \& {Percival}, W.~J. 2017, \mnras, 472, 4935

\bibitem[{{Izard} {et~al.}(2018){Izard}, {Fosalba}, \&
  {Crocce}}]{2018MNRAS.473.3051I}
{Izard}, A., {Fosalba}, P., \& {Crocce}, M. 2018, \mnras, 473, 3051

\bibitem[{{Jing}(2005)}]{2005ApJ...620..559J}
{Jing}, Y.~P. 2005, \apj, 620, 559

\bibitem[{{Kaiser}(1987)}]{1987MNRAS.227....1K}
{Kaiser}, N. 1987, \mnras, 227, 1

\bibitem[{{Kauffmann} {et~al.}(1997){Kauffmann}, {Nusser}, \&
  {Steinmetz}}]{1997MNRAS.286..795K}
{Kauffmann}, G., {Nusser}, A., \& {Steinmetz}, M. 1997, \mnras, 286, 795

\bibitem[{{Kerscher} {et~al.}(2000){Kerscher}, {Szapudi}, \&
  {Szalay}}]{2000ApJ...535L..13K}
{Kerscher}, M., {Szapudi}, I., \& {Szalay}, A.~S. 2000, \apjl, 535, L13

\bibitem[{Kitaura {et~al.}(2022)Kitaura, Balaguera-Antolínez, Sinigaglia, \&
  Pellejero-Ibáñez}]{10.1093/mnras/stac671}
Kitaura, F.-S., Balaguera-Antolínez, A., Sinigaglia, F., \&
  Pellejero-Ibáñez, M. 2022, Monthly Notices of the Royal Astronomical
  Society, 512, 2245

\bibitem[{Kitaura \& Hess(2013)}]{doi:10.1093/mnrasl/slt101}
Kitaura, F.-S. \& Hess, S. 2013, \mnras, 435, L78

\bibitem[{{Kitaura} {et~al.}(2016){Kitaura}, {Rodr{\'\i}guez-Torres}, {Chuang},
  {Zhao}, {Prada}, {Gil-Mar{\'\i}n}, {Guo}, {Yepes}, {Klypin}, {Sc{\'o}ccola},
  {Tinker}, {McBride}, {Reid}, {S{\'a}nchez}, {Salazar-Albornoz}, {Grieb},
  {Vargas-Magana}, {Cuesta}, {Neyrinck}, {Beutler}, {Comparat}, {Percival}, \&
  {Ross}}]{2016MNRAS.456.4156K}
{Kitaura}, F.-S., {Rodr{\'\i}guez-Torres}, S., {Chuang}, C.-H., {et~al.} 2016,
  \mnras, 456, 4156

\bibitem[{{Kitaura} {et~al.}(2023){Kitaura}, {Sinigaglia},
  {Balaguera-Antol{\'\i}nez}, \& {Favole}}]{2023arXiv230103648K}
{Kitaura}, F.~S., {Sinigaglia}, F., {Balaguera-Antol{\'\i}nez}, A., \&
  {Favole}, G. 2023, arXiv e-prints, arXiv:2301.03648

\bibitem[{{Kitaura} {et~al.}(2014){Kitaura}, {Yepes}, \&
  {Prada}}]{2014MNRAS.439L..21K}
{Kitaura}, F.-S., {Yepes}, G., \& {Prada}, F. 2014, \mnras, 439, L21

\bibitem[{{Klypin} \& {Prada}(2018)}]{2018MNRAS.478.4602K}
{Klypin}, A. \& {Prada}, F. 2018, \mnras, 478, 4602

\bibitem[{{Koda} {et~al.}(2016){Koda}, {Blake}, {Beutler}, {Kazin}, \&
  {Marin}}]{2016MNRAS.459.2118K}
{Koda}, J., {Blake}, C., {Beutler}, F., {Kazin}, E., \& {Marin}, F. 2016,
  \mnras, 459, 2118

\bibitem[{{Kravtsov} {et~al.}(2004){Kravtsov}, {Berlind}, {Wechsler}, {Klypin},
  {Gottl{\"o}ber}, {Allgood}, \& {Primack}}]{2004ApJ...609...35K}
{Kravtsov}, A.~V., {Berlind}, A.~A., {Wechsler}, R.~H., {et~al.} 2004, \apj,
  609, 35

\bibitem[{{Kreisch} {et~al.}(2021){Kreisch}, {Pisani}, {Villaescusa-Navarro},
  {Spergel}, {Wandelt}, {Hamaus}, \& {Bayer}}]{2021arXiv210702304K}
{Kreisch}, C.~D., {Pisani}, A., {Villaescusa-Navarro}, F., {et~al.} 2021, arXiv
  e-prints, arXiv:2107.02304

\bibitem[{{Lacasa}(2018)}]{2018A&A...615A...1L}
{Lacasa}, F. 2018, \aap, 615, A1

\bibitem[{{Lahav} {et~al.}(1991){Lahav}, {Lilje}, {Primack}, \&
  {Rees}}]{1991MNRAS.251..128L}
{Lahav}, O., {Lilje}, P.~B., {Primack}, J.~R., \& {Rees}, M.~J. 1991, \mnras,
  251, 128

\bibitem[{{Lan} {et~al.}(2022){Lan}, {Tojeiro}, {Armengaud}, {Prochaska},
  {Davis}, {Alexander}, {Raichoor}, {Zhou}, {Yeche}, {Balland}, {BenZvi},
  {Berti}, {Canning}, {Carr}, {Chittenden}, {Cole}, {Cousinou}, {Dawson},
  {Dey}, {Douglass}, {Edge}, {Escoffier}, {Glanville}, {Gontcho}, {Guy},
  {Hahn}, {Howlett}, {Hwang}, {Jiang}, {Kovacs}, {Mezcua}, {Moore}, {Nadathur},
  {Oh}, {Parkinson}, {Rocher}, {Ross}, {Ruhlmann-Kleider}, {Sabiu}, {Said},
  {Saulder}, {Sierra-Porta}, {Weiner}, {Yu}, {Zarrouk}, {Zhang}, {Zou},
  {Ahlen}, {Bailey}, {Brooks}, {Cooper}, {de la Macorra}, {Dey}, {Dhungana},
  {Doel}, {Eftekharzadeh}, {Fanning}, {Font-Ribera}, {Garrison}, {Gaztanaga},
  {Kehoe}, {Kisner}, {Kremin}, {Landriau}, {Le Guillou}, {Levi}, {Magneville},
  {Meisner}, {Miquel}, {Moustakas}, {Myers}, {Newman}, {Nie},
  {Palanque-Delabrouille}, {Percival}, {Poppett}, {Prada}, {Schubnell},
  {Tarle}, {Weaver}, {Zhang}, \& {Zhou}}]{2022arXiv220808516L}
{Lan}, T.-W., {Tojeiro}, R., {Armengaud}, E., {et~al.} 2022, arXiv e-prints,
  arXiv:2208.08516

\bibitem[{{Lazeyras} {et~al.}(2017){Lazeyras}, {Musso}, \&
  {Schmidt}}]{2017JCAP...03..059L}
{Lazeyras}, T., {Musso}, M., \& {Schmidt}, F. 2017, \jcap, 2017, 059

\bibitem[{{Lee} {et~al.}(2017){Lee}, {Primack}, {Behroozi},
  {Rodr{\'\i}guez-Puebla}, {Hellinger}, \& {Dekel}}]{2017MNRAS.466.3834L}
{Lee}, C.~T., {Primack}, J.~R., {Behroozi}, P., {et~al.} 2017, \mnras, 466,
  3834

\bibitem[{{Levi} {et~al.}(2013){Levi}, {Bebek}, {Beers}, {Blum}, {Cahn},
  {Eisenstein}, {Flaugher}, {Honscheid}, {Kron}, {Lahav}, {McDonald}, {Roe},
  {Schlegel}, \& {representing the DESI collaboration}}]{DESI}
{Levi}, M., {Bebek}, C., {Beers}, T., {et~al.} 2013, ArXiv e-prints

\bibitem[{{Li} {et~al.}(2021){Li}, {Ni}, {Croft}, {Di Matteo}, {Bird}, \&
  {Feng}}]{2021PNAS..11822038L}
{Li}, Y., {Ni}, Y., {Croft}, R. A.~C., {et~al.} 2021, Proceedings of the
  National Academy of Science, 118, e2022038118

\bibitem[{{Libeskind} {et~al.}(2018){Libeskind}, {van de Weygaert}, {Cautun},
  {Falck}, {Tempel}, {Abel}, {Alpaslan}, {Arag{\'o}n-Calvo}, {Forero-Romero},
  {Gonzalez}, {Gottl{\"o}ber}, {Hahn}, {Hellwing}, {Hoffman}, {Jones},
  {Kitaura}, {Knebe}, {Manti}, {Neyrinck}, {Nuza}, {Padilla}, {Platen},
  {Ramachandra}, {Robotham}, {Saar}, {Shandarin}, {Steinmetz}, {Stoica},
  {Sousbie}, \& {Yepes}}]{2018MNRAS.473.1195L}
{Libeskind}, N.~I., {van de Weygaert}, R., {Cautun}, M., {et~al.} 2018, \mnras,
  473, 1195

\bibitem[{{Lippich} {et~al.}(2019){Lippich}, {S{\'a}nchez}, {Colavincenzo},
  {Sefusatti}, {Monaco}, {Blot}, {Crocce}, {Alvarez}, {Agrawal}, {Avila},
  {Balaguera-Antol{\'{\i}}nez}, {Bond}, {Codis}, {Dalla Vecchia}, {Dorta},
  {Fosalba}, {Izard}, {Kitaura}, {Pellejero-Ibanez}, {Stein}, {Vakili}, \&
  {Yepes}}]{2019MNRAS.482.1786L}
{Lippich}, M., {S{\'a}nchez}, A.~G., {Colavincenzo}, M., {et~al.} 2019, \mnras,
  482, 1786

\bibitem[{{Maion} {et~al.}(2022){Maion}, {Angulo}, \&
  {Zennaro}}]{2022arXiv220403868M}
{Maion}, F., {Angulo}, R.~E., \& {Zennaro}, M. 2022, arXiv e-prints,
  arXiv:2204.03868

\bibitem[{{Maksimova} {et~al.}(2021){Maksimova}, {Garrison}, {Eisenstein},
  {Hadzhiyska}, {Bose}, \& {Satterthwaite}}]{2021MNRAS.508.4017M}
{Maksimova}, N.~A., {Garrison}, L.~H., {Eisenstein}, D.~J., {et~al.} 2021,
  \mnras, 508, 4017

\bibitem[{{Manera} {et~al.}(2013){Manera}, {Scoccimarro}, {Percival},
  {Samushia}, {McBride}, {Ross}, {Sheth}, {White}, {Reid}, {S{\'a}nchez}, {de
  Putter}, {Xu}, {Berlind}, {Brinkmann}, {Maraston}, {Nichol}, {Montesano},
  {Padmanabhan}, {Skibba}, {Tojeiro}, \& {Weaver}}]{2013MNRAS.428.1036M}
{Manera}, M., {Scoccimarro}, R., {Percival}, W.~J., {et~al.} 2013, \mnras, 428,
  1036

\bibitem[{{Mao} {et~al.}(2018){Mao}, {Zentner}, \&
  {Wechsler}}]{2018MNRAS.474.5143M}
{Mao}, Y.-Y., {Zentner}, A.~R., \& {Wechsler}, R.~H. 2018, \mnras, 474, 5143

\bibitem[{{Matsubara}(1999)}]{1999ApJ...525..543M}
{Matsubara}, T. 1999, \apj, 525, 543

\bibitem[{{McDonald} \& {Roy}(2009)}]{2009JCAP...08..020M}
{McDonald}, P. \& {Roy}, A. 2009, \jcap, 8, 020

\bibitem[{{Miller et al.}(2022)}]{DESIcorrector2022}
{Miller et al.} 2022, in preparation

\bibitem[{{Mohayaee} {et~al.}(2006){Mohayaee}, {Mathis}, {Colombi}, \&
  {Silk}}]{2006MNRAS.365..939M}
{Mohayaee}, R., {Mathis}, H., {Colombi}, S., \& {Silk}, J. 2006, \mnras, 365,
  939

\bibitem[{{Monaco}(2016)}]{2016Galax...4...53M}
{Monaco}, P. 2016, Galaxies, 4, 53

\bibitem[{{Monaco} {et~al.}(2013){Monaco}, {Sefusatti}, {Borgani}, {Crocce},
  {Fosalba}, {Sheth}, \& {Theuns}}]{2013MNRAS.433.2389M}
{Monaco}, P., {Sefusatti}, E., {Borgani}, S., {et~al.} 2013, \mnras, 433, 2389

\bibitem[{{Monaco} {et~al.}(2002){Monaco}, {Theuns}, \&
  {Taffoni}}]{2002MNRAS.331..587M}
{Monaco}, P., {Theuns}, T., \& {Taffoni}, G. 2002, \mnras, 331, 587

\bibitem[{{Montero-Dorta} {et~al.}(2017){Montero-Dorta}, {P{\'e}rez}, {Prada},
  {Rodr{\'\i}guez-Torres}, {Favole}, {Klypin}, {Cid Fernandes}, {Gonz{\'a}lez
  Delgado}, {Dom{\'\i}nguez}, {Bolton}, {Garc{\'\i}a-Benito}, {Jullo}, \&
  {Niemiec}}]{2017ApJ...848L...2M}
{Montero-Dorta}, A.~D., {P{\'e}rez}, E., {Prada}, F., {et~al.} 2017, \apjl,
  848, L2

\bibitem[{{Moustakas et al.}(2022)}]{sga}
{Moustakas et al.} 2022, in preparation

\bibitem[{{Munari} {et~al.}(2017){Munari}, {Monaco}, {Koda}, {Kitaura},
  {Sefusatti}, \& {Borgani}}]{2017JCAP...07..050M}
{Munari}, E., {Monaco}, P., {Koda}, J., {et~al.} 2017, \jcap, 2017, 050

\bibitem[{{Musso} {et~al.}(2018){Musso}, {Cadiou}, {Pichon}, {Codis},
  {Kraljic}, \& {Dubois}}]{2018MNRAS.476.4877M}
{Musso}, M., {Cadiou}, C., {Pichon}, C., {et~al.} 2018, \mnras, 476, 4877

\bibitem[{{Myers} {et~al.}(2022){Myers}, {Moustakas}, {Bailey}, {Weaver},
  {Cooper}, {Forero-Romero}, {Abolfathi}, {Alexander}, {Brooks}, {Chaussidon},
  {Chuang}, {Dawson}, {Dey}, {Dey}, {Dhungana}, {Doel}, {Fanning},
  {Gazta{\~n}aga}, {Gontcho}, {Gonzalez-Morales}, {Hahn}, {Herrera-Alcantar},
  {Honscheid}, {Ishak}, {Karim}, {Kirkby}, {Kisner}, {Koposov}, {Kremin},
  {Lan}, {Landriau}, {Levi}, {Magneville}, {Martini}, {Meisner}, {Napolitano},
  {Newman}, {Palanque-Delabrouille}, {Percival}, {Poppett}, {Prada},
  {Raichoor}, {Ross}, {Schlafly}, {Schubnell}, {Tan}, {Tarle}, {Wilson},
  {Y{\`e}che}, {Zhou}, {Zhou}, \& {Zou}}]{2022arXiv220808518M}
{Myers}, A.~D., {Moustakas}, J., {Bailey}, S., {et~al.} 2022, arXiv e-prints,
  arXiv:2208.08518

\bibitem[{{Nasirudin} {et~al.}(2019){Nasirudin}, {Iliev}, \&
  {Ahn}}]{2019arXiv191012452N}
{Nasirudin}, A., {Iliev}, I.~T., \& {Ahn}, K. 2019, arXiv e-prints,
  arXiv:1910.12452

\bibitem[{{Navarro} {et~al.}(1996){Navarro}, {Frenk}, \&
  {White}}]{1996ApJ...462..563N}
{Navarro}, J.~F., {Frenk}, C.~S., \& {White}, S. D.~M. 1996, \apj, 462, 563

\bibitem[{{Neyrinck}(2013)}]{2013MNRAS.428..141N}
{Neyrinck}, M.~C. 2013, \mnras, 428, 141

\bibitem[{{O'Connell} \& {Eisenstein}(2019)}]{2019MNRAS.487.2701O}
{O'Connell}, R. \& {Eisenstein}, D.~J. 2019, \mnras, 487, 2701

\bibitem[{{Paranjape} {et~al.}(2018){Paranjape}, {Hahn}, \&
  {Sheth}}]{2018MNRAS.476.3631P}
{Paranjape}, A., {Hahn}, O., \& {Sheth}, R.~K. 2018, \mnras, 476, 3631

\bibitem[{{Paz} \& {S{\'a}nchez}(2015)}]{2015MNRAS.454.4326P}
{Paz}, D.~J. \& {S{\'a}nchez}, A.~G. 2015, \mnras, 454, 4326

\bibitem[{Peacock \& Heavens(1985)}]{Peacock1985}
Peacock, J.~A. \& Heavens, A.~F. 1985, Monthly Notices of the Royal
  Astronomical Society, 217, 805

\bibitem[{{Pearson} \& {Samushia}(2016)}]{2016MNRAS.457..993P}
{Pearson}, D.~W. \& {Samushia}, L. 2016, \mnras, 457, 993

\bibitem[{{Peebles}(1980)}]{1980lssu.book.....P}
{Peebles}, P.~J.~E. 1980, {The large-scale structure of the universe}

\bibitem[{{Percival} {et~al.}(2014){Percival}, {Ross}, {S{\'a}nchez},
  {Samushia}, {Burden}, {Crittenden}, {Cuesta}, {Magana}, {Manera}, {Beutler},
  {Chuang}, {Eisenstein}, {Ho}, {McBride}, {Montesano}, {Padmanabhan}, {Reid},
  {Saito}, {Schneider}, {Seo}, {Tojeiro}, \& {Weaver}}]{2014MNRAS.439.2531P}
{Percival}, W.~J., {Ross}, A.~J., {S{\'a}nchez}, A.~G., {et~al.} 2014, \mnras,
  439, 2531

\bibitem[{{Piras} {et~al.}(2022){Piras}, {Joachimi}, \&
  {Villaescusa-Navarro}}]{2022arXiv220507898P}
{Piras}, D., {Joachimi}, B., \& {Villaescusa-Navarro}, F. 2022, arXiv e-prints,
  arXiv:2205.07898

\bibitem[{{Pollack} {et~al.}(2012){Pollack}, {Smith}, \&
  {Porciani}}]{2012MNRAS.420.3469P}
{Pollack}, J.~E., {Smith}, R.~E., \& {Porciani}, C. 2012, \mnras, 420, 3469

\bibitem[{{Porciani} \& {Giavalisco}(2002)}]{2002ApJ...565...24P}
{Porciani}, C. \& {Giavalisco}, M. 2002, \apj, 565, 24

\bibitem[{{Potter} {et~al.}(2017){Potter}, {Stadel}, \&
  {Teyssier}}]{2017ComAC...4....2P}
{Potter}, D., {Stadel}, J., \& {Teyssier}, R. 2017, Computational Astrophysics
  and Cosmology, 4, 2

\bibitem[{{Press} {et~al.}(2002){Press}, {Teukolsky}, {Vetterling}, \&
  {Flannery}}]{2002nrca.book.....P}
{Press}, W.~H., {Teukolsky}, S.~A., {Vetterling}, W.~T., \& {Flannery}, B.~P.
  2002, {Numerical recipes in C++ : the art of scientific computing}

\bibitem[{{Pujol} {et~al.}(2017){Pujol}, {Hoffmann}, {Jim{\'e}nez}, \&
  {Gazta{\~n}aga}}]{2017A&A...598A.103P}
{Pujol}, A., {Hoffmann}, K., {Jim{\'e}nez}, N., \& {Gazta{\~n}aga}, E. 2017,
  \aap, 598, A103

\bibitem[{Raichoor {et~al.}(2020)Raichoor, Eisenstein, Karim, Newman,
  Moustakas, Brooks, Dawson, Dey, Duan, Eftekharzadeh, Gazta{\~{n}}aga, Kehoe,
  Landriau, Lang, Lee, Levi, Meisner, Myers, Palanque-Delabrouille, Poppett,
  Prada, Ross, Schlegel, Schubnell, Staten, Tarl{\'{e}}, Tojeiro, Y{\`{e}}che,
  \& Zhou}]{Raichoor_2020}
Raichoor, A., Eisenstein, D.~J., Karim, T., {et~al.} 2020, Research Notes of
  the {AAS}, 4, 180

\bibitem[{{Raichoor} {et~al.}(2022){Raichoor}, {Moustakas}, {Newman}, {Karim},
  {Ahlen}, {Alam}, {Bailey}, {Brooks}, {Dawson}, {de la Macorra}, {de Mattia},
  {Dey}, {Dey}, {Dhungana}, {Eftekharzadeh}, {Eisenstein}, {Fanning},
  {Font-Ribera}, {Garcia-Bellido}, {Gaztanaga}, {Gontcho}, {Guy}, {Honscheid},
  {Ishak}, {Kehoe}, {Kisner}, {Kremin}, {Lan}, {Landriau}, {Le Guillou},
  {Levi}, {Magneville}, {Martini}, {Meisner}, {Myers}, {Nie},
  {Palanque-Delabrouille}, {Percival}, {Poppett}, {Prada}, {Ross},
  {Ruhlmann-Kleider}, {Sabiu}, {Schlafly}, {Schlegel}, {Tarle}, {Weaver},
  {Yeche}, {Zhou}, {Zhou}, \& {Zou}}]{2022arXiv220808513R}
{Raichoor}, A., {Moustakas}, J., {Newman}, J.~A., {et~al.} 2022, arXiv
  e-prints, arXiv:2208.08513

\bibitem[{{Raichoor et al.}(2022)}]{fba}
{Raichoor et al.} 2022, in preparation

\bibitem[{{Rimes} \& {Hamilton}(2006)}]{2006MNRAS.371.1205R}
{Rimes}, C.~D. \& {Hamilton}, A. J.~S. 2006, \mnras, 371, 1205

\bibitem[{{Rodr{\'\i}guez-Torres} {et~al.}(2016){Rodr{\'\i}guez-Torres},
  {Chuang}, {Prada}, {Guo}, {Klypin}, {Behroozi}, {Hahn}, {Comparat}, {Yepes},
  {Montero-Dorta}, {Brownstein}, {Maraston}, {McBride}, {Tinker},
  {Gottl{\"o}ber}, {Favole}, {Shu}, {Kitaura}, {Bolton}, {Scoccimarro},
  {Samushia}, {Schlegel}, {Schneider}, \& {Thomas}}]{2016MNRAS.460.1173R}
{Rodr{\'\i}guez-Torres}, S.~A., {Chuang}, C.-H., {Prada}, F., {et~al.} 2016,
  \mnras, 460, 1173

\bibitem[{{Romano-D{\'\i}az} \& {van de Weygaert}(2007)}]{2007MNRAS.382....2R}
{Romano-D{\'\i}az}, E. \& {van de Weygaert}, R. 2007, \mnras, 382, 2

\bibitem[{{Ruiz-Macias} {et~al.}(2020){Ruiz-Macias}, {Zarrouk}, {Cole},
  {Norberg}, {Baugh}, {Brooks}, {Dey}, {Duan}, {Eftekharzadeh}, {Eisenstein},
  {Forero-Romero}, {Gazta{\~n}aga}, {Hahn}, {Kehoe}, {Landriau}, {Lang},
  {Levi}, {Lucey}, {Meisner}, {Moustakas}, {Myers}, {Palanque-Delabrouille},
  {Poppett}, {Prada}, {Raichoor}, {Schlegel}, {Schubnell}, {Tarl{\'e}},
  {Weinberg}, {Wilson}, \& {Y{\`e}che}}]{2020RNAAS...4..187R}
{Ruiz-Macias}, O., {Zarrouk}, P., {Cole}, S., {et~al.} 2020, Research Notes of
  the American Astronomical Society, 4, 187

\bibitem[{{Satpathy} {et~al.}(2019){Satpathy}, {A C Croft}, {Ho}, \&
  {Li}}]{2019MNRAS.484.2148S}
{Satpathy}, S., {A C Croft}, R., {Ho}, S., \& {Li}, B. 2019, \mnras, 484, 2148

\bibitem[{{Schlafly et al.}(2022)}]{ops}
{Schlafly et al.} 2022, in preparation

\bibitem[{{Schlegel et al.}(2022)}]{dr9}
{Schlegel et al.} 2022, in preparation

\bibitem[{{Scoccimarro} \& {Sheth}(2002)}]{2002MNRAS.329..629S}
{Scoccimarro}, R. \& {Sheth}, R.~K. 2002, \mnras, 329, 629

\bibitem[{{Sheth}(2005)}]{2005MNRAS.364..796S}
{Sheth}, R.~K. 2005, \mnras, 364, 796

\bibitem[{{Sheth} {et~al.}(2013){Sheth}, {Chan}, \&
  {Scoccimarro}}]{2013PhRvD..87h3002S}
{Sheth}, R.~K., {Chan}, K.~C., \& {Scoccimarro}, R. 2013, \prd, 87, 083002

\bibitem[{{Sheth} {et~al.}(2005){Sheth}, {Connolly}, \&
  {Skibba}}]{2005astro.ph.11773S}
{Sheth}, R.~K., {Connolly}, A.~J., \& {Skibba}, R. 2005, arXiv e-prints, astro

\bibitem[{Sheth \& Tormen(2004)}]{10.1111/j.1365-2966.2004.07733.x}
Sheth, R.~K. \& Tormen, G. 2004, Monthly Notices of the Royal Astronomical
  Society, 350, 1385

\bibitem[{{Sigad} {et~al.}(2000){Sigad}, {Branchini}, \&
  {Dekel}}]{2000ApJ...540...62S}
{Sigad}, Y., {Branchini}, E., \& {Dekel}, A. 2000, \apj, 540, 62

\bibitem[{{Silber} {et~al.}(2022){Silber}, {Fagrelius}, {Fanning}, {Schubnell},
  {Aguilar}, {Ahlen}, {Ameel}, {Ballester}, {Baltay}, {Bebek}, {Beard},
  {Besuner}, {Cardiel-Sas}, {Casas}, {Castander}, {Claybaugh}, {Dobson},
  {Duan}, {Dunlop}, {Edelstein}, {Emmet}, {Elliott}, {Evatt}, {Gershkovich},
  {Guy}, {Harris}, {Heetderks}, {Heetderks}, {Honscheid}, {Illa}, {Jelinsky},
  {Jelinsky}, {Jimenez}, {Karcher}, {Kent}, {Kirkby}, {Kneib}, {Lambert},
  {Lampton}, {Leitner}, {Levi}, {McCauley}, {Meisner}, {Miller}, {Miquel},
  {Mundet}, {Poppett}, {Rabinowitz}, {Reil}, {Roman}, {Schlegel}, {Serrano},
  {Van Shourt}, {Sprayberry}, {Tarl{\'e}}, {Sien Tie}, {Weaverdyck}, {Zhang},
  {Azzaro}, {Bailey}, {Becerril}, {Blackwell}, {Bouri}, {Brooks},
  {Buckley-Geer}, {Pe{\~n}ate Castro}, {Derwent}, {Dey}, {Dhungana}, {Doel},
  {Eisenstein}, {Fahim}, {Garcia-Bellido}, {Gazta{\~n}aga}, {Gontcho},
  {Gutierrez}, {H{\"o}rler}, {Kehoe}, {Kisner}, {Kremin}, {Kronig}, {Landriau},
  {Le Guillou}, {Martini}, {Moustakas}, {Palanque-Delabrouille}, {Peng},
  {Percival}, {Prada}, {Allende Prieto}, {Gonzalez de Rivera}, {Sanchez},
  {Sanchez}, {Sharples}, {Soares-Santos}, {Schlafly}, {Weaver}, {Zhou}, {Zhu},
  \& {Zou}}]{2022arXiv220509014S}
{Silber}, J.~H., {Fagrelius}, P., {Fanning}, K., {et~al.} 2022, arXiv e-prints,
  arXiv:2205.09014

\bibitem[{{Simon}(2005)}]{2005A&A...430..827S}
{Simon}, P. 2005, \aap, 430, 827

\bibitem[{{Sinigaglia} {et~al.}(2021){Sinigaglia}, {Kitaura},
  {Balaguera-Antol{\'\i}nez}, {Nagamine}, {Ata}, {Shimizu}, \&
  {S{\'a}nchez-Benavente}}]{2021ApJ...921...66S}
{Sinigaglia}, F., {Kitaura}, F.-S., {Balaguera-Antol{\'\i}nez}, A., {et~al.}
  2021, \apj, 921, 66

\bibitem[{{Sinigaglia} {et~al.}(2022){Sinigaglia}, {Kitaura},
  {Balaguera-Antol{\'\i}nez}, {Shimizu}, {Nagamine}, {S{\'a}nchez-Benavente},
  \& {Ata}}]{2022ApJ...927..230S}
{Sinigaglia}, F., {Kitaura}, F.-S., {Balaguera-Antol{\'\i}nez}, A., {et~al.}
  2022, \apj, 927, 230

\bibitem[{{Skibba} {et~al.}(2006){Skibba}, {Sheth}, {Connolly}, \&
  {Scranton}}]{2006MNRAS.369...68S}
{Skibba}, R., {Sheth}, R.~K., {Connolly}, A.~J., \& {Scranton}, R. 2006,
  \mnras, 369, 68

\bibitem[{{Skibba} \& {Macci{\`o}}(2011)}]{2011MNRAS.416.2388S}
{Skibba}, R.~A. \& {Macci{\`o}}, A.~V. 2011, \mnras, 416, 2388

\bibitem[{{Smith} {et~al.}(2007){Smith}, {Scoccimarro}, \&
  {Sheth}}]{2007PhRvD..75f3512S}
{Smith}, R.~E., {Scoccimarro}, R., \& {Sheth}, R.~K. 2007, \prd, 75, 063512

\bibitem[{{Somerville} {et~al.}(2001){Somerville}, {Lemson}, {Sigad}, {Dekel},
  {Kauffmann}, \& {White}}]{2001MNRAS.320..289S}
{Somerville}, R.~S., {Lemson}, G., {Sigad}, Y., {et~al.} 2001, \mnras, 320, 289

\bibitem[{{Sousbie} {et~al.}(2008){Sousbie}, {Courtois}, {Bryan}, \&
  {Devriendt}}]{2008ApJ...678..569S}
{Sousbie}, T., {Courtois}, H., {Bryan}, G., \& {Devriendt}, J. 2008, \apj, 678,
  569

\bibitem[{{Spergel} {et~al.}(2015){Spergel}, {Gehrels}, {Baltay}, {Bennett},
  {Breckinridge}, {Donahue}, {Dressler}, {Gaudi}, {Greene}, {Guyon}, {Hirata},
  {Kalirai}, {Kasdin}, {Macintosh}, {Moos}, {Perlmutter}, {Postman},
  {Rauscher}, {Rhodes}, {Wang}, {Weinberg}, {Benford}, {Hudson}, {Jeong},
  {Mellier}, {Traub}, {Yamada}, {Capak}, {Colbert}, {Masters}, {Penny},
  {Savransky}, {Stern}, {Zimmerman}, {Barry}, {Bartusek}, {Carpenter}, {Cheng},
  {Content}, {Dekens}, {Demers}, {Grady}, {Jackson}, {Kuan}, {Kruk}, {Melton},
  {Nemati}, {Parvin}, {Poberezhskiy}, {Peddie}, {Ruffa}, {Wallace}, {Whipple},
  {Wollack}, \& {Zhao}}]{2015arXiv150303757S}
{Spergel}, D., {Gehrels}, N., {Baltay}, C., {et~al.} 2015, arXiv e-prints,
  arXiv:1503.03757

\bibitem[{{Springel}(2005)}]{2005MNRAS.364.1105S}
{Springel}, V. 2005, \mnras, 364, 1105

\bibitem[{{Springel} {et~al.}(2021){Springel}, {Pakmor}, {Zier}, \&
  {Reinecke}}]{2021MNRAS.506.2871S}
{Springel}, V., {Pakmor}, R., {Zier}, O., \& {Reinecke}, M. 2021, \mnras, 506,
  2871

\bibitem[{{Stein} {et~al.}(2019){Stein}, {Alvarez}, \&
  {Bond}}]{2019MNRAS.483.2236S}
{Stein}, G., {Alvarez}, M.~A., \& {Bond}, J.~R. 2019, \mnras, 483, 2236

\bibitem[{{Takada} \& {Hu}(2013)}]{2013PhRvD..87l3504T}
{Takada}, M. \& {Hu}, W. 2013, \prd, 87, 123504

\bibitem[{{Tassev} {et~al.}(2013){Tassev}, {Zaldarriaga}, \&
  {Eisenstein}}]{2013JCAP...06..036T}
{Tassev}, S., {Zaldarriaga}, M., \& {Eisenstein}, D.~J. 2013, \jcap, 6, 036

\bibitem[{{Taylor} {et~al.}(2013){Taylor}, {Joachimi}, \&
  {Kitching}}]{2013MNRAS.432.1928T}
{Taylor}, A., {Joachimi}, B., \& {Kitching}, T. 2013, \mnras, 432, 1928

\bibitem[{{Tegmark} \& {Peebles}(1998)}]{1998ApJ...500L..79T}
{Tegmark}, M. \& {Peebles}, P.~J.~E. 1998, \apjl, 500, L79

\bibitem[{{Tinker} {et~al.}(2010){Tinker}, {Robertson}, {Kravtsov}, {Klypin},
  {Warren}, {Yepes}, \& {Gottl{\"o}ber}}]{2010ApJ...724..878T}
{Tinker}, J.~L., {Robertson}, B.~E., {Kravtsov}, A.~V., {et~al.} 2010, APJ,
  724, 878

\bibitem[{{Valageas}(2011)}]{2011A&A...525A..98V}
{Valageas}, P. 2011, \aap, 525, A98

\bibitem[{{Vale} \& {Ostriker}(2004)}]{2004MNRAS.353..189V}
{Vale}, A. \& {Ostriker}, J.~P. 2004, \mnras, 353, 189

\bibitem[{{van de Weygaert} {et~al.}(2009){van de Weygaert}, {Aragon-Calvo},
  {Jones}, \& {Platen}}]{2009arXiv0912.3448V}
{van de Weygaert}, R., {Aragon-Calvo}, M.~A., {Jones}, B.~J.~T., \& {Platen},
  E. 2009, ArXiv e-prints

\bibitem[{{Vargas-Magana} {et~al.}(2019){Vargas-Magana}, {Brooks}, {Levi}, \&
  {Tarle}}]{2019arXiv190101581V}
{Vargas-Magana}, M., {Brooks}, D.~D., {Levi}, M.~M., \& {Tarle}, G.~G. 2019,
  arXiv e-prints, arXiv:1901.01581

\bibitem[{{Variu et al.}(2023)}]{Variu2023}
{Variu et al.} 2023, in preparation

\bibitem[{{Villaescusa-Navarro} {et~al.}(2021){Villaescusa-Navarro},
  {Angl{\'e}s-Alc{\'a}zar}, {Genel}, {Spergel}, {Somerville}, {Dave},
  {Pillepich}, {Hernquist}, {Nelson}, {Torrey}, {Narayanan}, {Li}, {Philcox},
  {La Torre}, {Maria Delgado}, {Ho}, {Hassan}, {Burkhart}, {Wadekar},
  {Battaglia}, {Contardo}, \& {Bryan}}]{2021ApJ...915...71V}
{Villaescusa-Navarro}, F., {Angl{\'e}s-Alc{\'a}zar}, D., {Genel}, S., {et~al.}
  2021, \apj, 915, 71

\bibitem[{{Wang} \& {Zhao}(2020)}]{2020RAA....20..158W}
{Wang}, Y. \& {Zhao}, G.-B. 2020, Research in Astronomy and Astrophysics, 20,
  158

\bibitem[{{Wechsler} \& {Tinker}(2018)}]{2018ARA&A..56..435W}
{Wechsler}, R.~H. \& {Tinker}, J.~L. 2018, \araa, 56, 435

\bibitem[{{Wechsler} {et~al.}(2006){Wechsler}, {Zentner}, {Bullock},
  {Kravtsov}, \& {Allgood}}]{2006ApJ...652...71W}
{Wechsler}, R.~H., {Zentner}, A.~R., {Bullock}, J.~S., {Kravtsov}, A.~V., \&
  {Allgood}, B. 2006, \apj, 652, 71

\bibitem[{Weinberg(1992)}]{doi:10.1093/mnras/254.2.315}
Weinberg, D.~H. 1992, \mnras, 254, 315

\bibitem[{{Weinberger} {et~al.}(2020){Weinberger}, {Springel}, \&
  {Pakmor}}]{2020ApJS..248...32W}
{Weinberger}, R., {Springel}, V., \& {Pakmor}, R. 2020, \apjs, 248, 32

\bibitem[{{White}(2014)}]{2014MNRAS.439.3630W}
{White}, M. 2014, \mnras, 439, 3630

\bibitem[{{White} {et~al.}(2014){White}, {Tinker}, \&
  {McBride}}]{2014MNRAS.437.2594W}
{White}, M., {Tinker}, J.~L., \& {McBride}, C.~K. 2014, \mnras, 437, 2594

\bibitem[{{Xavier} {et~al.}(2016){Xavier}, {Abdalla}, \&
  {Joachimi}}]{2016MNRAS.459.3693X}
{Xavier}, H.~S., {Abdalla}, F.~B., \& {Joachimi}, B. 2016, \mnras, 459, 3693

\bibitem[{{Xu} {et~al.}(2021){Xu}, {Zehavi}, \&
  {Contreras}}]{2021MNRAS.502.3242X}
{Xu}, X., {Zehavi}, I., \& {Contreras}, S. 2021, \mnras, 502, 3242

\bibitem[{{Yang} {et~al.}(2017){Yang}, {Zhang}, {Lu}, {Wang}, {Shi}, {Tweed},
  {Li}, {Luo}, {Lu}, \& {Yang}}]{2017ApJ...848...60Y}
{Yang}, X., {Zhang}, Y., {Lu}, T., {et~al.} 2017, \apj, 848, 60

\bibitem[{{Y{\`e}che} {et~al.}(2020){Y{\`e}che}, {Palanque-Delabrouille},
  {Claveau}, {Brooks}, {Chaussidon}, {Davis}, {Dawson}, {Dey}, {Duan},
  {Eftekharzadeh}, {Eisenstein}, {Gazta{\~n}aga}, {Kehoe}, {Landriau}, {Lang},
  {Levi}, {Meisner}, {Myers}, {Newman}, {Poppett}, {Prada}, {Raichoor},
  {Schlegel}, {Schubnell}, {Staten}, {Tarl{\'e}}, \&
  {Zhou}}]{2020RNAAS...4..179Y}
{Y{\`e}che}, C., {Palanque-Delabrouille}, N., {Claveau}, C.-A., {et~al.} 2020,
  Research Notes of the American Astronomical Society, 4, 179

\bibitem[{{Yu} {et~al.}(2015){Yu}, {Zhang}, {Jing}, \&
  {Zhang}}]{2015PhRvD..92h3527Y}
{Yu}, Y., {Zhang}, J., {Jing}, Y., \& {Zhang}, P. 2015, \prd, 92, 083527

\bibitem[{{Zehavi} {et~al.}(2019){Zehavi}, {Kerby}, {Contreras}, {Jim{\'e}nez},
  {Padilla}, \& {Baugh}}]{2019ApJ...887...17Z}
{Zehavi}, I., {Kerby}, S.~E., {Contreras}, S., {et~al.} 2019, \apj, 887, 17

\bibitem[{{Zel'dovich}(1970)}]{1970A&A.....5...84Z}
{Zel'dovich}, Y.~B. 1970, \aap, 5, 84

\bibitem[{{Zentner}(2007)}]{2007IJMPD..16..763Z}
{Zentner}, A.~R. 2007, International Journal of Modern Physics D, 16, 763

\bibitem[{{Zhang} {et~al.}(2015){Zhang}, {Zheng}, \&
  {Jing}}]{2015PhRvD..91d3522Z}
{Zhang}, P., {Zheng}, Y., \& {Jing}, Y. 2015, \prd, 91, 043522

\bibitem[{{Zhang} {et~al.}(2019){Zhang}, {Wang}, {Zhang}, {Sun}, {He},
  {Contardo}, {Villaescusa-Navarro}, \& {Ho}}]{2019arXiv190205965Z}
{Zhang}, X., {Wang}, Y., {Zhang}, W., {et~al.} 2019, arXiv e-prints,
  arXiv:1902.05965

\bibitem[{{Zhao} {et~al.}(2021){Zhao}, {Chuang}, {Bautista}, {de Mattia},
  {Raichoor}, {Ross}, {Hou}, {Neveux}, {Tao}, {Burtin}, {Dawson}, {de la
  Torre}, {Gil-Mar{\'\i}n}, {Kneib}, {Percival}, {Rossi}, {Tamone}, {Tinker},
  {Zhao}, {Alam}, \& {Mueller}}]{2021MNRAS.503.1149Z}
{Zhao}, C., {Chuang}, C.-H., {Bautista}, J., {et~al.} 2021, \mnras, 503, 1149

\bibitem[{{Zhao} {et~al.}(2015){Zhao}, {Kitaura}, {Chuang}, {Prada}, {Yepes},
  \& {Tao}}]{2015MNRAS.451.4266Z}
{Zhao}, C., {Kitaura}, F.-S., {Chuang}, C.-H., {et~al.} 2015, \mnras, 451, 4266

\bibitem[{{Zheng} {et~al.}(2019){Zheng}, {Song}, \& {Oh}}]{2019JCAP...06..013Z}
{Zheng}, Y., {Song}, Y.-S., \& {Oh}, M. 2019, \jcap, 2019, 013

\bibitem[{{Zheng} {et~al.}(2013){Zheng}, {Zhang}, {Jing}, {Lin}, \&
  {Pan}}]{2013PhRvD..88j3510Z}
{Zheng}, Y., {Zhang}, P., {Jing}, Y., {Lin}, W., \& {Pan}, J. 2013, \prd, 88,
  103510

\bibitem[{{Zhou} {et~al.}(2022){Zhou}, {Dey}, {Newman}, {Eisenstein}, {Dawson},
  {Bailey}, {Berti}, {Guy}, {Lan}, {Zou}, {Aguilar}, {Ahlen}, {Alam}, {Brooks},
  {de la Macorra}, {Dey}, {Dhungana}, {Fanning}, {Font-Ribera}, {Gontcho},
  {Honscheid}, {Ishak}, {Kisner}, {Kov{\'a}cs}, {Kremin}, {Landriau}, {Levi},
  {Magneville}, {Martini}, {Meisner}, {Miquel}, {Moustakas}, {Myers}, {Nie},
  {Palanque-Delabrouille}, {Percival}, {Poppett}, {Prada}, {Raichoor}, {Ross},
  {Schlafly}, {Schlegel}, {Schubnell}, {Tarl{\'e}}, {Weaver}, {Wechsler},
  {Y{\`e}che}, \& {Zhou}}]{2022arXiv220808515Z}
{Zhou}, R., {Dey}, B., {Newman}, J.~A., {et~al.} 2022, arXiv e-prints,
  arXiv:2208.08515

\bibitem[{{Zhou} {et~al.}(2020){Zhou}, {Newman}, {Dawson}, {Eisenstein},
  {Brooks}, {Dey}, {Dey}, {Duan}, {Eftekharzadeh}, {Gazta{\~n}aga}, {Kehoe},
  {Landriau}, {Levi}, {Licquia}, {Meisner}, {Moustakas}, {Myers},
  {Palanque-Delabrouille}, {Poppett}, {Prada}, {Raichoor}, {Schlegel},
  {Schubnell}, {Staten}, {Tarl{\'e}}, \& {Y{\`e}che}}]{2020RNAAS...4..181Z}
{Zhou}, R., {Newman}, J.~A., {Dawson}, K.~S., {et~al.} 2020, Research Notes of
  the American Astronomical Society, 4, 181

\bibitem[{{Zou} {et~al.}(2017){Zou}, {Zhou}, {Fan}, {Zhang}, {Zhou}, {Nie},
  {Peng}, {McGreer}, {Jiang}, {Dey}, {Fan}, {He}, {Jiang}, {Lang}, {Lesser},
  {Ma}, {Mao}, {Schlegel}, \& {Wang}}]{2017PASP..129f4101Z}
{Zou}, H., {Zhou}, X., {Fan}, X., {et~al.} 2017, \pasp, 129, 064101

\end{thebibliography}

%\newpage
\appendix

\section{Hierarchy of properties}\label{ap:hier}
\begin{figure}
\includegraphics[trim = 0cm 0cm 0cm 0cm ,clip=true, width=0.48\textwidth]{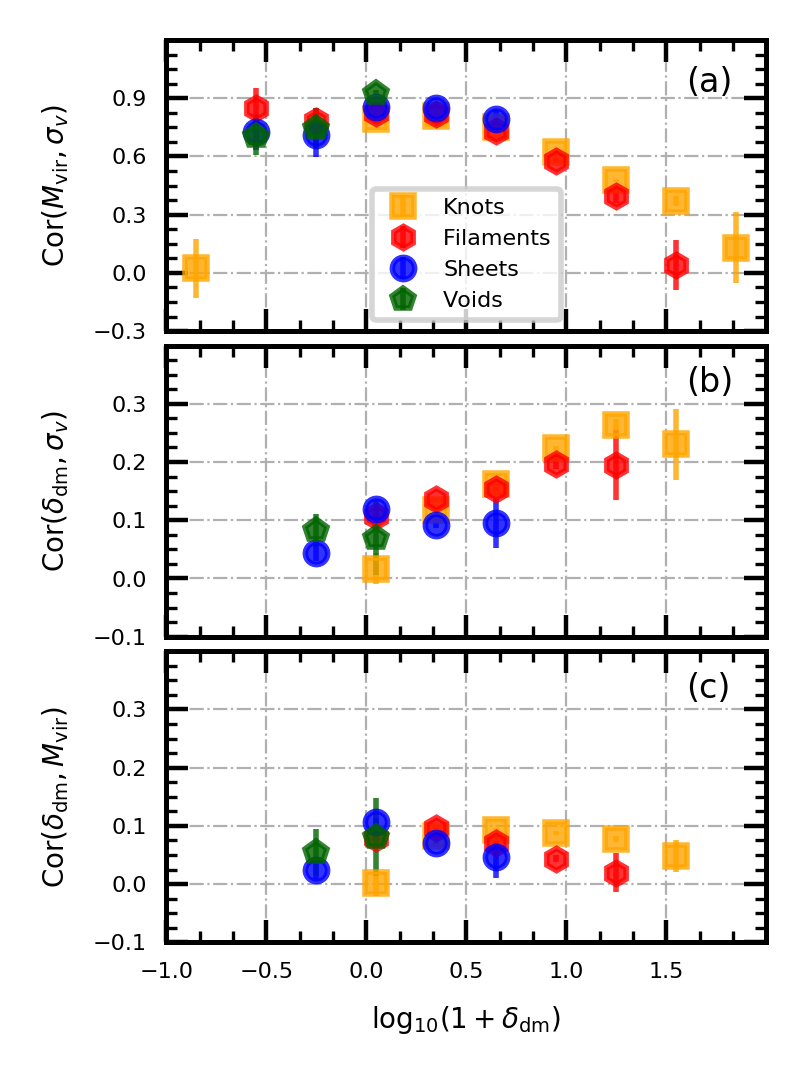}
    \caption{Pearson correlation coefficients between halo properties (panel a), the local dark matter density with halo virial mass (panel b) and velocity dispersion (panel c), measured in bins of local density for different cosmic-web types, as obtained from the \texttt{SLICS} reference simulation. The error bars denote the standard deviation obtained from the set of catalogues.}
\label{fig:hcorr}
\end{figure}

To motivate the hierarchical property assignment used in \S\ref{sec:hprops}, we compute the Pearson correlation coefficient using the local dark matter density, virial mass, and velocity dispersion obtained in bins of local density and different cosmic-web type. The results shown in Fig.~\ref{fig:hcorr} indicate that (i) the correlation between halo properties in different cosmic-web types (panel (a)) follows a similar trend in low and intermediate densities, reaching a maximum at $\delta_{\rm dm}\sim 0$, and then decreasing towards high densities, where the correlation in knots becomes significantly dominant as compared to that in filaments; (ii) the correlation between velocity dispersion and dark matter displays (panel (b)) higher values, that is, of $\sim 30\%$, especially at high densities, and (iii) the correlations between virial mass and local density (panel (c)) are rather weak, with $\leq 15\%$ on average over the density range, and no strong dependence on cosmic-web type. We note that the underlying dark matter density field used to compute these correlations is an approximated version of the original field (which has $512$ times more particles) whose dark matter particles are used to define these halos. Hence, these measurements must not be taken as general claims as to the behaviour of halo properties, but instead a characterisation of the current setup within the \texttt{BAM} machinery. With this in mind, we adopt underlying dark matter density field as the main property with which to start the assignment procedure. 
% ======================================================

\section{Multi-scaling assignment}\label{sec:multi}
One key aspect to take into account when sampling the halo number counts field with discrete tracers is, on one side, the need to preserve the large-scale clustering pattern according to that of the reference, and on the other hand, to avoid halo exclusion \cite[see e.g.][]{2002ApJ...565...24P,2013PhRvD..88h3507B,2019MNRAS.489.4170G}. The former is ensured by an appropriate choice of the gravity solver, while the latter can be addressed at the
moment of assigning properties. To that aim, we envisaged a {multi-scale assignment scheme}, in which a given property (in this case, the velocity dispersion) is assigned in an ordered fashion, ensuring that the highest values are linked to tracers that are sufficiently far away to avoid exclusion \cite[see also][]{2015MNRAS.451.4266Z}. Let us briefly describe the steps designed within the \texttt{BAM} algorithm to assign halo properties:
\begin{enumerate}
    \item We randomly select one reference halo catalogue and define a threshold $\sigma_{\rm th}$, above which the multi-scaling assignment will be performed. 
    
    \item For values of velocity dispersion above the threshold $\sigma_{\rm th}$, we divide the $\sigma_{\rm v}$ range probed by the halo population into $\mathcal{N}$ intervals. Each of these intervals is defined by a minimum and maximum value of velocity dispersion, $\sigma^{\ell}_{\rm v}\in [\sigma^{\rm min}_{\rm v},\sigma^{\rm max}_{\rm v})_{\ell}$ ($\ell=1, \cdots \mathcal{N}$).

    \item Let $N_{\ell}$ denote the number of tracers read from one reference simulation in each of these intervals. We construct a number of $\mathcal{N}$ spatial meshes covering the cosmological volume of the \texttt{BAM} mocks, each of which is characterised by $N_{\ell}$ cells, such that, on average, each cell is populated by one halo within the corresponding interval of $\sigma_{v}$. 

    \item We then select a number $\tilde{N}_{ref}$ of reference halo catalogues. This number can be smaller than the figure used for the generation of halo number counts (i.e. $N_{ref}=27$). For the current setup, we randomly selected $\tilde{N}_{ref}=5$ from the available set of \texttt{SLICs} catalogues.
    \item From the set of $\tilde{N}_{ref}$ catalogues, we identify the values of velocity dispersion $\{\sigma_{\rm v}\}_{\ell}(\{\Theta\}_{\rm ref})$ as a function of the dark matter properties $\{\Theta\}_{\rm ref}$, and sort these values top-down in each bin of $\{\Theta\}_{\rm ref}$ for all values above the minimum threshold $\sigma^{\rm th}_{\rm v}$.
    \item We then proceed to the assignment of properties read from the reference to tracers in a \texttt{BAM} halo catalogue. The assignment starts with the interval containing the highest values of $\sigma_{\rm v}$ (this mesh is the least populated). 
    \item We then loop over the $N_{\ell}$ cells in the mesh built over the \texttt{BAM} mock. In each step of this loop, the algorithm randomly selects one fiducial cell (i.e. defined by the nominal resolution $192^{3}$). To avoid the assignment of properties to tracers in adjacent cells, we previously randomise the order of the $N_{\ell}$ cells. 
    \item Using the dark matter field of the current mock catalogue, we identify the corresponding set of properties $\{\Theta\}_{\rm mock}$ of the chosen fiducial cell.
    \item We randomly choose one mock tracer in that fiducial cell (if any) and assign to it a value of $\sigma_{\rm v}$ from the list $\{\sigma_{\rm v}\}_{\ell}(\{\Theta\}_{\rm mock}=\{\Theta\}_{\rm ref})$. Given that this list is sorted, the first assignment (i.e. the first step in the loop over the $N_{\ell}$-cells) will correspond to the largest value of $\sigma_{\rm v}$ in the reference catalogue.
    \item We then repeat the full loop until the number of requested tracers in the level are assigned with properties.
\end{enumerate}
This procedure is repeated for the different levels $\mathcal{M}_{\ell}$. For values below the minimum threshold, the assignment is performed in the following way:
\begin{enumerate}
    \item We randomly select one tracer (without assigned property) from the mock, and then identify the fiducial cell where it resides and the corresponding set of properties $\{\Theta\}_{\rm mock}$.
    \item Returning to the list $\{\sigma_{\rm v}\}(\{\Theta\}_{\rm ref})$, we randomly assign any of the corresponding values still available (i.e. with $\sigma_{v}<\sigma^{\rm th}_{\rm v}$).
\end{enumerate}
Finally, if there are still tracers with no assigned properties (e.g. due to cosmic variance), \texttt{BAM} statistically assigns values of velocity dispersion by sampling the global halo abundance measured from one reference catalogue. This last case represents $\sim 5\%$ of the total assignment, depending on the realisation. For the current case, we implemented $\mathcal{N}=3$ levels with thresholds at $\sigma^{\rm th}_{\rm v}=4,\, 8$ and $10$ km/s.

Once velocity dispersion is assigned, we measure the scaling relation $\mathcal{P}(M_{\rm vir}|\sigma_{\rm v},\{\Theta\})$ (using one realisation of the reference set) to assign virial masses with 
\begin{equation}
M^{ij}_{\rm vir}\curvearrowleft \mathcal{P}(M_{\rm vir}|\sigma^{i}_{\rm v}=\sigma^{\rm ref}_{\rm v},\{\Theta\}_{\rm mock\,j}=\{\Theta_{\rm ref}\}).
\end{equation}
For this step, we implemented, along with the information of the velocity dispersion, the local dark matter density and the cosmic-web classification. The scheme can be generalised to any other set of halo properties tabulated in the reference catalogue (e.g. spin, concentration, maximum circular velocity).

\section{Calibration using galaxy catalogues}\label{ap:gala}

\begin{figure*}[t]
\includegraphics[trim = 0cm 0cm 0cm 0cm, clip=true, width=1.03\textwidth]{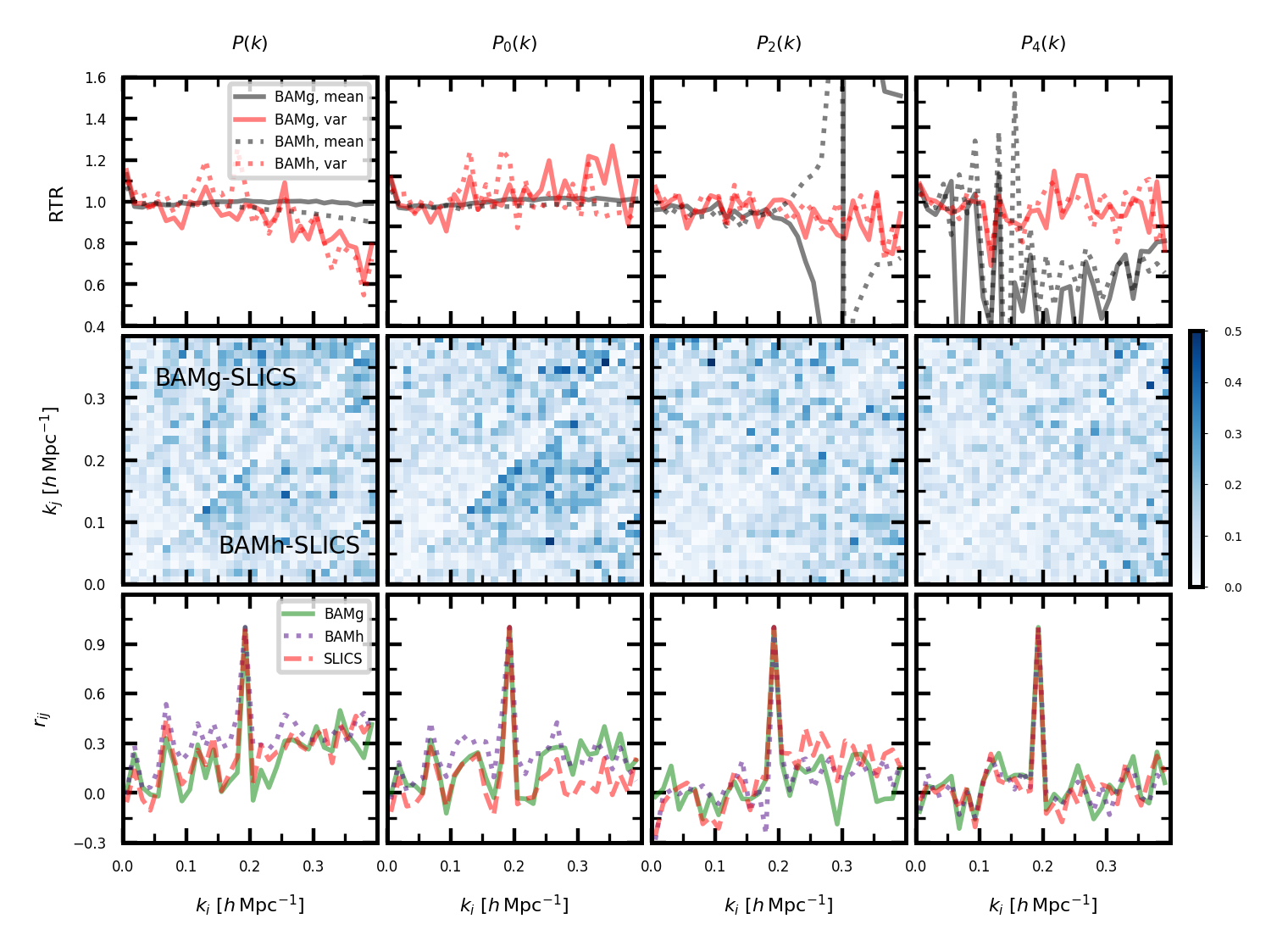}
\caption{Comparison of the summary statistics of the set of \texttt{BAM} galaxies generated from the halo catalogues (\texttt{BAMh}) with those of the \texttt{BAM} galaxies generated from calibration (\texttt{BAMg}) obtained from the \texttt{SLICS} galaxy catalogue. The top row shows the ratios to the references (RTF) from the mean and variance of the power spectrum (real and redshift space). The middle row shows the absolute value of the difference between the correlation coefficients $r_{ij}$ of the \texttt{BAM} sets (\texttt{BAMg}, upper diagonal, \texttt{BAMh}, lower diagonal) and those from the \texttt{SLICS}. The third row shows the elements of the correlation coefficients (in only one Fourier bin, to avoid clutter) from the three sets. }
\label{fig:galc}
\end{figure*}

In this Appendix, we assess the precision of mock galaxy catalogues built using the set of ELG catalogues described in \S\ref{sec:gals}   as a reference (i.e. \texttt{SLICS} halos plus HOD). Following the procedures described in \S\ref{sec:iter} and \S~\ref{sec:counts} and using the \texttt{TkWEB} model to characterise the galaxy bias in \texttt{BAM}, we generated $770$ realisations of galaxy number counts. Coordinates of DM particles are used to define the positions of the galaxy-type tracers, while random tracers are similarly introduced, collapsing them towards their closest DM particles. Numerical tests indicate that for the current galaxy population, a fraction $f_{\rm col}=0.05$ has to be used to obtain a per cent accuracy in the real space mean power spectrum on small scales. We note that this collapsing fraction is smaller than that used for halos ($\sim 0.35$), and is expected as galaxies populate smaller scales than their parent halos, thus generating the need for a stronger collapse of the random set. The assignment of velocities to this set of tracers is not evident, as no clear identification of centrals or satellites is available, In order to match the large-scale clustering signal, bulk velocities can be assigned as shown in \S\ref{sec:vels} with a constant velocity bias of $\sim 20\%$. On the other hand, small-scale clustering can be replicated by adding a random velocity component to the components of the bulk velocities from a normal distribution, with  a width that can vary among the different cosmic-web types (see  \S\ref{sec:vels}). Such a random component can be added to each component of the velocity (as would be the case of galaxies inside dark matter halos) or to the magnitude of the velocity, a situation that can be linked to parent halos in the mildly non-linear regime \citep[][]{2016MNRAS.455L..77H,2019JCAP...06..013Z}. Either of these options can be used to replicate the small-scale redshift space clustering signal up the $k\sim 0.4 h$Mpc$^{-1}$. For example, assigning a random component to each velocity component in knots demands a velocity dispersion of $\sigma\sim 350$ km$/$s, while adding the noise and keeping  the direction of the velocity of each tracer fixed demands $\sigma\sim 740$ km$/$s. Here, we  do not aim to precisely determine the best scenario for the assignment of random components to galaxy-like tracers. For the current test, we used $\sigma= 350$ km$/$s applied only to galaxies in knots. Figure \ref{fig:galc} shows the comparison of different summary statistics in Fourier space for the set of \texttt{BAM} galaxies generated from halo catalogues (\texttt{BAMh}) and the \texttt{BAM} set generated directly from the calibration to the \texttt{SLICS} galaxy catalogues (\texttt{BAMg}) described in \S\ref{sec:gals}. We can summarise the results of this comparison as follows:
\begin{itemize}
    \item In real space (first column of Fig.~\ref{fig:galc}), the \texttt{BAMg} set yields a more accurate description of the mean power spectrum, especially towards small scales  ($k\sim 0.4h$Mpc$^{-1}$). The variance of the power spectrum is also improved. With respect to the reference correlation matrix, this quantity displays less mode coupling for the \texttt{BAMg} than the  \texttt{BAMh} set. In general, the improvement in real space of the set  \texttt{BAMg} is due to the lack of inaccuracies in the galaxy distribution associated with the assignment of halo properties, as in the  \texttt{BAMh} case.
    \item The behaviour in redshift space from the \texttt{BAMg} is consistent with that from \texttt{BAMh}. Noticeable differences are (i) the mean of the quadrupole on scales $k\geq 0.2h$Mpc$^{-1}$ and (ii) the correlation matrix, where the set \texttt{BAMg} displays less extra coupling than seen in \texttt{BAMh}.
\end{itemize}
These results are remarkable for two reasons: On one hand, the fact that the \texttt{BAMh} set is consistent with the \texttt{BAMg} highlights the ability of the method to generate galaxy catalogues with information on the hosting halos without a significant decrement in the precision of two-point statistics. On the other hand, the behaviour of the \texttt{BAMg} set shows how the method can be further adapted to generate galaxy catalogues by directly learning from a reference galaxy catalogue containing galaxy velocities based on theoretical models.

% ======================================================
% ======================================================

\section{Forthcoming applications: Experiments with paired-fixed amplitude cosmological simulations}\label{sec:unit}

\begin{figure}
\includegraphics[trim = 0cm 0cm 0cm 0cm ,clip=true, width=0.48\textwidth]{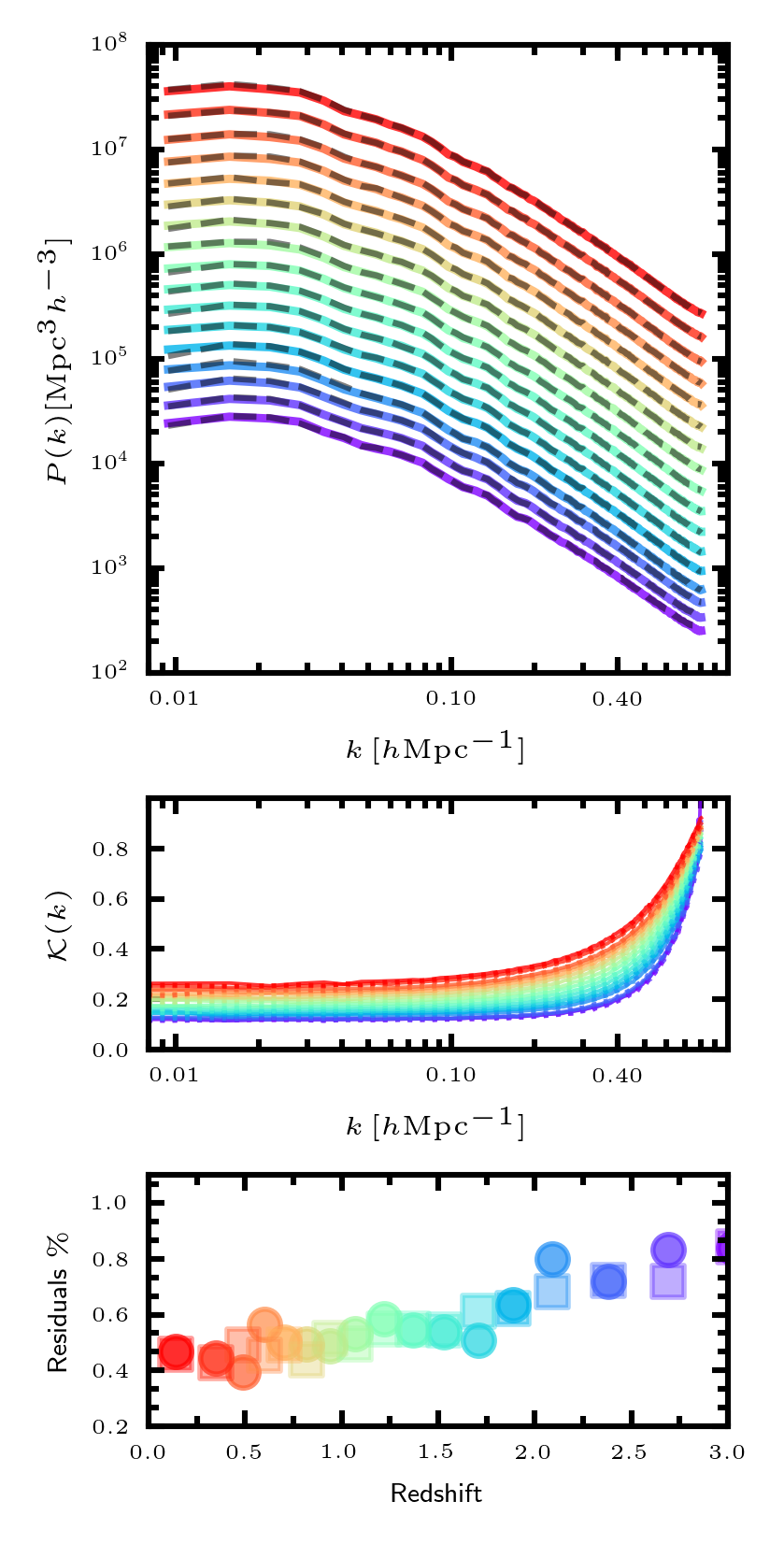}
\caption{Summary statistics of mock halo catalogues generated with \texttt{BAM} using the \texttt{UNIT}sim as reference. Top panels: Power spectrum $P(k)$ obtained from learning phase (\S\ref{sec:iter}) (solid lines) compared to the same statistics from the reference (dashed lines). Colours indicate different snapshots of the \texttt{UNIT} simulation, ranging from $z=0$ (red, upper curves) up to $z=3$ (violet, bottom curves). The outputs and references at each redshift have been shifted by the same amount to facilitate the comparison. The middle panel shows the kernel $\mathcal{K}$ and the bottom panels show the relative residuals in the power spectrum at the different redshifts, obtained from the normal (filled circles) and the phase-inverted (filled squares) reference catalogues.}
    \label{fig:calibration_unit}
\end{figure}

\begin{figure*}
\includegraphics[trim = 0cm 0cm 0cm 0cm ,clip=true, width=1\textwidth]{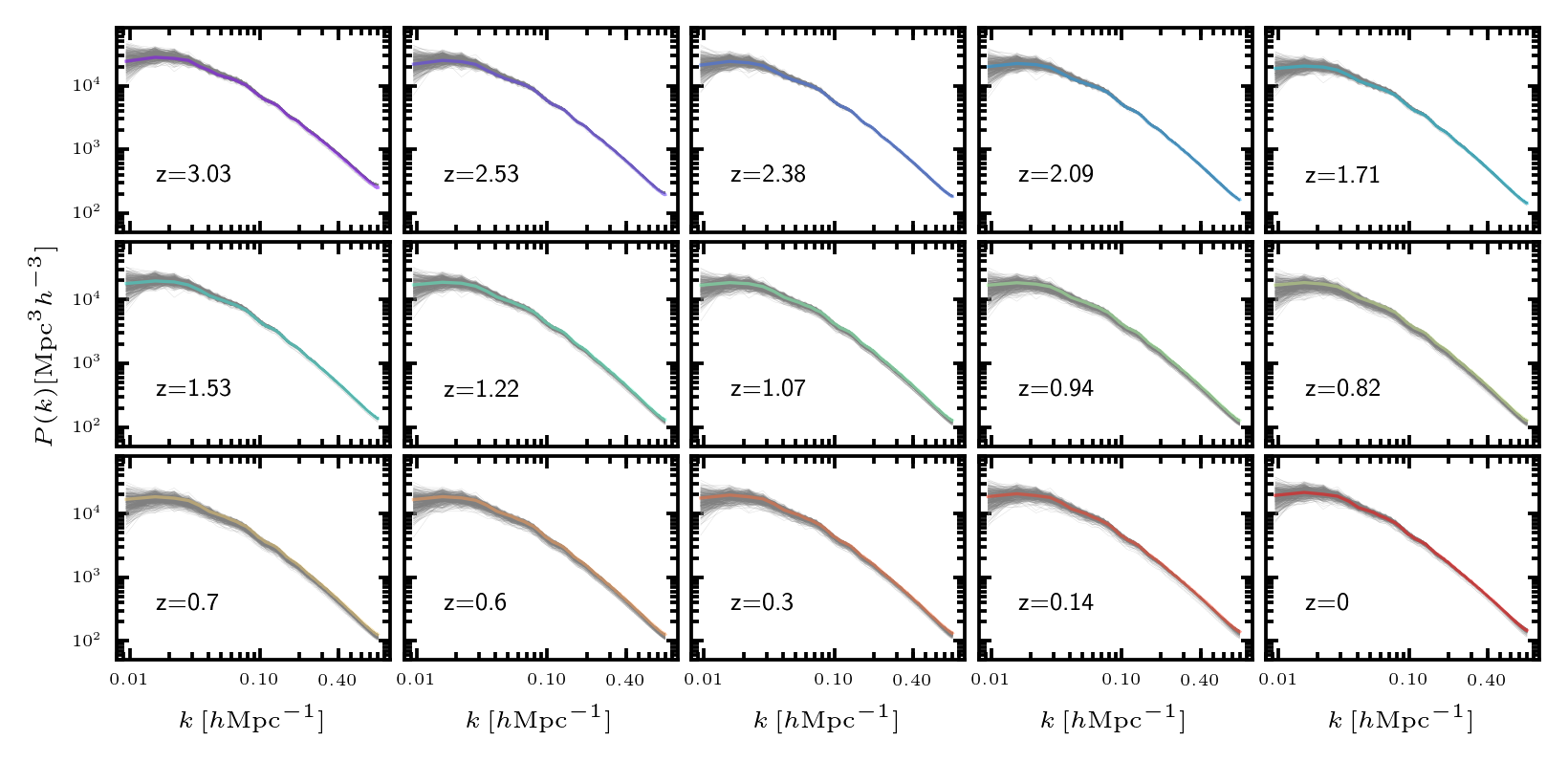}
\caption{Halo power spectra of $N_{\rm sim}=400$ \texttt{BAM-UNITsim}-mock catalogues (grey lines) at different redshifts, obtained from the assignment of number counts. In each panel, the coloured lines represent the power spectrum of the reference halo catalogue at the respective redshift.}
    \label{fig:upower}
\end{figure*}

We show how the cosmological volume probed by the \texttt{SLICS} suite has imposed a limit on the precision and accuracy in the different summary statistics of the \texttt{BAM} halo catalogues when learning from one realisation. To circumvent this situation, we pushed the algorithm to learn from more than one reference simulation, effectively increasing the volume of the training set (see \S\ref{sec:using}). Such an enlargement of the training set yields products (i.e halo bias and kernel) that, from a statistical perspective,  are not fully compatible, as the averaged kernel defined in Eq.~(\ref{eq:btot}) effectively tries to reproduce the behaviour from a fixed-amplitude reference, that is, with reduced cosmic variance, while the stacked version of the halo bias enlarges the effective volume.

As pointed out in \S\ref{sec:using}, a more convenient scenario for the calibration procedure and the generation of mock catalogues with \texttt{BAM} is the implementation of {phase-inverted} pairs of large cosmological simulations generated with fixed-amplitude initial conditions \citep[see e.g.][]{2016MNRAS.462L...1A,2019MNRAS.487...48C,2022arXiv220403868M}. In this section, we present results from numerical tests to motivate the implementation of \texttt{BAM} with that specific type of training set.

In our context, the advantages of $N$-body simulations with suppressed variance are straightforward. On one hand, the calibration with this type of IC allows the generation of kernels without any uncertainty from cosmic variance, which in turn allows an accurate extrapolation of this quantity towards smaller wavenumbers, with the aim being to generate halo catalogues in larger cosmological volumes (while learning from smaller ones). On the other hand, calibrating the method using the set of two-paired simulations (dubbed as the `{normal}' and the `{phase-inverted}') allows us to consistently extract halo bias, as in Eq.~(\ref{eq:btot}), which can then be used in conjunction with an average kernel, which (being phase-independent) is the same for the set of two paired references. This, complemented with larger cosmological volumes (and high mass resolution), provides a good statistical description of the dark matter density field \citep[e.g.][]{2006PhRvD..73f3520C,2018MNRAS.478.4602K}, and in general, of the outputs from the learning phase.

To assess the performance of \texttt{BAM} in this scenario, we used the \texttt{UNIT} simulation \cite[][]{2019MNRAS.487...48C}. This suite is represented by a set of two pair-fixed realisations in a volume of $1({\rm Gpc} h ^{-1})^{3}$, run with $4096^{3}$ dark matter particles and mass resolution of $1.2\times 10^{9}M_{\odot}h^{-1}$ and halo catalogues with a minimum (virial) mass of $3.6\times 10^{11}M_{\odot} h^{-1}$. We use a number of snapshots in the redshift range $0\leq z \leq 3$ to generate the same number of kernels and halo bias for the normal and phase-inverted realisations. 
Following the procedure outlined in \S\ref{sec:ic}, the IC of the \texttt{UNITsim} is down-sampled to a lower resolution ---in this case, $256^{3}$. For each of the snapshots, a kernel and a halo bias are produced following the steps of \S\ref{sec:iter}, based on the \texttt{TkWEB} model of halo bias (see \S\ref{sec:dmprops}). Figure \ref{fig:calibration_unit} shows the power spectrum, the kernel, and residuals at different cosmological redshifts, as obtained after the learning phase. The solid (dotted) lines in the middle panel show the kernel obtained from the normal (phase-inverted) realisations, while the bottom panel shows the residuals derived from the process with the two references.

The procedure described in \S\ref{sec:counts} is repeated to generate independent halo number counts. A number ($\sim 1000$) of ICs are generated using the \texttt{FastPM} code \citep[][]{2016MNRAS.463.2273F} with the same cosmological parameters and input power spectrum as the \texttt{UNITsim}, this time with allowing cosmic variance in the form of random Gaussian realisations.  The same number of approximated dark matter density fields are accordingly generated. Figure \ref{fig:upower} shows the power spectrum of a number of halo number-count distributions at different redshifts compared with the reference power spectrum from (one of the two) paired references. As pointed out above, the results from the calibrations of the paired set are used to generate a total bias and an averaged kernel, as dictated by Eq.~(\ref{eq:btot}). An exhaustive analysis of the summary statistics from this set of mocks is out of the scope of this paper and will be presented in forthcoming publications. However, a simple visual analysis of the different spectra at different redshifts reveals good behaviour up to the Nyquist frequency (for this setup, $\sim 0.8\,h$Mpc$^{-1}$).
Forthcoming work will be dedicated to applying the strategy described in this paper using larger simulations such as the \texttt{Abacus-summit} \citep[][]{2018ApJS..236...43G}.

\end{document}